\documentclass[twoside,12pt]{article}
\usepackage{graphicx}

\usepackage{amssymb}
\usepackage{dcolumn}  
\usepackage{bm}       
\usepackage{amsmath}
\usepackage{sidecap}  %
\usepackage{url}

\newcommand{\beqn}{\begin{equation}}
\newcommand{\eeqn}{\end{equation}}
\newcommand{\beq}{\begin{equation}}
\newcommand{\eeq}{\end{equation}}
\newcommand{\bea}{\begin{eqnarray}}
\newcommand{\eea}{\end{eqnarray}}

\newcommand{\wh}{\widehat}

\newcommand{\vlowk}{V_{{\rm low}\,k}}

\newcommand{\fmi}{\, \text{fm}^{-1}}
\newcommand{\mev}{\, \text{MeV}}

\newcommand{\la}{\langle}
\newcommand{\ra}{\rangle}

\newcommand{\wt}{\widetilde}
\newcommand{\bdelta}{\rho}

\newcommand{\Gamint}{\widetilde{\Gamma}_{\rm int}}

\newcommand{\kf}{k_{\text{F}}}

\renewcommand{\vector}[1]{{\bf #1}}
\newcommand{\vext}{v_{\rm ext}}   

\newcommand{\xvec}{\vector{\bf{x}}}

\newcommand{\rvec}{{\bf x}}  
\newcommand{\rovec}{\vector{r}}

\newcommand{\Hhat}{\widehat H}
\newcommand{\That}{\widehat T}
\newcommand{\Phibra}{\langle \Phi_0 |}
\newcommand{\Phiket}{| \Phi_0 \rangle}
\newcommand{\Psibra}{\langle \Psi_0 |}
\newcommand{\Psiket}{| \Psi_0 \rangle}
\newcommand{\PhibraKS}{\langle \Phi_{\rm KS} |}
\newcommand{\PhiketKS}{| \Phi_{\rm KS} \rangle}

\newcommand{\bfx}{{\bf x}}
\newcommand{\bfy}{{\bf y}}
\newcommand{\imax}{i_{{\rm max}}}
\newcommand{\Mstar}{M^{\ast}}
\newcommand{\Rvec}{{\bf R}}

\newcommand{\bi}{\begin{itemize}}
\newcommand{\ei}{\end{itemize}}
\newcommand{\I}{\item}
\newcommand{\yvec}{{\bf y}}
\newcommand{\Vext}{v_{\rm ext}}
\newcommand{\Vextop}{\wh V_{\rm ext}}

\newcommand{\efermi}{\varepsilon_{{\scriptscriptstyle \rm F}}}

\newcommand{\abinitio}{\textit{ab initio}}

\newcommand{\pairj}{j}
\newcommand{\psiup}{\psi_{\uparrow}}
\newcommand{\psidagup}{\psi^\dagger_{\uparrow}}
\newcommand{\psidown}{\psi_{\downarrow}}
\newcommand{\psidagdown}{\psi^\dagger_{\downarrow}}
\newcommand{\grounds}{{\rm gs}}
\newcommand{\GKS}{G_{\ks}}
\newcommand{\FKS}{F_{\ks}}
\newcommand{\ks}{{\rm ks}}
\newcommand{\VKS}{\wh V_{\rm KS}}
\newcommand{\vKS}{v_{\rm KS}}
\newcommand{\Egs}{E_{\rm gs}}
\newcommand{\rhogs}{\rho_{\rm gs}}
\newcommand{\Wint}{E_{\rm int}}
\newcommand{\rhoSL}{\rho_{\rm SL}}
\newcommand{\rhoNM}{\rho_{\rm NM}}
\newcommand{\svec}{{\bf s}}

\newcommand{\WHF}{E_{\rm HF}}
\newcommand{\openone}{\leavevmode\hbox{ \small1\normalsize\kern-.33em1}} 
\newcommand{\antisymV}{\mbox{\boldmath $\mathcal{V}$}}
\newcommand{\rone}{{\bf r}_1}
\newcommand{\rtwo}{{\bf r}_2}
\newcommand{\rthree}{{\bf r}_3}
\newcommand{\rfour}{{\bf r}_4}

\newcommand{\tr}{{\rm Tr}}
\newcommand{\rhobold}{\mbox{\boldmath $\rho$}}

\newcommand{\bDel}{\bm{\Delta}}
\newcommand{\bSig}{\bm{\Sigma}}
\newcommand{\kvec}{{\bf k}}
\newcommand{\kpvec}{{\bf k'}}
\newcommand{\qvec}{{\bf q}}
\newcommand{\pvec}{{\bf p}}
\newcommand{\qb}{\bar{q}}
\newcommand{\pb}{\bar {p}}
\newcommand{\bfcdot}{\bm{\cdot}}

\newcommand{\cc}{\mbox{c.c.}}

\newcommand{\Leuclid}{{\cal L}_{E}}
\newcommand{\nab}{\overrightarrow{\nabla}}

\newcommand{\galnab}{\stackrel{\leftrightarrow}{\nabla}}

\newcommand{\etadag}{\eta^{\dagger}}
\newcommand{\Geucl}{{\cal G}_0}

\newcommand{\tauE}{t_E}

\newcommand{\VN}{V}

\begin{document}

\title{Toward ab initio density functional theory for nuclei}

\author{J.E.\ Drut, R.J.\ Furnstahl, L. Platter\\
  Department of Physics, Ohio State University, Columbus, OH 43210\\
}

\maketitle

\begin{abstract}
We survey approaches to nonrelativistic 
density functional theory (DFT) for nuclei using
progress toward \abinitio\ DFT for Coulomb systems as a guide. 
\textit{Ab initio} DFT starts with a
microscopic Hamiltonian and is naturally formulated using
orbital-based functionals, which generalize the conventional 
local-density-plus-gradients form.
The orbitals satisfy single-particle equations with 
multiplicative (local) potentials.
The DFT functionals can be developed starting
from internucleon forces 
using
wave-function based methods or by Legendre
transform via effective actions.
We describe known and unresolved issues for applying
these formulations to the nuclear many-body
problem and
discuss how \abinitio\ approaches can help improve
empirical energy density functionals.
\end{abstract}

\emph{Keywords:} Density functional theory, nuclear structure,
  many-body perturbation theory

\tableofcontents


\section{Introduction}
 \label{sec:intro}

\subsection{Overview} \label{subsec:overview}
 
Density functional theory (DFT) has been applied to the Coulomb
many-body problem with
great phenomenological success  in predicting properties of atoms, molecules, 
and solids~\cite{DREIZLER90,Parr:1994,Koch:2001,Fiolhais:2003,Martin:2004,Kohanoff:2006}.  
DFT calculations are comparatively simple to implement yet often very
accurate and have
a computational cost that makes them at present
the only choice for systems with
large numbers of electrons~\cite{Head-Gordon:2008}.  
For these same reasons (with nucleons rather than electrons), large-scale 
collaborations of
nuclear physicists in the SciDAC UNEDF~\cite{unedf:2007,unedfweb} 
(``Universal Nuclear Energy
Density Functional'') and FIDIPRO~\cite{fidiproweb} projects, as well
as many other
individuals, 
are working on further developing DFT for the nuclear many-body problem.
Questions in astrophysics
and the advent of new experimental facilities to study nuclei at the
limits of existence, as well as societal needs, are driving multi-pronged
efforts to calculate nuclear properties
and reactions across the full table of the nuclides more accurately and reliably 
than what is currently possible
with existing energy density functional (EDF) methods 
(e.g., those based on Skyrme, Gogny, or relativistic mean-field 
functionals~\cite{Bender:2003jk}).

A principal strategy is  to
exploit the substantial and ongoing progress in \abinitio\ nuclear
structure calculations, which are primarily based on approximating
the many-nucleon wave function.  This progress is the consequence of 
synergistic advances in the
construction of internucleon interactions, in methods to calculate
properties of many-nucleon systems, and in the
ability to effectively use growing computational power~\cite{unedf:2007}.  
Because these approaches will be limited in scope for the foreseeable future,
a natural goal is to develop \abinitio\ DFT for nuclei.
In this
context, ``\abinitio'' is taken to mean a formalism
based directly on a microscopic nuclear Hamiltonian that
describes two-nucleon and few-body scattering and bound-state
observables,
in analogy to calculations in quantum chemistry or
condensed matter physics that start from the Coulomb interaction.
This contrasts with many nuclear EDF approaches~\cite{Bender:2003jk} that 
fit a functional without relying on an explicit underlying
Hamiltonian~\cite{Lacroix:2008rj}. 
Efforts to construct
bridges between \abinitio\ few-body calculations and the largely empirical
nuclear EDF's 
are bringing together diverse theorists and formal techniques
using insights from other fields.
The language and formalism differences are a barrier to progress.
We hope to lessen this barrier with this review by setting up
\abinitio\ DFT as an intermediary.
  
There are multiple possible paths to \abinitio\ DFT
and the optimal choice for describing nuclei is not clear. 
In confronting the limitations of the most widely used conventional Coulomb DFT 
implementations (such as so-called ``generalized gradient approximation''
or GGA functionals), condensed matter physicists and quantum chemists
have made extensive developments toward \abinitio\ Coulomb DFT
based on wave-function methods.
We would like to exploit these advances.
This means understanding what can be borrowed directly for nuclei and where
modifications are needed.
At the same time, DFT based on effective actions may suggest alternative
approximations as well as connections to effective field theory (EFT).
The goal of this review is to outline various strategies that are
being adopted (or may be explored soon), identify common features
and challenges,
and generally make them more accessible to the
various communities of nuclear physicists attacking these problems.
We restrict ourselves to a definition of \abinitio\ DFT 
that is consistent with 
usage in Coulomb systems (see Section~\ref{subsec:features})
but which is a subset of the full range of efforts pointing toward 
non-empirical nuclear EDF (e.g., see
Refs.~\cite{Duguet:2006jg,Duguet:2007be,Hebeler:2009dy,Duguet:2009gc}).
 
The focus on \abinitio\ DFT does not mean 
we propose abandoning the successes of the empirical EDF's, which
already achieve an accuracy for known nuclear masses that 
will be hard to reach directly with \abinitio\ functionals. 
Furthermore, it will only be possible in the near future to make
\abinitio\ calculations of
a limited subset of all nuclei.
DFT was originally formulated and is still typically described in 
terms of existence proofs.  
These proofs imply that it is \emph{possible} to find a functional (or functionals)
that depends only explicitly on the density and which is minimized
at the ground state energy with the ground state density.
While these proofs are not constructive, they can be taken to
justify empirical nuclear EDF approaches.
An important prong of the nuclear DFT effort 
seeks to make the EDF's less empirical and therefore more
reliable for extrapolation to unmeasured nuclear properties
by generalizing or constraining the functionals based on \abinitio\ input.
This can be done directly using constraints from accurate \abinitio\
nuclear structure calculations (e.g., fitting the theoretical neutron matter
equation of state) but also through insights 
from \abinitio\ DFT about the form and characteristics of
the functionals.

\subsection{Basic features/ingredients of DFT} \label{subsec:features}

Our discussion is based on 
the nuclear many-body problem formulated in terms of a
nonrelativistic Schr\"odinger equation for protons and
neutrons,
with a Hamiltonian of the form
\beqn
  \wh H_N = \wh T + \wh \VN \equiv \wh T + \wh V_{\rm NN} 
  + \wh V_{\rm NNN} + \ldots
  \;,
  \label{eq:HN}
\eeqn
where $\wh T$ is the kinetic energy and $\wh \VN$ is the sum of two-
and three- and higher-body forces in a decreasing hierarchy,
which is truncated at three-body forces in the most complete
present-day calculations.
(Effects from relativity and other degrees of freedom 
are absorbed into the potentials either explicitly or implicitly.)
Such Hamiltonians are derived in low-energy effective theories
of quantum chromodynamics (QCD) with varying degrees of
model dependence.
The development of better Hamiltonians, and of many-body forces
in particular, is an on-going enterprise~\cite{Epelbaum:2008ga}.
We emphasize that
``Hamiltonians'' is plural because there is not a unique or even preferred
form of the short-distance parts of the potentials
(the longest-ranged part, pion exchange, \emph{does} have a common,
local form in almost all potentials).
Contrast this with the electronic case, where the long-range
Coulomb potential is for many systems the entire story.

For most of our discussion it is irrelevant whether the Hamiltonian
being used 
results from a systematic effective field theory (EFT) 
expansion~\cite{Epelbaum:2008ga}
or a more phenomenological form~\cite{Wiringa:1994wb}, 
as long as it reproduces few-body observables.
What \emph{is} relevant is that the initial Hamiltonian can be transformed
(e.g., with renormalization group methods) to maintain observables
while making it more suitable for particular many-body methods (see
below).  We argue that
transformations to soften the potential
will be critical in making DFT a feasible framework
for nuclei; that is, DFT implies an organization of the many-body
problem that will not work well with all nuclear Hamiltonians.

We can classify microscopic nuclear structure methods into two broad
categories, wave function and Green's function methods.
In the former, one solves in some approximation the $A$-body
Schr\"odinger equation for the $A$-body wave function
$\Psi(x_1,\cdots,x_A)$, where $x_i$ is shorthand for all
of the variables of nucleon $i$ (e.g., ${\bf x}_i$, spin, isospin).
If the operators are known, this allows the calculation
of any nuclear observable.  Methods in this category include
Green's function%
\footnote{Despite the name, GFMC is not a Green's function method in the
sense it is used here.} and auxiliary field Monte Carlo 
(GFMC/AFMC)~\cite{Pieper:2001mp,Pieper:2004qh},
no-core shell model (NCSM)~\cite{Navratil:2009ut}, 
and coupled cluster (CC)~\cite{Hagen:2007ew,Hagen:2007hi}.
The computational cost of such calculations rises rapidly with
$A$.  Nevertheless, most of the recent progress in 
\abinitio\ nuclear structure physics has come from pushing
these techniques to higher
$A$~\cite{Pieper:2007ax,Navratil:2009ut,Hagen:2008iw}. 

Our \abinitio\ DFT discussion will connect to wave-function based formulations
that use a single-particle basis and can handle non-local interactions 
(e.g., NCSM and CC but not GFMC).
In principle, such a formulation solves
the problem of finding the ground-state energy $\Egs$ of a given
$\Hhat_N$ (for a specified number of nucleons $A$) by minimizing 
$\la \Psi | \Hhat_N | \Psi \ra$ over all normalized
anti-symmetric $A$-particle wave functions~\cite{Fiolhais:2003}:
\beqn
           \Egs = \min_{\Psi}\, \la \Psi | \Hhat_N | \Psi \ra
           \;.
           \label{eq:Egsmin}
\eeqn
In practice, of course, 
the Hilbert space (i.e., the basis size) is finite
and $\Egs$ is found approximately.
(The calculation is also not variational in many cases, such as
CC, but that is not an issue here.) 

An alternative to working with the many-body
wave function is density functional theory (DFT)
\cite{DREIZLER90,Fiolhais:2003,Argaman:2000xx}, 
which as the 
name implies, has fermion densities as the fundamental ``variables''.
We will start with
DFT as it is typically introduced, citing 
a theorem of Hohenberg and Kohn (HK)~\cite{Hohenberg:1964zz}:
There \emph{exists} an energy functional   $E_{v}[\rho]$ 
of the density $\rho(\bfx)$,%
\footnote{As discussed below, we will have multiple densities in
practice but considering the fermion density only suffices for now.}
 labeled by a (static) external
potential $\Vext(\bfx)$ such that
 \beqn
   E_{v}[\rho] 
      = F[\rho] + \int\!d{\bfx}\, \Vext({\bf x}) \rho({\bf x})
      \; ,
      \label{eq:HKtheorem}
 \eeqn
which is minimized at the ground-state energy $\Egs$ with the ground-state
density $\rhogs(\bfx)$.
An example of $\Vext$ is the electrostatic potential from ions in
atoms and molecules, as in Eq.~\eqref{eq:vextion}.
The functional $F$,  often designated $F_{\rm HK}$ in the literature,
is independent of the external potential $\Vext$ and has the same form
for any $A$.  In this sense it is said to be {\em universal\/}.
The HK theorem offers no help in constructing $F$, but is useful
in that it gives a license to search for (or guess)
approximate energy functionals. 
This would serve as justification for nuclear EDF's except for the disquieting
feature that there is no $\Vext$ for self-bound nuclei,
which makes the meaning of Eq.~\eqref{eq:HKtheorem} unclear.

A constrained search derivation~\cite{Levy:1982zz} is 
a more illuminating alternative to the proof-by-contradiction approach
to DFT used by Hohenberg and Kohn.
We start as before with the minimization
in Eq.~\eqref{eq:Egsmin}, adding an external potential,
but now we separate the minimization into two steps~\cite{Fiolhais:2003}:
\begin{enumerate}
 \I First minimize over all $\Psi$ that yield a given density $\rho(\bfx)$:
 \beqn
   \min_{\Psi\rightarrow\rho}\, \la \Psi | \Hhat_N + \wh V_{\rm ext} | \Psi \ra
   = \min_{\Psi\rightarrow\rho}\,
   \la \Psi | \That + \wh \VN | \Psi \ra
   + \int\!d{\bfx}\, \vext(\bfx) \rho(\bfx)
   \;,  
 \eeqn
 where $\wh \VN$ is the full internucleon interaction
 and $\wh V_{\rm ext} = \int\!d\rvec\, \vext(\bfx)\wh\rho(\bfx)$.
 Define $F[\rho]$ as the resulting contribution of the first term:
 \beqn
    F[\rho] \equiv 
   \la \Psi^{\rm min}_\rho | \That + \wh \VN | \Psi^{\rm min}_\rho \ra
   \;.
 \eeqn
 
 \I Then minimize over $\rho(\bfx)$:
 \beqn
   E = \min_\rho E_{v}[\rho]
     \equiv 
     \min_\rho \bigl\{ F[\rho] + \int\!d{\bfx}\, \vext(\bfx) \rho(\bfx) \bigr\}
     \;.
     \label{eq:Emin}
 \eeqn
 where the external potential $\vext(\bfx)$ is held fixed.
\end{enumerate}
In principle one works at fixed $A$ by introducing a chemical
potential $\mu$:
 \beqn
  \delta \bigl\{
  F[\rho] + \int\!d{\bfx}\, \vext(\bfx) \rho(\bfx) -
  \mu \int\!d{\bfx}\, \rho(\bfx)
   \bigr\} = 0
   \;,
 \eeqn
which implies
\beqn
   \frac{\delta F[\rho]}{\delta \rho(\bfx)} + \vext(\bfx) = \mu
   \;.
   \label{eq:mueq}
\eeqn
The chemical potential is adjusted 
until the density  
$\rho$ resulting from solving Eq.~\eqref{eq:mueq} 
yields the desired particle number  
$A$.
Or one minimizes only over $\rho$ that satisfy
$\int\!d{\bfx}\, \rho(\bfx) = A$.

As noted by Kutzelnigg~\cite{Kutzelnigg:2006aa},
the presentation of the HK theorem as an existence proof 
is often accompanied by
misleading statements such as ``all information about a
quantum mechanical ground state is contained in its electron density
$\rho$'' or that ``the energy is completely expressible in terms of
the density alone.''  
These claims seem at odds with the observation
that while the external potential energy is expressible in terms of $\rho$
(if there is a $\Vext$),
the kinetic energy is given in terms of the one-particle density matrix
and interaction energies require two-particle and higher density
matrices.  How do we reconcile this?
The key is that the
usual wave-function treatment of the many-body problem as in
Eq.~\eqref{eq:Egsmin} has in mind a \emph{single,
fixed Hamiltonian}.  
In that case,
to make a variational calculation of the ground-state
wave function $\Psi$, the energy $E$ must be made stationary to
variations in the relevant density matrices and not just the
density.  This corresponds
to variations of the normalized $A$-body wave function $\Psi$.

To understand DFT we should consider instead a \emph{family} of
Hamiltonians $\Hhat[v]$, each characterized by a potential $v$ for which we
know the corresponding ground state energy $E[v]$.  We might ask,
if we know $E[v]$, why not just evaluate at $v = \Vext$ and avoid
more complications?  But if we do know $E[v]$, then we can 
construct the functional Legendre transform~\cite{Kutzelnigg:2006aa},
\beqn
   -F[\rho] = \min_{v} \bigl\{
       \int\!d{\bfx}\, v(\bfx) \rho(\bfx) - E[v] 
      \bigr\}
      \;,
\eeqn
where the minimization is over an appropriate domain of $v$
(really the infimum or greatest lower bound)
rather than just considering a fixed $\Vext$~\cite{Kutzelnigg:2006aa}.
Thus we obtain the dependence of the internal energy on the
density.
This justifies the suggestive notation of Eq.~\eqref{eq:HKtheorem}:
If $v(\bfx)$ is set to a constant, it acts as a chemical potential
and the equation expresses a Legendre transform between two
thermodynamic potentials.
(An expanded version of the thermodynamic analogy is given
in Section~\ref{sec:lt}.)
If the Legendre transform is possible
(see Ref.~\cite{Eschrig:2003,Kutzelnigg:2006aa}), 
we can also obtain with a second Legendre
transformation that
\beqn
  E[v] = \min_{\rho}
     \bigl\{ \int\!d{\bfx}\, v(\bfx) \rho(\bfx)  + F[\rho] \bigr\}
     \equiv \min_{\rho} \bigl\{ E_v[\rho] \bigr\}
     \;,
\eeqn
which is the energy for fixed $v$ expressed as a minimization
over a trial set of densities.
Thus we reproduce Eq.~\eqref{eq:Emin} and show the origin of the HK expression in
Eq.~\eqref{eq:HKtheorem}.

This perspective shows that DFT and 
Eq.~\eqref{eq:HKtheorem} are really in the spirit of the 
other major category of microscopic nuclear structure methods,
namely Green's functions.
Instead of the many-body
wave function, the Green's function approach considers  the response of the
ground state to adding or removing 
particles~\cite{FETTER71,Negele:1988vy,Dickhoff:2005aa}.  The underlying
idea is that knowing the most general response of the ground
state (or the partition function in the presence of the most
general sources) gives a complete specification of the many-body problem.
Observables such as the ground state energy, densities,
single-particle excitations, and more can be expressed in terms
of Green's functions (with one-body operators needing the
single-particle Green's function, two-body operators generally
needing the two-particle Green's function, and so on).
The special case considered here starts with the response 
of the energy to a ``source'' $v(\bfx)$, 
which is coupled to the density.
Instead of non-local sources that individually create particles in one 
place and destroy them elsewhere, here the perturbation by the
source is a local shift in the density.
Because it is a more limited response than the usual Green's functions, 
the corresponding observables
probed are also limited, but include the ground-state energy.
A natural mathematical framework for such responses and
other Legendre transforms is the
effective action formalism using path integrals.
We give more details in Section~\ref{sec:lt} on how this works.
(This perspective also shows that taking $v=0$ is not a problem in
principle; such sources are usually set equal to zero at the end,
although in this case there are related issues with self-bound systems
such as nuclei.)

\begin{figure}[t]
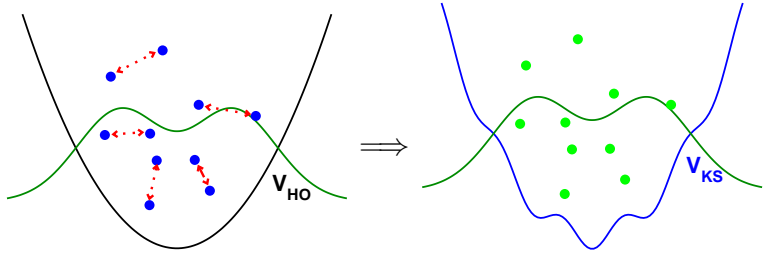

\centering
 \includegraphics*[width=1.8in,angle=0.]{kohn_sham1_new}
 \raisebox{0.5in}{$\Longrightarrow$}
 \includegraphics*[width=1.8in,angle=0.]{kohn_sham2_new}
 \caption{Kohn-Sham DFT for a $\Vext = V_{\rm HO}$ harmonic trap.  On
 the left is the interacting system and on the right the Kohn-Sham
 system.  The density profile is the same in each.}
 \label{fig:3}       
\end{figure}

In practice, DFT is rarely implemented as a pure functional of the
density,  
such as a generalized Thomas-Fermi functional, because
no one has succeeded in constructing one that yields the desired accuracy
(an immediate problem is finding an adequate functional of density
only for the kinetic energy).
Instead, the most successful procedure is to introduce single-particle 
orbitals that are used in what appears to be an auxiliary
problem, but which still leads to the minimization of the energy 
functional for the ground-state energy $E_{gs}$ and density $\rho_{gs}$.
This is called Kohn-Sham (KS) DFT, and is illustrated schematically
in Fig.~\ref{fig:3} for fermions in a harmonic trap.
The characteristic feature is that the interacting density for $A$ fermions 
in the external potential $\Vext$ is equal (by construction)
to the non-interacting density in another single-particle potential.
This is achieved by orbitals $\{\phi_i(\bfx)\}$
in the \emph{local} potential $\vKS([\rho],\bfx) \equiv \vKS(\bfx)$,
which are solutions to 
 \beqn
   [-\bm{\nabla}^2/2m + \vKS(\bfx)]\phi_i(\bfx)
   = \varepsilon_i\phi_i(\bfx)
   \;,
   \label{eq:kseq}
\eeqn
and determine the density by
\beqn    
     \rho(\bfx) = \sum_{i} n_i |\phi_i(\bfx)|^2 
     = \sum_{i=1}^A |\phi_i(\bfx)|^2
   \;,
 \eeqn
where the sum is over the lowest $A$ states with $n_i =
\theta(\efermi-\varepsilon_i)$ here.
When we include pairing, the sum is generalized to be over all
orbitals with appropriate occupation numbers (see
Section~\ref{subsec:pairing}).
The magic Kohn-Sham potential $\vKS([\rho],\bfx)$ 
is in turn determined from  $\delta E_{v}[\rho]/\delta \rho(\bfx)$
(see below).
Thus the Kohn-Sham orbitals depend on the potential, which depends
on the density, which depends on the orbitals, so we must 
solve self-consistently (for example, by iterating until convergence).
We will return later to address the meaning of the KS eigenvalues
$\varepsilon_i$.  The ground-state energy is $E_{\rm gs} = E_{v_{\rm
ext}}[\rho_{\rm gs}]$. 
 
We will define orbital-based
density functional theory (DFT) broadly as \emph{any} many-body 
method based on a local (``multiplicative'') background potential
(what we called $\vKS$ above)
used to calculate the ground-state energy
and density of inhomogeneous systems in the manner just described.
That is, there will be a single-particle, non-interacting component of the
problem that involves solving for orbitals with a local (diagonal
in coordinate space) potential.
We will also require that there are no corrections to the
density obtained from these occupied orbitals (as in usual
Kohn-Sham DFT).
It is not obvious at this point that this is a necessary
feature, because it is not essential to the numerical simplicity
or good scaling behavior. 
We will see in later sections how it arises.
This characterization of DFT can be realized in seemingly very
different approaches, such as a particular organization of
(possibly resummed)
many-body perturbation theory (MBPT) and
effective actions for composite
operators (based on functional
Legendre transformations).  
The DFT formalism is often said to be a mean-field approach because
of the Kohn-Sham potential and this applies to our general
definition as well.  The point is that it is not a mean-field
\emph{approximation} but an organization that takes a mean-field
state as a reference state, which if solved completely includes
all many-body correlations.  (The real issue is how much correlation
is included in a given approximation to the exact functional.)

\begin{figure}[tbh-]
\centering
 \includegraphics*[width=2.8in,angle=0.]{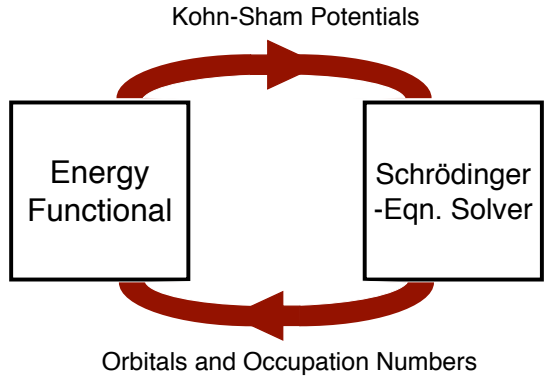}
 \caption{Generic self-consistency cycle
 for Kohn-Sham DFT.  The energy functional takes orbital wave
 functions and eigenvalues (with occupation numbers) as inputs.  
 The outputs are the local Kohn-Sham potentials from the functional
 derivative of the energy functional.  These could be directly
 evaluated as with a Skyrme or density matrix expansion functional,
 or solved from OEP equations.}
 \label{fig:KScycle}       
\end{figure}

Implementations of orbital-based DFT will have a self-consistency
cycle of the form shown in Fig.~\ref{fig:KScycle}.
The code that solves the KS single-particle Schr\"odinger equations
(``Schr\"odinger-Eqn. Solver'') can
be generic (if generalized to include pairing) because the
same equations are solved for different energy functionals
(only the potentials change).
This part of the calculation is generally
the key to the computational scaling because the cost goes
up gently with $A$ and it also means that we can adapt
the well-developed tools used for Skyrme calculations.
The ``Energy Functional'' will be particular to the implementation, 
ranging from simple
function evaluation to solving complicated integral equations.
If the cost of evaluating this box, which includes calculating
the functional derivative defining $\vKS$, 
can be kept under control, the computational advantage
will hold. 
(Conversely, in full calculations of orbital-based DFT, one
has to consider seriously the computational scaling when comparing
to alternative strategies.) 
It is often the case that the energy functional is taken to be of the
local or semi-local%
\footnote{``Semi-local'' in this context means that the energy density
at $\rvec$ depends only on the electron density and orbitals in an infinitesimal
neighborhood of $\rvec$~\cite{Perdew:2009}.
So $\mathcal{E}$ has only a finite number of gradients.}
form:
\beqn
  E[\rho_i,\tau_i,\ldots] = \int\!d{\bfx}\, 
      \mathcal{E}(\rho_i(\bfx),\tau_i(\bfx),\ldots) 
      \;,
      \label{eq:Edensity} 
\eeqn
where we have allowed different types of densities such as the
kinetic energy density $\tau_i$ and
where dependence on gradients of the densities is also allowed.
For example, the Skyrme functional in Eq.~\eqref{eq:ESHF} has this
structure.
We emphasize that 
\emph{this is not a general form}, but requires
significant approximations to be
derived from a non-local
orbital-based functional
(such as by applying the density matrix expansion, see
Section~\ref{subsec:dme}).

We also emphasize that the Kohn-Sham 
\emph{potentials} $\vKS$ are always local no matter
how non-local the energy density becomes.
This is different from some nuclear EDF
approaches that feature finite-range effective interactions
in the form of a Hartree-Fock functional (e.g., Gogny), 
for which the single-particle
equations have non-local exchange potentials~\cite{Bender:2003jk}.
While the locality of Kohn-Sham potentials eases the computational
burden, it is a constraint that may ultimately prove to be too limiting
for nuclear energy functionals~\cite{Duguet:2006jg,Duguet:2009gc}.
 
We can separate our subsequent discussion of \abinitio\
DFT into two parts:
\begin{enumerate}
 \I Given an energy functional, what is the associated Kohn-Sham
 potential?
 \I How do we construct an \abinitio\ energy functional systematically?
\end{enumerate} 
The second question is more fundamental but the first one 
is more generic, so we begin with it in
Section~\ref{sec:dftoep};
two approaches to  
the second question are described in  Sections~\ref{sec:abinitio} and
\ref{sec:lt}. 
We stress that Kohn-Sham DFT, with orbitals, is in fact a natural
 development in each of these approaches, rather than simply a ``trick''
 to better approximate the kinetic energy.
Further, we find that keeping the densities 
fixed entirely from the Kohn-Sham orbitals either follows as a consequence
of the DFT minimization conditions
or can be used as an imposed condition to derive a functional.
In our working definition of DFT
we allow complete freedom in defining the Kohn-Sham system,
which permits more physics to be shifted into the potential (``mean
field''), making the DFT functional more effective
(e.g., a perturbative expansion will converge more rapidly).
One way of doing this is to
allow \emph{any} local densities paired with corresponding sources.

 \subsection{Coulomb vs.\ nuclear DFT} \label{subsec:coulomb} 

We will rely heavily on the progress made in \abinitio\ DFT for
Coulomb systems, but we should always
keep in mind the differences between
Coulomb and nuclear many-body problems, which will introduce
substantial challenges.
For most Coulomb applications, the Hamiltonian is well known and takes a simple,
two-body local form (that is, it is diagonal in coordinate
representation and has no spin dependence).  
While in principle one could modify the interaction
at short distances, e.g., with unitary transformations, the original
local form is clearly preferred.  
Because the interaction is to good approximation
$1/r$, it does not make sense to transform it.

The $1/r$ potential follows in a straightforward manner from the
underlying theory of quantum electrodynamics (QED).%
\footnote{If heavier atoms are being considered there will be 
in practice a more
involved potential and possibly the need to treat the system
fully relativistically.} 
A directly analogous \abinitio\ calculation
of the strong interaction would have to start with the quark and gluon
interactions of quantum chromodynamics (QCD).  
But because
quarks and gluons are not efficient low-energy degrees of freedom because of
confinement, 
low-energy effective theories of QCD are used to construct
interactions between protons and neutrons.
These interactions may be systematic (e.g., using EFT) or more
phenomenological.
But as effective interactions they are not unique and transformations
may result in forms more amenable to the DFT formulation.

The weak strength and long range
of the Coulomb potential means that the binding energies
of atoms and molecules are numerically dominated by the Hartree 
contribution~\cite{Kohanoff:2006}.
This dominance would make the problem simple except that while
the exchange-correlation energy (what is beyond Hartree) is a 
often a small
fraction of the total binding energy of atoms, molecules, and solids,
it is of the same size as the chemical bonding or atomization
energy~\cite{Fiolhais:2003}.
Thus an accurate DFT functional for this contribution
(called $E_{\rm xc}$) is essential and this is the principal
challenge of Coulomb DFT,
although there are complications in certain electron systems (e.g.,
from pseudopotentials, relativity, etc.).

It might be supposed that the dominance of 
Hartree-Fock contributions to the energy,
so that correlations are small corrections treatable in
(possibly resummed) perturbation theory, is an important reason why
Coulomb DFT works so well.  In contrast, for typical realistic NN interactions,
correlations are much greater than the Hartree-Fock contribution!
This may mean that DFT fails for these Hamiltonians.
But we can use the possibility of modifying the interactions
by renormalization group methods~\cite{Bogner:2003wn,Bogner:2006pc} 
(and related methods~\cite{Roth:2005pd,Roth:2005ah})
to change the ``perturbativeness'' of nuclei.    
There are several sources of nonperturbative physics for 
typical nucleon-nucleon interactions:%
\footnote{There is also pairing, which we consider separately.} 
\begin{enumerate}    
 \item
  Strong short-range repulsion (``hard-core'' for short).
 \item Iterated tensor interactions (e.g., from pion
 exchange).
 \item
 Near zero-energy bound states (e.g., the deuteron and near bound
 state
 in the $^1$S$_0$ channel).
\end{enumerate}
However, the first two sources depend on 
the resolution (i.e., the degree of coupling to high-energy physics), and
the third one is affected by Pauli blocking.
Thus we can use the freedom of low-energy theories to simplify
calculations by lowering the resolution, which softens
the potential and makes the nonperturbative nuclear physics
more perturbative.
That is, the convergence rate of perturbative
expansions such as the Born series for free-space scattering or the
in-medium sum of particle-particle ladder diagrams improves at
lower resolution. This is demonstrated using a quantitative measure
of the convergence in Ref.~\cite{Bogner:2005sn}.

In general, transformations to low-resolution interactions that are
compatible with the natural resolution scale of nuclear bound
states make the nuclear problem look in some ways more like the Coulomb
problem.
Not entirely, of course, but generally much more perturbative with weaker
tensor correlations.
For example, Hartree-Fock plus second-order many-body perturbation
theory may be well converged for the bulk energy
of the uniform system~\cite{Bogner:2005sn,Bogner:2009un}.
Residual issues include  how to handle
contributions from mixing into the ground
state of low-lying excitations in finite
nuclei, which could require a nonperturbative
treatment.
In addition, we might expect that such contributions to the energy
functional are significantly non-local, which may be why empirical
nuclear EDF's (which are semi-local) 
have problems incorporating this physics and typically rely on
additional procedures to handle these corrections~\cite{Bender:2005vy}.  
Testing whether
orbital-based DFT can accommodate this physics will be an important
area for study.

The conventional DFT for Coulomb systems 
takes as its starting point
precision calculations of the uniform electron gas.
These are \abinitio\ numerical calculations (e.g., using GFMC) combined 
with well-controlled analytic limits. 
This solid starting point for a local density approximation (LDA)
is combined with constrained gradient terms to construct
semi-local functionals (GGA) that are not microscopic
but are also not fit to data.%
\footnote{These are called
``non-empirical''~\cite{Perdew:1986,Perdew:2003} because the
additional parameters are determined by constraints rather than
data; quantum
chemists also construct empirical Kohn-Sham DFT functionals that
are less microscopic.} 
For any DFT, the 
uniform system is special, because $E$ vs.\ $\rho$ is a limiting
case for the functional that is an observable function.
(In general, the functional evaluated with a density that is not
at the minimum is not an observable, see Section~\ref{subsec:analogy}.)  
At present, \abinitio\ nuclear calculations
of the uniform system with equal numbers of protons and neutrons
(``symmetric nuclear matter'') are much less controlled than
the electron gas counterpart.
As a result, present-day
nuclear EDF's are often purely empirical (e.g., most Skyrme
interactions): a parametrized form
has constants fit to properties such as binding energies and
radii of a set of nuclei.
Indeed, the most stringent constraints on the saturation properties
of symmetric nuclear matter come from taking the uniform limit
of the fitted EDF's.
Other nuclear EDF's take guidance from the best possible
nuclear matter calculations but
then fine tune to yield quantitatively accurate properties of
finite nuclei (e.g., see Refs.~\cite{Chappert:2008zz,Goriely:2009zz} 
and references cited therein).
  
The calculation of nuclei also has features without direct parallel
in the basic Coulomb systems, which complicates the
formulation of DFT.  Perhaps most apparent is the role of symmetries.
Atoms and molecules can generally be treated in 
Born-Oppenheimer approximation,
 where the slow nuclear degrees of
freedom provide an external (Coulomb) potential (this is $\Vext$!) for
the fast electronic degrees of freedom. 
The basic Coulomb problem for DFT is to find the minimal
energy given a configuration of nuclei, treated as fixed
in space, and the related
distribution of electrons (the density or, more generally,
the spin densities).
This external potential
means that symmetries of the Hamiltonian such as translational
and rotational invariance are not realized in the physical ground
state.  In contrast, ordinary nuclei are self-bound and should
reflect these symmetries.  

But the nuclear Kohn-Sham potential $\vKS(\bfx)$ breaks these
symmetries, so that the system being calculated is an
intrinsic ``deformed'' state.
This is familiar from any mean-field based calculation of
nuclei~\cite{Blaizot:1985,Ring:2005};
in general, the organization of the problem about a mean-field
inevitably breaks symmetries.
Handling the restoration of symmetries is well known from a wave-function
point of view through the use of projection.  
How we should deal with it for \abinitio\ DFT has only recently 
been considered
and may require significant 
developments~\cite{Lacroix:2008rj,Bender:2008rn,Duguet:2008rr}.
The fact that nuclei are self-bound presents not only practical
problems but conceptual
problems, because there is no external field in the
case of interest.
In the face of symmetry breaking, is the Kohn-Sham approach
even well defined? 
This issue has been considered by various authors recently and
will be reviewed in Section~\ref{sec:symmetry}.  This is a situation where an
alternative formulation, in this case using effective actions, lends
a somewhat complementary interpretation that can suggest different ideas
for approximation.

Bulk nuclear matter is superfluid and the treatment of pairing is
found to be crucial in the accurate reproduction
 of experimental trends
in finite nuclei by empirical energy density functionals.
For Skyrme-type EDF's, 
pairing has been accomodated by generalizing from zero-range
Hartree-Fock equations  (equivalent to Hartree) to zero-range
Hartree-Fock-Bogoliubov equations~\cite{Dobaczewski:2001ed}, with
empirical fits of the pairing parameters.
This requires the inclusion of ``anomalous'' densities,
which is not physics that arises for atoms and molecules,
and the use of \emph{local} anomalous potentials leads to divergences.
Bulgac and collaborators~\cite{Bulgac:2001ai,Bulgac:2001ei} have clarified
the corresponding renormalization issues  and developed procedures that 
could be incorporated into \abinitio\ orbital-based DFT (see
Section~\ref{subsec:pairing}).
An alternative path has been followed for Coulomb DFT applied to
bulk supercondutivity, where non-local Kohn-Sham potentials
avoid the new divergences.
Similarly, nuclear EDF's with finite-range pairing potentials (e.g., Gogny) 
and recent non-empirical pairing based on 
low-momentum microscopic
interactions in non-local form~\cite{Hebeler:2009dy,Duguet:2009gc}
have natural high-momentum cutoffs.
While we focus in this review on methods with local Kohn-Sham potentials,
we stress that the best way to incorporate pairing is an open question.

All of these features of nuclei
impact the energy functional at the same
level of accuracy as we are trying to achieve, so they cannot be
ignored. This further motivates a multi-pronged attack to be able
to have complementary calculations and to cross-check results,
as well as the need to be open to alternatives to strict orbital-based
DFT.

 \subsection{Scope and plan of review}  \label{subsec:roadmap}

Our target audience for this review
spans several communities of nuclear physicists.
Those who do wave-function based many-body theory, such as
coupled cluster, are likely most familiar with second quantization
formalism and many-body perturbation theory (MBPT).
The practitioners of effective field theory, who often come from a
particle physics background, are generally more fluent in the language
of path integrals and effective actions (although probably
not in the form directly analogous to DFT).
The users of EDF's are experts in the language and techniques of
mean-field approximations and broken symmetries,  dealing with
pairing and the like.
Rather than concentrate on only one of these formalisms, we consider
both effective action and  many-body
perturbation theory perspectives as well as discuss
how to make empirical approaches less empirical.
We believe this will make each more understandable, help focus on
the key issues, and suggest approximations and generalizations. 

However,
it is clear that a complete treatment of these various formalisms would
require detailed reviews on each topic.  
Because of space limitations,
our discussions will necessarily be rather schematic, but we will
indicate where to find more details in the literature.  
Fortunately, there are many
fine articles targeted at the Coulomb many-body problem that can
fill in details.
We hope that our treatment here will make these articles more accessible to
a nuclear physics audience (or, more precisely, audiences).
 
Our intention is to provide a guide to possible pathways to
improved DFT for nuclei based on the \abinitio\ ideas that result in
local Kohn-Sham potentials
rather than to make an assessment of the current status
of alternative approaches. 
Therefore
we provide limited details on the successes and problems with
present-day DFT (or EDF) for nuclei, except for pointers to the literature.  
We also omit various topics, including
the many developments in covariant density functional theory,
which builds upon the phenomenologically successful relativistic
mean-field calculations,
time-dependent DFT, the superfluid local density approximation (SLDA),
and
efforts to analyze and improve EDF's without reference to an underlying
Hamiltonian.
At the end we return
to provide brief perspectives on some of these topics.  

The plan of the review is as follows.
In Section~\ref{sec:dftoep}, the formalism for orbital-based
DFT is derived several ways to help clarify its
nature. 
We also present simplifying
approximations that have proven notably effective in Coulomb
applications.
In Section~\ref{sec:abinitio}, the connection to \abinitio\ wave 
function methods is explored through conventional nuclear MBPT
and an improved perturbation approach for Coulomb DFT.
The critical role of low-momentum potentials to make MBPT for
nuclei viable and how this can be exploited in a semi-local expansion
is also examined.
The idea of DFT based on Legendre transformations as
formulated using effective actions is reviewed in 
Section~\ref{sec:lt}, with connections to effective field theory (EFT)  
for DFT and conventional Green's function methods.
The issues of symmetry breaking for self-bound systems, pairing, and
single-particle energies 
are addressed in Section~\ref{sec:nuclear},  
along with an overview of efforts to make empirical nuclear EDF's 
closer to \abinitio.
Section~\ref{sec:summary} summarizes our perspective on
the paths to \abinitio\ DFT and points out some alternative routes.


\section{Orbital-based DFT}
  \label{sec:dftoep}

In this section, we introduce the general motivation for orbital-based
DFT using the experience of Coulomb systems as a guide
and derive in several ways how to go from an energy functional
to multiplicative (local) Kohn-Sham potentials within
the optimized effective potential (OEP) method. 
We rely heavily on the reviews
by Engel, which appears in Ref.~\cite{Fiolhais:2003},
and by K\"ummel and Kronik~\cite{Kummel:2008}
(see also Refs.~\cite{Gorling:2005aa,Kohanoff:2006} and Kurth/Pittalis in
Ref.~\cite{Grotendorst:2006}), but annotate the presentation
with comments on differences expected for applications to nuclei.
We defer discussion of the \emph{construction} of 
appropriate energy functionals to later sections.
To provide a point of comparison for the OEP method,
we start with a brief review of the Hartree-Fock (HF) approximation
in coordinate representation, which 
is the simplest version of the wave-function microscopic approach.
While the exchange-only OEP has close similarities to HF, there are important
distinctions that persist when OEP includes correlations. 

For simplicity, we present formulas as if the microscopic
interactions were always finite-ranged but local
(i.e., diagonal in coordinate representation), spin-isospin
independent,
and two-body only.
We caution the reader 
that for the nuclear case
we will have to deal with non-local forces and (at least)
three-body forces, all with spin-isospin dependence (see
Section~\ref{subsec:dme} for examples of these generalizations).

\subsection{Hartree-Fock in coordinate representation}

The distinction between the usual wave-function description based
on many-body perturbation theory (resummed or not) and orbital-based
DFT is already evident with Hartree-Fock.
As noted earlier, HF is a natural starting place for any Coulomb
calculation and, with low-momentum interactions, the same is now
true for nuclei (although Hartree-Fock plus second-order may be
a fairer comparison).
Further, most nuclear EDF's have been viewed at least originally
as arising from Hartree-Fock calculations of effective interactions
(which might be zero-ranged as with Skyrme or finite-ranged as with
Gogny)~\cite{Bender:2003jk}.

Hartree-Fock is the simplest approximate 
realization of Eq.~\eqref{eq:Egsmin}, with
single Slater determinants of orbitals $\phi_i$:
   \beqn
     | \Psi_{\rm HF} \rangle \longrightarrow 
     \det\{\phi_i(\xvec), i=1\cdots A  \}
     \; 
   \eeqn
as the wave functions over which the energy is minimized.   
The spin-isospin dependence of the nuclear interaction is critical
to the physics but not to the structure of the equations, where
it is a distraction.  So   
in order to keep the notation from getting cumbersome, through
much of this review we let $\bfx$ represent 
the coordinate variable and, when relevant, also the spin-isospin
indices.  (Similarly, $\int\!d\rvec$ includes a summation over
spin and isospin when relevant.)

The Hartree-Fock energy in coordinate representation 
for a \emph{local} two-body potential $V(\xvec,\yvec)$
in the presence of an external potential
is~\cite{Ring:2005}
  \bea
   \langle  \Psi_{\rm HF} | \wh H | \Psi_{\rm HF} \rangle
     &=&
     \sum_{i=1}^A \frac{\hbar^2}{2M} \int\!\! d\xvec\,
        \bm{\nabla}\phi^\dagger_i\bm{\cdot}\bm{\nabla}\phi_i
   {+ \frac12\sum_{i,j=1}^A
     \int\!\! d\xvec \! \int\!\! d\yvec \, 
      |\phi_i(\xvec)|^2 V(\xvec,\yvec) |\phi_j(\yvec)|^2} 
   \nonumber \\ & & 
   \null {- \frac12\sum_{i,j=1}^A
     \int\!\! d\xvec \! \int\!\! d\yvec \, 
      \phi^\dagger_i(\xvec) \phi_i(\yvec) V(\xvec,\yvec) 
      \phi^\dagger_j(\yvec)\phi_j(\xvec) } 
    + \sum_{i=1}^A  \int\!\! d\yvec\, \Vext(\yvec) |\phi_j(\yvec)|^2
    \; , 
    \label{eq:HFenergy}
  \eea
where the sums are over occupied states.
Note that this energy functional of the orbitals is non-local
in that there are integrals over $\xvec$ \emph{and} $\yvec$.
The minimization of the energy is achieved by a variation 
of Eq.~\eqref{eq:HFenergy} with respect to the $\phi_i$:
\beqn
        \frac{\delta}{\delta \phi_i^\ast({\bf x})} 
	\Bigl(
	    \langle  \Psi_{\rm HF} | \wh H | \Psi_{\rm HF} \rangle
           - \sum_{j=1}^A \varepsilon_j 
	   \int\! d\yvec\, |\phi_j(\yvec)|^2
	\Bigr) = 0 \;,
\eeqn 
with the HF eigenvalues $\varepsilon_j$ introduced as Lagrange multipliers
to constrain the orbitals to be normalized.
There are no other subsidiary conditions, such as imposed below in the OEP.
The result is the familiar coordinate-space equation for $\phi_i(\bfx)$:
\beqn
  -\frac{\hbar^2}{2M} \bm{\nabla}^2\phi_i(\bfx)
    + \sum_{j=1}^A
     \int\! d{\bfy}\, V(\bfx,\bfy)
     \phi^\ast_j(\bfy)\bigl\{
      \phi_j(\bfy)\phi_i(\bfx) - \phi_j(\bfx)\phi_i(\bfy)
     \bigr\}
     = \varepsilon_i \phi_i(\bfx)
     \;,
     \label{eq:HFeq}
\eeqn
or, after defining the local Hartree potential
and the non-local exchange or Fock potential,
\bea
  \Gamma_{\rm H}(\bfx) &\equiv&
    \int\!d{\bfy}\, V(\bfx,\bfy) \sum_{j=1}^A |\phi_j(\bfy)|^2
       = \int\!d{\bfy}\, V(\bfx,\bfy) \rho(\bfy) 
   \\
  \Gamma_{\rm F}(\bfx) &\equiv& 
  - V(\bfx,\bfy)  \sum_{j=1}^A \phi_j^\ast(\bfy) \phi_j(\bfx)
     = - V(\bfx,\bfy) \rho(\bfx,\bfy)
     \;,
\eea
we find the non-local Schr\"odinger equation:
\beqn
  \Bigl\{
   -\frac{\hbar^2}{2M} \bm{\nabla}^2 + \Gamma_{\rm H}(\bfx) 
   \Bigr\}\phi_i(\bfx)
    +  \int\! d{\bfy}\, \Gamma_{\rm F}(\bfx,\bfy)
     \phi_i(\bfy)
     = \varepsilon_i \phi_i(\bfx)
     \;.
     \label{eq:nlhf}
\eeqn
The equations for the Hartree-Fock orbitals are solved with the
same self-consistency cycle as in Fig.~\ref{fig:KScycle}.
This is more involved than solving the
Kohn-Sham equations \eqref{eq:kseq} because
of the non-local Fock potential, but it is drastically simpler than
solving for the full wave function.
(A typical numerical solution method is to introduce a basis, 
e.g. the harmonic oscillator basis, which
reduces the calculation to a straightforward linear algebra problem.)

The coordinate-space HF equations 
become significantly
more complicated with non-local NN potentials,
such as the soft low-momentum NN potentials, and with three-body potentials.
With non-local potentials,
even the Hartree piece is no longer a multiplicative potential.
Indeed, working in coordinate representation is not particularly
natural in this case, which may raise questions about the
appropriateness of the DFT focus on locality~\cite{Bertolli:2008uq}.
The treatment of non-local, momentum-space potentials at HF is
outlined below in Section~\ref{subsec:dme}.

At the opposite extreme, if the interaction is taken to be zero ranged
(contact interactions) so that $V(\bfx,\bfy) \rightarrow
V(\bfx)\delta(\bfx-\bfy)$, then the Hartree and Fock
terms reduce to the same multiplicative form.
This is why ``Skyrme-Hartree-Fock'' is not only equivalent to
a Hartree functional, but has the same form as Kohn-Sham functionals
that include correlations beyond exchange.
This is also why a ladder of approximations leading to full
orbital-based DFT (as described in Section~\ref{sec:summary})
naturally involves Skyrme functionals on the lower rungs.
(However, we emphasize that the microscopic functionals described
below \emph{do not} assume zero-range interactions.)

Finally, we comment on the interpretation of orbital eigenvalues.
The Kohn-Sham eigenvalues in Eq.~\eqref{eq:kseq} appear to be analogs of the 
HF orbital energies in Eq.~\eqref{eq:nlhf}.  The latter are well-defined
approximations to the separation energies:
  \bea
      E_\alpha^{A+1} - E_0^{A}  \quad&\mbox{for}&\quad \alpha > A
        \quad \mbox{(particle)}
        \\
      E_0^{A} - E_\alpha^{A-1}  \quad&\mbox{for}&\quad
      \alpha\leq A
        \quad \mbox{(hole)}
      \;.  
  \eea
The interpretation of the corresponding KS eigenvalues
that appear in the orbital-based DFT treatment is less clear,
as there is no Koopman's theorem to identify the difference of
$A+1$ and $A$ body ground-state energies with the 
eigenvalues~\cite{Kummel:2008}.
However, G\"orling~\cite{Gorling:1996aa} has shown that 
differences of Kohn-Sham eigenvalues are well-defined approximations 
to excitation energies.
In addition, Janak's theorem 
holds that the ionization potential
equal to the chemical potential is given by the highest
occupied KS eigenvalue~\cite{Kohanoff:2006}.
(This follows in most cases simply from the large-distance fall-off of the 
physical density as dictated by the contribution of the
least bound particle,
whose wave function
falls off according to its energy.)
The viability of a physical interpretation for the KS eigenvalues
is an important topic for nuclear DFT and we return to it below.

\subsection{Motivation for orbital-dependent functionals}

The motivation given in the literature
for orbital-dependent DFT functionals for Coulomb
systems is based on some characteristic failures of
semi-local functionals as well as the need for greater
accuracy in some applications.  
Both of these points are relevant to nuclear DFT, where
EDF's of the Skyrme type are also semi-local and where
greater accuracy is sought globally and greater reliability
is sought for extrapolations.
Before considering the motivation
in more detail,
we first review some of the standard Coulomb DFT formalism, indicating
where our treatment for nuclei will differ.
Following Engel~\cite{Fiolhais:2003},
we restrict our discussion to non-relativistic, time-independent,
spin-saturated systems;  this is for simplicity and is
not a limitation of the formalism.  

The total energy functional is generally decomposed for Coulomb
systems as
(except that $n$ is typically used for density instead of $\rho$;
other details of the notation also vary in the literature) 
\beqn
  E_{\rm tot}[\rho] = T_s[\rho] + E_{\rm ext}[\rho] + E_{\rm H}[\rho]
     + E_{\rm xc}[\rho]
     \;,
    \label{eq:Etot} 
\eeqn
where $T_s$ is the KS kinetic energy, $E_{\rm ext}$ is the external 
potential energy, $E_{\rm H}$ is the Hartree energy, and
$E_{\rm xc}$ (``xc'' stands for ``exchange-correlation'')
is defined to be everything that is left over; i.e.,
it is \emph{implicitly} defined by specifying the other pieces.
(Note: We have omitted a piece describing the energy of the ions themselves.)  
For Coulomb systems with just a $1/r$ potential there are explicit expressions
in terms of KS orbitals for all but $E_{\rm xc}$, namely
\beqn
  T_s[\rho] = - \frac{1}{2m} \sum_{i=1}^A 
     \int\! d{\bfx} \, \phi_i^\dagger(\bfx) \nabla^2 \phi_i(\bfx)
     \;,
\eeqn
\beqn
  E_{\rm ext}[\rho] = \int\! d{\bfx} \, v_{\rm ext}(\bfx)\, \rho(\bfx)
    \;,
\eeqn
\beqn
  E_{\rm H}[\rho] =  \frac{1}{2}\int\! d{\bfx} \int\! d{\bfy} \,
       \rho(\bfx) V_c(\bfx,\bfy) \rho(\bfy)
     \quad\mbox{where}\quad  
     V_c(\bfx,\bfy) = \frac{e^2}{|\bfx - \bfy|}
     \;.
\eeqn
As before (and throughout this review), these orbitals satisfy 
\beqn
  \left[
    -\frac{\nabla^2}{2m} + \vKS(\bfx)
  \right] \phi_i(\bfx) = \varepsilon_i \phi_i(\bfx)
   \;,
   \label{eq:KSorbital}
\eeqn
where the KS potential $\vKS(\bfx) = \delta (E_{\rm ext}+E_{\rm H} + E_{\rm
xc})/\delta \rho(\bfx)$.%
\footnote{The KS potential, density, etc.\ are often denoted
$v_s$, $\rho_s$, and so on in the Coulomb DFT literature.}
The signature features for \abinitio\ DFT (in our broad definition)
is that $\vKS(\bfx)$ appears multiplicatively in the
KS equation and that the density is given completely by summing up
the occupied (defined as the energetically lowest here) KS states:
\beqn
   \rho(\bfx) = \sum_i  n_i \, |\phi_i(\bfx)|^2
   \;.
\eeqn
At finite temperature or when pairing is introduced, the sum
will be extended to all orbitals with appropriate occupation
numbers $n_i$ (see Section~\ref{subsec:pairing}).

Given Eq.~\eqref{eq:Etot},
the KS potential $\vKS(\bfx)$ is the sum of three pieces,
\beqn
   \vKS(\bfx) = v_{\rm ext}(\bfx) + v_{\rm H}(\bfx) 
                  + v_{\rm xc}(\bfx)
   \;,
\eeqn
where
\beqn
  \Vext(\bfx) = -\sum_{\alpha=1}^{N_{\rm ion}} \frac{Z_\alpha
  e^2}{|\bfx - \bfx_\alpha|}
  \;,
  \label{eq:vextion}
\eeqn
\beqn
   v_{\rm H}(\bfx) = \frac{\delta E_{\rm H}[\rho]}{\delta \rho(\bfx)}
     = e^2 \int\! d{\bfx}' \frac{\rho(\bfx')}{|\bfx-\bfx'|}
   \;,
\eeqn
and
\beqn
  v_{\rm xc}(\bfx) = \frac{\delta E_{\rm xc}[\rho]}{\delta \rho(\bfx)}
  \;.
  \label{eq:vxc}
\eeqn
These formulas for $\Vext(\bfx)$ and $v_{\rm H}(\bfx)$ are particular
to the Coulomb problem, 
but the structure of these terms is more general, as is the expression
of $v_{\rm xc}(\bfx)$ as a functional derivative.

Some comments about these formulas:
\bi
  \I
While real nuclei usually do not have external potentials,
it can be useful to theoretically put the nucleus in a trap for comparisons of
empirical functionals to \abinitio\ calculations.
However, the external potential should be viewed more generally
as a \emph{source} that is varied and then set to zero at the end,
such as those used in field theory with path integrals.
Therefore we can add more general sources coupled to local
densities, such as the kinetic energy, spin-orbit, and pairing
densities found in Skyrme EDF theory and expect to have better
energy functionals.
In this case, we have a series of Kohn-Sham potentials, each equal
to a functional derivative of the corresponding density.
We will see how this works
explicitly for the kinetic energy and anomalous (pairing) densities
in Sections~\ref{subsec:effact} and \ref{subsec:pairing}, respectively.

  \I
If a functional in the semi-local form of Eq.~\eqref{eq:Edensity} is 
constructed (e.g., from a local density approximation plus gradient
corrections), then $v_{\rm xc}$ can be evaluated trivially 
from Eq.~\eqref{eq:vxc} and the
self-consistency cycle of Fig.~\ref{fig:KScycle} can be carried
out directly.   This is the form of Kohn-Sham DFT that is most
widely used (e.g., GGA and its variations for the Coulomb problem
and Skyrme, SLDA, etc.\ for the nuclear problem).
 \I
The success of Kohn-Sham compared to Thomas-Fermi 
(for which the kinetic energy is a functional of the density) followed from the
introduction of orbitals to treat the kinetic
energy, with manifest improvements such as the
reproduction of oscillations from shell structure~\cite{Argaman:2000xx}.
Thus, in this sense orbital-based functionals are already inevitable and 
making the potential energy orbital dependent is not a major 
step conceptually~\cite{Fiolhais:2003}.
Note that the kinetic energy contribution  here is just a definition
and, while a good approximation, 
it is \emph{not} the many-body kinetic energy (which would be
expressed in
terms of the full single-particle Green's function or the
one-body density matrix).  The
exchange-correlation part has to absorb the difference between these
kinetic energies.

 \I
For Coulomb systems the Hartree contribution is dominant
and the Hartree energy functional takes a special form that
depends \emph{explicitly} on the density only.  These are the reasons
for singling it out.  
For nuclear systems 
the Hartree piece is neither the dominant contribution
nor simple in general with low-momentum potentials,
which are non-local and require integrals over quantities
that do not reduce to the density (e.g., the one-particle density
matrix).
So singling it out is not generally helpful.
It \emph{is} advantageous in some approximations such as the
density matrix expansion (DME) to isolate the long-range part
of the potential, which is local, and to treat this part of the Hartree potential
explicitly.

\I
The exchange part can also be singled out for Coulomb
because of the possibility of a precise formula in terms of
the KS orbitals and the Coulomb potential.
But isolating particular parts of the functional
is not \emph{necessary}, because
\beqn
    v_{\rm H}(\rvec) + v_{\rm xc}(\rvec) 
    \equiv v_{\rm Hxc}(\rvec)
    = \frac{\delta }{\delta \rho(\bfx)}
     \bigl\{ E_{\rm H}[\rho] + E_{\rm xc}[\rho] \bigr\}
    \equiv \frac{\delta E_{\rm Hxc}[\rho]}{\delta\rho(\bfx)}
    \;,
    \label{eq:vHplusvxc}
\eeqn
so all the interaction terms can simply be combined (and
we use the ``Hxc'' subscript to indicate when this is done).
This is the form we generally use for the application to nuclei.

\ei

As is well documented, the conventional Kohn-Sham DFT with semi-local
functionals has been extremely successful for Coulomb
systems.  However, there
are failures and these  
help motivate the development of orbital-dependent functionals.
It is not yet clear to what extent these failures have analogs for 
nuclear-based system, but we summarize the list of arguments 
(taken largely from Engel's review in Ref.~\cite{Fiolhais:2003};
look there for specific references)
along with
some speculations about nuclei: 

\begin{itemize}
  \item \textbf{Heavy Elements.}
  For heavy constituents (e.g., gold), 
  the local density approximation (LDA) 
  tends to work better than the generalized gradient approximation
  (GGA) and this is not attributable to relativistic effects.  
  The suggestion is that GGA has
  trouble with higher angular momentum ($d$ and $f$).  If this
  is a generic problem it would certainly be relevant for nuclei.
  The fact that GGA is not a systematic improvement over LDA is
  also motivation for \abinitio\ DFT --- to develop a hierarchy of
  approximations that do systematically improve, as for the coupled
  cluster method~\cite{Bartlett:2005aa,Bartlett:2006aa}.

  \item \textbf{Negative Ions.}
  The fall-off of the Kohn-Sham
  potential does not have the $1/r$ asymptotic
  form needed for negative ions and Rydberg states.  The physical
  picture is that if one electron is far away from the others, it should
  see the net charge of the remaining system of $N-1$ electrons
  and $N$ protons.  But because $v_{\rm H}$ still always has the
  Coulomb repulsion of the far electron, it has to be removed by
  $v_{\rm x}$, but this does not happen with LDA or GGA functionals.
  Engel emphasizes that the exchange functional needs to be rather
  non-local to cancel the self-interaction in the Hartree 
  potential~\cite{Fiolhais:2003}.
  For the nuclear case, the impact of the self-interaction problem
  was considered long ago in Ref.~\cite{Stringari:1978zz} but
  only recently reconsidered along with the additional problem
  of self-pairing in Refs.~\cite{Lacroix:2008rj,Bender:2008rn,Duguet:2008rr}.

  \item \textbf{Dispersion Forces.}
  Dispersion forces are a type of van der Waals force.
  The problem here is the locality of the exchange-correlation
  functional.  If two atoms are so separated such that there is no overlap
  in the densities, then the density is the sum of the two
  atomic densities.  But we expect virtual dipole excitations leading
  to molecular bonding (this is called the London dispersion force).
  This does not work for LDA because the binding energy from the correlation
  functional is
  \beqn
    E_b = E_c^{\rm LDA}[\rho_A + \rho_B] - E_c^{\rm LDA}[\rho_A]
      - E_c^{\rm LDA}[\rho_B] \;,
  \eeqn
  which vanishes because only regions with non-zero density contribute
  to the correlation energy (so the first term on the right side
  is the sum of the other two terms).
  The same result holds for GGA because the density only in the near
  vicinity of $\rvec$ contributes to the energy density at $\rvec$.
  So we need non-locality for virtual excitations. Analogous issues
  for nuclei may arise from coupling to low-lying vibrations.
  
  \item \textbf{Strongly Correlated Systems.}
  Here there are failures for certain systems, such as some $3d$
  transition metal monoxides, which the LDA and GGA either predict
  are metallic when they are actually Mott insulators or else greatly
  underestimate the band gap.  Indications are that the incorrect
  treatment of self-interaction correction is the problem (although
  this is not proven).

\end{itemize}

The self-interaction problem can be illustrated by considering
the ``exact exchange'' functional $E_{\rm x}$ of DFT, which is defined
to be the Fock term as in Eq.~\eqref{eq:HFenergy}
written with KS orbitals.  So for a local
potential $\VN(\rvec,\rvec')$, this is simply
\beqn
  E_{\rm x} \equiv -\frac12 \sum_{kl} n_k n_l
    \int\! d{\rvec} \int\! d{\rvec}' \,
     {\phi_k^\dagger(\rvec)\phi_l(\rvec)\VN(\rvec,\rvec')
     \phi^\dagger_l(\rvec')\phi_k(\rvec')}
     \;.
     \label{eq:Ex}
\eeqn
This is \emph{not} the usual HF exchange contribution, 
because while it agrees in \emph{form} with Eq.~\eqref{eq:HFenergy},
it does not have orbitals that satisfy 
the non-local HF equations.  Just like the difference in the kinetic part,
the difference between HF and KS exchange 
is absorbed into the correlation functional
by construction.  The functional $E_{\rm x}$ is a density functional
in that the orbitals are uniquely determined by the density, but
it is implicit in the density dependence.  
The more general $E_{\rm Hxc}[\rho]$ will
also depend on the KS eigenvalues.

The form of $E_{\rm x}$ ensures exact
cancellation of the self-interaction energy in 
$E_{\rm H}$~\cite{Fiolhais:2003}.  
Suppose we split $E_{\rm x}$ into two pieces according to whether
or not $k=l$ in the double sum. Then
\bea
  E_{\rm x} &=& -\frac12 \sum_{k\neq l} n_k n_l
    \int\! d{\rvec} \int\! d{\rvec}' \,
     {\phi_k^\dagger(\rvec)\phi_l(\rvec)\VN(\rvec,\rvec')
     \phi^\dagger_l(\rvec')\phi_k(\rvec')}
     \nonumber \\
     && \null -\frac12  \sum_{k} n_k
    \int\! d{\rvec} \int\! d{\rvec}' \,
     {\phi_k^\dagger(\rvec)\phi_k(\rvec)\VN(\rvec,\rvec')
     \phi^\dagger_k(\rvec')\phi_k(\rvec')}
          \;.
\eea
Note that the integrand in the second term \emph{does not} reduce to the product
of densities because there is only one $k$ sum.  
However, this second term does cancel the corresponding 
part of the Hartree functional if
we rewrite the latter in a similar form:
\bea
  E_{\rm H} &=&  \frac12 \sum_{k\neq l} n_k n_l
    \int\! d{\rvec} \int\! d{\rvec}' \,
     {\phi_k^\dagger(\rvec)\phi_k(\rvec)\VN(\rvec,\rvec')
     \phi^\dagger_l(\rvec')\phi_l(\rvec')}
     \nonumber \\
     && \null + \frac12\sum_{k} n_k
    \int\! d{\rvec} \int\! d{\rvec}' \,
     {\phi_k^\dagger(\rvec)\phi_k(\rvec)\VN(\rvec,\rvec')
     \phi^\dagger_k(\rvec')\phi_k(\rvec')}
          \;.
\eea
This cancellation is familiar from ordinary Hartree-Fock.
But if $E_{\rm x}$ is expanded in semi-local form as in the LDA
or GGa for Coulomb systems or the DME for the nuclear
case (see Sect.~\ref{subsec:dme}), this cancellation
is lost.

K\"ummel and Kronik~\cite{Kummel:2008} emphasize the formal deficiencies
that can lead to qualitative failures of the LDA and GGA predictions. 
These include not only the non-cancellation of self-interaction but
the lack of a derivative discontinuity in $E_{\rm xc}$.  The latter
issue starts with the DFT definition of a chemical potential $\mu$:
\beqn
   \mu \equiv \frac{\delta E_{\rm tot}[\rho]}{\delta\rho(\rvec)}
   \;,
\eeqn
which is position independent when evaluated 
at the ground-state density.  Perdew
et al.\ argued~\cite{Perdew:1982} that this chemical potential
must have a discontinuity: if the integer number of electrons
is approached from below its absolute value should be the ionization
potential while if from above it should equal the electron affinity. 
The derivative discontinuities at integer particle numbers in general
comes from both the noninteracting kinetic energy and $E_{\rm xc}$.
But the LDA and GGA $E_{\rm xc}$'s are continuous in the density and
its gradient, so there is no particle number discontinuity.  
Ref.~\cite{Kummel:2008} has more details on why this is an important
issue for Coulomb DFT.  We know of no investigation, however, into its impact
on nuclear EDF's.

In the end, a basic issue is that any full DFT energy functional 
\emph{must}
have non-localities and it may be problematic for nuclear structure 
to expand in a semi-local 
functional~\cite{Lacroix:2008rj,Bender:2008rn,Duguet:2008rr}.   
The only way we know to test this is
by comparing such functionals derived from microscopic interactions
(e.g., through the density matrix expansion, see
Section~\ref{subsec:dme}) to a full
orbital-based functional.
A program to carry out such comparisons was recently initiated
as part of the UNEDF project~\cite{unedf:2007,unedfweb}.

\subsection{Derivation of the optimized effective potential}

The fundamental problem in extracting the Kohn-Sham potential
is to calculate the functional derivative with respect to the density as
in Eq.~\eqref{eq:vHplusvxc}.  (More generally, we will need to
take functional derivatives with respect to other densities, such
as the kinetic energy density and the anomalous density for pairing
for nuclear applications.)
Starting from an \abinitio\ energy functional, one way to proceed
is an expansion such as the density matrix expansion (DME), which
results in functionals with the densities explicit
(see Section~\ref{subsec:dme}). 
In the absence of such approximations,
the density is naturally explicit only in 
the Hartree energy functional (and in this case only for local
interactions). 
Therefore, we need an \emph{implicit} calculation of the 
derivatives~\cite{Fiolhais:2003}.

The most direct procedure is to 
use the chain rule~\cite{Fiolhais:2003}.  
As reviewed in later sections, the energy
functional can be built from the KS orbitals and eigenvalues (e.g.,
with Kohn-Sham MBPT as in Section~\ref{sec:abinitio}), so
we express the density derivative in terms of those.
As an intermediate derivative we vary the KS potential $\vKS$:
\bea
  \frac{\delta E_{\rm xc}[\phi_k,\varepsilon_k]}{\delta \rho(\rvec)}
  &=& \int\! d{\rvec}' \,
  \frac{\delta E_{\rm xc}}{\delta \vKS(\rvec')}
  \frac{\delta \vKS(\rvec')}{\delta \rho(\rvec)}
  \nonumber \\
  &=&
  \int\! d{\rvec}' \,
  \frac{\delta \vKS(\rvec')}{\delta \rho(\rvec)}
  \sum_k \biggl\{
    \int\! d{\rvec}'' \biggl[
      \frac{\delta \phi_k^\dagger(\rvec'')}{\delta \vKS(\rvec')}
      \frac{\delta E_{\rm xc}}{\delta \phi_k^\dagger(\rvec'')}
    + \cc \biggr]
    + \frac{\delta\varepsilon_k}{\delta \vKS(\rvec')}
      \frac{\partial E_{\rm xc}}{\partial \varepsilon_k}
  \biggr\}  
  \;.
\eea
Note that in all of the expressions given here
the sum over orbitals $k$ is \emph{not} restricted to occupied
states unless $n_k$ explicitly appears.
(We assume filled shells here and refer the reader to the
cited literature for the case of unfilled shells.)
Now we need to find the functional derivatives introduced in this
expression.%
\footnote{In general one also needs to vary the occupation
numbers, although this is not needed if there is
fixed particle number~\cite{Kurth:2009ab}.}

For the exchange functional defined in Eq.~\eqref{eq:Ex} 
with a local potential, we obtain
\beqn
  \frac{\delta E_{\rm x}}{\delta \phi_k^\dagger(\rvec')}
  =  - n_k \sum_l n_l\, \phi_l(\rvec')
  \int\! d{\rvec}\, 
  {\phi_l^\dagger(\rvec)\phi_k(\rvec)}V(\rvec,\rvec')
  \;,
\eeqn
and $\partial E_{\rm x}/\partial \varepsilon_k = 0$.
The functional derivatives
$\delta\phi_k^\dagger/\delta \vKS$ and $\delta \epsilon_k/\delta \vKS$
follow from simple first-order (non-degenerate)
perturbation theory with
$\delta \vKS$ as the perturbation.  That is, we treat the KS equation
for the orbitals just like in conventional quantum mechanics.
So we start with the change in the eigenvalue:
\beqn
    \delta \varepsilon_k[\delta \vKS] = \int\! d{\rvec}'\, 
    \phi_k^\dagger(\rvec')\,
       \delta \vKS(\rvec')\, \phi_k(\rvec')
       \;,
\eeqn
which directly implies the functional derivative
\beqn
   \frac{\delta\varepsilon_k}{\delta \vKS(\rvec)}
      = \phi_k^\dagger(\rvec) \phi_k(\rvec)
      \;.
\eeqn
Similarly, for the wave function,
\beqn
   \delta \phi_k(\rvec) = \sum_{l \neq k} \frac{\phi_l(\rvec)}
     {\varepsilon_k - \varepsilon_l}
     \int\! d^3\rvec''\, \phi_l^\dagger(\rvec'') \delta \vKS(\rvec'')
                         \phi_k(\rvec'')
   \;,                
\eeqn
and
\beqn
   \delta \phi_k^\dagger(\rvec) = \sum_{l \neq k} 
     \int\! d^3\rvec''\, \phi_k^\dagger(\rvec'') \delta \vKS(\rvec'')
                         \phi_l(\rvec'')
       \frac{\phi_l^\dagger(\rvec)}                  
     {\varepsilon_k - \varepsilon_l}
   \;,                
\eeqn
which imply the functional derivatives
\beqn
  \frac{\delta\phi_k(\rvec)}{\delta \vKS(\rvec')}
    =  \sum_{l \neq k} \frac{\phi_l(\rvec)\phi_l^\dagger(\rvec')}
            {\varepsilon_k - \varepsilon_l}
       \phi_k(\rvec')
       = - G_k(\rvec,\rvec') \phi_k(\rvec')
     \;,  
\eeqn
and
\beqn
  \frac{\delta\phi_k^\dagger(\rvec)}{\delta \vKS(\rvec')}
    =  \phi_k^\dagger(\rvec') 
     \sum_{l \neq k} \frac{\phi_l(\rvec')\phi_l^\dagger(\rvec)}
            {\varepsilon_k - \varepsilon_l}       
       = -  \phi_k^\dagger(\rvec') G_k(\rvec',\rvec)
     \;,  
\eeqn
where (note the sign change in the denominator, which can differ
in the literature)
\beqn
  G_k(\rvec,\rvec') \equiv 
     \sum_{l \neq k} \frac{\phi_l(\rvec)\phi_l^\dagger(\rvec')}
            {\varepsilon_l - \varepsilon_k}
        \;.    
\eeqn
Note that to construct the Green's function for every
state requires explicit solutions for all the orbitals. 
Below we discuss approximations that avoid the complication
of unoccupied orbitals.

To complete the chain rule we need to evaluate 
$\delta \vKS/\delta\rho$.
To do so, we focus first on the inverse function, which is
the static response function of the KS system,
\beqn
  \frac{\delta \rho(\rvec)}{\delta \vKS(\rvec')}
    = \chi_s(\rvec,\rvec')
    = - \sum_k n_k\, \phi_k^\dagger(\rvec) G_k(\rvec,\rvec')
       \phi_k(\rvec') + \cc
       \;.
       \label{eq:chisdef}
\eeqn
This follows directly by applying our functional derivative formulas
to the expression for $\rho(\rvec)$ in terms of orbitals.
Note that only the terms with $l$ unoccupied in $G_k$ will contribute
to $\chi_s$, because those with $l$ occupied will cancel (e.g., pick an
$l$ and a $k$ and note that the complex conjugate term is the same
but with the opposite sign energy denominator).

By multiplying $\delta E_{\rm xc}/\delta \rho(\rvec)$ by $\chi_s$ and
integrating over $\rvec$, the OEP or 
OPM (optimized potential method) integral
equation is obtained:%
\footnote{The names OEP and OPM are used in the literature by different
authors to describe either this equation or more generally the
orbital-based approach.}
\beqn
  \int\! d{\rvec}'\, \chi_s(\rvec,\rvec') v_{\rm xc}(\rvec')
    = \Lambda_{\rm xc}(\rvec)
    \;,
    \label{eq:OPM}
\eeqn
where
\beqn
  \Lambda_{\rm xc}(\rvec) = \sum_k \biggl\{
   -\int\! d{\rvec}'\, \Bigl[ \phi_k^\dagger(\rvec) G_k(\rvec,\rvec')
     \frac{\delta E_{\rm xc}}{\delta \phi_k^\dagger(\rvec')}
     + \cc \Bigr]
  + |\phi_k(\rvec)|^2 \frac{\partial E_{\rm xc}}{\partial \varepsilon_k}
  \biggr\}
  \;.
  \label{eq:Lambdaxc}
\eeqn
This is a Fredholm integral equation of the first kind.
Note that because Eqs.~\eqref{eq:OPM} and \eqref{eq:Lambdaxc}
are linear in $E_{\rm xc}$, we can consider separate equations
for different pieces of  $E_{\rm xc}$.
It is common to introduce the ``orbital shifts''~\cite{Kummel:2008}
\beqn
  \psi_k^\dagger(\rvec) \equiv
  \int\! d{\rvec}'\, \phi_k^\dagger(\rvec')
    \bigl[
    u_{{\rm xc},k}(\rvec') - v_{\rm xc}(\rvec')
    \bigr] G_k(\rvec',\rvec)
    \;,
\eeqn
where
\beqn
  u_{{\rm xc},k} \equiv \frac{1}{\phi^\dagger_k(\rvec')}
          \frac{\delta E_{\rm xc}}{\delta\phi_k(\rvec's)}
          \;,
\eeqn
so that we can write the OPM integral equation compactly as
\beqn
  \sum_k  (\psi_k^\dagger(\rvec) \phi_k(\rvec) + \cc ) = 0
  \;.
\eeqn
In carrying  out the self-consistency loop (Fig.~\ref{fig:KScycle}),
the solution for the orbitals when given a Kohn-Sham potential proceeds
the same as always.  Given the new orbitals and eigenvalues,
we first find $G_k(\rvec,\rvec')$ and $u_{{\rm xc},k}$,
and then the new KS potential  
by solving the OPM equation, and the loop repeats.
As discussed in Section~\ref{subsec:approx}, this is numerically difficult and
comparatively inefficient, but there are
also good approximations that simplify the solution significantly.


\subsection{OEP from total energy minimization or density invariance}
\label{subsec:oepdensity}  

To get more insight into
the physics of the OEP,
we turn to the original derivation of the OPM integral equation,
which is based on the minimization of the energy functional with respect
to the density~\cite{Fiolhais:2003}.
Without explicit dependence on the density we do not know how to
do that minimization, but 
the Hohenberg-Kohn theorem that tells us
that $\vKS$ and $\rho$ are directly related.  
In particular, this implies that
we can replace the minimization with respect to $\rho$ by a minimization
with respect to $\vKS$ (for fixed particle number),
\beqn
   \frac{\delta E_{\rm tot}[\phi_k,\varepsilon_k]}{\delta \vKS(\rvec)} 
     = 0
     \;.
\eeqn
Now we apply the chain rule as before:
\beqn
  \frac{\delta E_{\rm tot}[\phi_k,\varepsilon_k]}{\delta \vKS(\rvec)}
  =
  \sum_k \biggl\{
    \int\! d{\rvec}' \biggl[
      \frac{\delta \phi_k^\dagger(\rvec')}{\delta \vKS(\rvec)}
      \frac{\delta E_{\rm tot}}{\delta \phi_k^\dagger(\rvec')}
    + \cc \biggr]
    + \frac{\delta\varepsilon_k}{\delta \vKS(\rvec')}
      \frac{\partial E_{\rm xc}}{\partial \varepsilon_k}
  \biggr\}  
  \;.
  \label{eq:EtotVKS}
\eeqn
The functional derivatives of $E_{\rm tot}$ with respect to 
the orbitals and eigenvalues are given by
\bea
      \frac{\delta E_{\rm tot}}{\delta \phi_k^\dagger(\rvec)}
  &=& n_k \left[ -\frac{\nabla^2}{2M} + \Vext(\rvec) + v_H(\rvec)
  \right]\phi_k(\rvec) + 
   \frac{\delta E_{\rm xc}}{\delta \phi_k^\dagger(\rvec)} 
   \;,  \label{eq:Etotphik}
  \\
  \frac{\partial E_{\rm tot}}{\partial\varepsilon_k}
  &=&
  \frac{\partial E_{\rm xc}}{\partial\varepsilon_k}
  \;,
\eea
with the remaining derivatives given by previous expressions.
The first term on the right side of Eq.~\eqref{eq:Etotphik} can
be rewritten because $\phi_k(\rvec)$ satisfies the KS orbital equation,
\beqn
      \frac{\delta E_{\rm tot}}{\delta \phi_k^\dagger(\rvec)}
  = n_k \left[\varepsilon_k - v_{\rm xc}(\rvec)
  \right]\phi_k(\rvec) + 
   \frac{\delta E_{\rm xc}}{\delta \phi_k^\dagger(\rvec)} 
   \;. 
\eeqn
Plugging everything into Eq.~\eqref{eq:EtotVKS} gives the minimization
condition:
\beqn
  \sum_k  \int\! d{\rvec}' \biggl\{
   \phi_k^\dagger(\rvec) G_k(\rvec,\rvec')
   \biggl[
   n_k \phi_k(\rvec)(v_{\rm xc} - \varepsilon_k)
   + \frac{\delta E_{\rm xc}}{\delta \phi_k^\dagger(\rvec')}
   \biggr] + \cc \biggr\}
    + \sum_k |\phi_k(\rvec)|^2
    \frac{\partial E_{\rm xc}}{\partial\varepsilon_k}
    = 0 \;.
\eeqn
Finally, the OPM integral equation is recovered after identifying 
$\chi_s(\rvec,\rvec')$ and $\Lambda_{\rm xc}(\rvec)$ and using
\beqn
  \int\! d{\rvec}\, \phi_k^\dagger(\rvec) G_k(\rvec,\rvec')
    = \int\! d{\rvec}'\,  G_k(\rvec,\rvec') \phi_k^\dagger(\rvec') = 0
    \;.
    \label{eq:identity}
\eeqn
(Note that Eq.~\eqref{eq:identity} means there will be complications
with inverting $\chi_s$.)

If we just include the direct and exchange terms in the functional
then $E_{\rm tot}$ looks just like an HF functional. 
The key difference is that the HF approach corresponds to a \emph{free}
(unconstrained) minimization of the total energy functional with respect
to the $\phi_k$ and $\varepsilon_k$.  But the minimization here
is not free; rather, the $\phi_k$ and $\varepsilon_k$ have to
satisfy the KS equations with a multiplicative potential.  This
is a \emph{subsidiary} condition to the minimization of $E_{\rm tot}$,
which is implemented by the OPM equation.
Because this means the variational calculation is over a more limited
set of states, the OEP applied to exchange only will always give
a higher energy than Hartree-Fock~\cite{Fiolhais:2003}.

The OPM equation can be derived yet another way,
which is based on the point-by-point
equivalence of the KS and interacting 
densities~\cite{Sham:1983aa,Casida:1995aa,Kummel:2003}:
\beqn
  \rho_s(\bfx) - \rho(\bfx) = 0
  \;.
  \label{eq:rhoequal}
\eeqn
This is perhaps the least obvious feature of Kohn-Sham DFT.
It might appear in the context of a perturbative expansion
(see Section~\ref{sec:abinitio}) to be simply a \emph{choice} of the Kohn-Sham
potential that makes higher order corrections to the density
vanish.  However, in fact it implies the energy minimization
that is common to all DFT applications, that is, that
the KS exchange-correlation potential is the variationally
best local approximation to the exchange-correlation
energy~\cite{Casida:1995aa}.  
The basic demonstration starts with expressing Eq.~\eqref{eq:rhoequal}
in terms of traces over
the KS and full single-particle Green's functions evaluated
at equal times and the same coordinate arguments:
\beqn
  -i\tr\{G_s(\rvec t,\rvec t^+) - G(\rvec t, \rvec t^+)\} = 0
  \;,
\eeqn
where 
the Green's functions are defined as usual as time-ordered products
of the field operators in the KS or fully interacting ground states:
\bea
   G_s(\rvec t,\rvec' t') &=& -i \PhibraKS  T\psi(\rvec t)
                          \psi^\dagger(\rvec' t') \PhiketKS
                          \\
   G(\rvec t,\rvec' t') &=& -i \Psibra  T\psi(\rvec t)
                          \psi^\dagger(\rvec' t') \Psiket
                          \;.
\eea
The specification $t^+$ serves to order the field operators as
$\psi^\dagger\psi$. 
The full Green's function is related  
to the KS Green's function by a Dyson equation
[using a four-vector notation $x = (\bfx,t)$]:
\beqn
  G(x,x') = G_s(x,x') + \int\!dy\, dy'\,
    G_s(x,y) \Bigl[ \Sigma_{\rm Hxc}(y,y') - \delta(y-y') v_{\rm Hxc}({\bf
    y})
    \Bigr] G(y',x')
    \;,
   \label{eq:shamshuluter} 
\eeqn
which follows from the separate Dyson equations
for $G_s$ and $G$
by forming $G^{-1}$ and $G_s^{-1}$ and subtracting, solving
for $G^{-1}$, and then taking the inverse~\cite{Fiolhais:2003}.
Note that the Hartree parts of the irreducible self-energy $\Sigma_{\rm
Hxc}(y,y')$
and the Kohn-Sham self-energy $\delta(y-y') v_{\rm Hxc}(\yvec)$ 
cancel for local potentials, leaving
the difference of $\Sigma_{\rm xc}$ and $v_{\rm xc}$.
This is known as the Sham-Schl\"uter equation and is a nonlinear
equation for $v_{\rm xc}$,
which can be shown to be equivalent to the OPM equation.
We leave the details of showing this equivalence to the references,
but give a schematic demonstration in Section~\ref{sec:lt}.


\subsection{Approximations}
\label{subsec:approx} 

While the preceding demonstrations illustrate  formal constructions
that suffice to carry out orbital-based DFT (given appropriate
energy functionals), solving the equations can be
numerically difficult and there are also efficiency issues.
(For recent work on numerically stable methods to solve the
OEP equations using Gaussian basis sets, see
Refs.~\cite{Bartlett:2006aa,Hesselmann:2007aa}.)
In Coulomb applications, OPM calculations are found to take one
or two orders of magnitude longer than corresponding GGA calculations.
The source of this inefficiency is the need for the
Kohn-Sham Green's function, which
requires knowledge of all the unoccupied as well as occupied orbitals.
An approximation (and subsequent variations) by Krieger, Li
and Iafrate (KLI)~\cite{Krieger:1990aa} based on a closure
approximation avoids this problem and seems in practice (for 
Coulomb cases) to be
quite accurate, at least for the exchange part of the functional.  

The key is to replace the energy denominator in the Green's function
by an averaged difference:
\beqn
  G_k(\rvec,\rvec') \approx \sum_{l\neq k} \frac{\phi_l(\rvec)
  \phi_l^\dagger(\rvec')}{\Delta \overline\varepsilon}
  = \frac{1}{\Delta \overline\epsilon}
  [\delta^3(\rvec-\rvec') - \phi_k(\rvec)\phi_k^\dagger(\rvec')]
  \;.
\eeqn
Upon substituting into the OPM integral equation, one obtains
\bea
  v_{\rm xc}(\rvec) &=&
  \frac{1}{2\rho(\rvec)}
  \sum_k \biggl\{
  \biggl[
   \phi_k^\dagger(\rvec)
 \frac{\delta E_{\rm xc}}{\delta \phi_k^\dagger(\rvec)} + \cc
  \biggr]
  + |\phi_k(\rvec)|^2 \biggl[ \Delta v_k - \Delta \overline\epsilon
  \frac{\partial E_{\rm xc}}{\partial \varepsilon_k} \biggr]
  \biggr\}
  \\
  \Delta v_k &=&
  \int\! d{\rvec}\,
  \biggl\{
  n_k |\phi_k(\rvec)|^2 v_{\rm xc}(\rvec)
    - \phi_k^\dagger(\rvec)
     \frac{\partial E_{\rm xc}}{\partial \varepsilon_k} 
  \biggr\}
  + \cc
  \;.
\eea
Consistent with the closure approximation, the derivative
$\partial E_{\rm xc}/\partial \varepsilon_k$ is neglected,
which finally yields
\beqn
  v^{\rm KLI}_{\rm xc}(\rvec) =
  \frac{1}{2\rho(\rvec)}
  \sum_k \biggl\{
  \biggl[
   \phi_k^\dagger(\rvec)
 \frac{\delta E_{\rm xc}}{\delta \phi_k^\dagger(\rvec)} + \cc
  \biggr]
  + |\phi_k(\rvec)|^2 \Delta v^{\rm KLI}_k 
  \biggr\}
  \;.
\eeqn  
Although $v^{\rm KLI}_{\rm xc}$ appears on both sides, one can either
iterate to self-consistency starting from (for example) an LDA
approximation to $\Delta v^{\rm KLI}_k$ or recast these as 
a set of linear equations allowing $\Delta v^{\rm KLI}_k$ to be
determined without knowing $v^{\rm KLI}_{\rm xc}$.

Results from representative calculations comparing OPM and KLI 
to LDA, GGA, and HF 
using exchange-only functionals have been studied systematically
and are summarized in Refs.~\cite{Fiolhais:2003} and \cite{Kummel:2008}.  
Note that in such comparisons the Hartree-Fock (HF) result serves
as the ``exact'' answer.
In most cases, exchange-only OPM is found to be a very good
approximation to HF, e.g., agreement at one part in $10^{-6}$
for the ground-state energies of heavy closed-subshell atoms. 
Because exchange-only OPM is a more
restricted variational
minimization than Hartree-Fock, the HF results must always
be more bound than OPM, but the small difference implies that 
the greater variational freedom for HF has
little effect in practice.  
Furthermore, the KLI energies
are very good approximations to the full OPM results; for this
same example the KLI--OPM difference is systematically
about 1/3 that of HF--OPM.  
The KLI energies are
above OPM in all cases,
as required because the full OPM energy is a minimum for local potentials.
The deviation for even the largest atom is still very small.  The level
of agreement can
be calibrated by comparison to LDA and GGA results; the latter
are much better than the former, but sometimes an order of magnitude
worse than KLI and without a systematic sign.

Other approximations related to KLI but with improvements 
have been proposed.  The Common Energy Denominator Approximation
(CEDA)~\cite{Gritsenko:2001aa} and the Localized Hartree-Fock (LHF)
approximation~\cite{DellaSala:2001aa} are the same as KLI except
that only the energy differences for occupied-unoccupied pairs
of states are approximated by an average, while the exact
differences are kept for occupied-occupied pairs.
CEDA is invariant under unitary transformations of the occupied
orbitals, which is a plus, but in practice the results are
similar to KLI~\cite{Kummel:2008}.
Yet another approximation is the effective local
potential proposed by Staroverov et al.~\cite{Staroverov:2006ab},
which is efficient to implement numerically.

The good results that have been found using the KLI and similar
approximations (for total energies) are encouraging because 
the corresponding potentials are far easier to calculate.
KLI has been shown to be a sort of mean-field approximation
to the full OEP, which accounts for its success~\cite{Kummel:2008}.
However, more work is needed on applications beyond exchange-only
functionals.
There are no calculations yet that compare these approximations
(and more extreme approximations such as the density
matrix expansion, see Section~\ref{subsec:dme})
for nuclear systems, so drawing conclusions on the prospects
would be premature.  
  


\section{DFT and ab initio wave function methods}
 \label{sec:abinitio}

The development of accurate \abinitio\ functionals 
for orbital-based DFT that go beyond 
exchange-only (what is called the ``correlation functional'') for
Coulomb systems is the subject of much recent activity.  Progress
is being made although it is still 
an open question whether the most ambitious
accuracy goals can be met (e.g., chemical accuracy).
The extensive review by K\"ummel and
Kronik~\cite{Kummel:2008} provides a good idea of the state of the art
in 2008 but we note that there are subsequent and ongoing advances.

For the nuclear problem, 
the development of analogous \abinitio\ functionals suitable for
orbital-based DFT is still in its infancy.
Thus our ``review'' in this section will mainly look toward the future,
considering first how to 
formulate an exchange-correlation functional in MBPT,
neglecting some serious issues such as symmetry breaking.
We then examine a specific implementation by Bartlett and
collaborators
for quantum chemistry that is based on requiring density invariance
and which highlights how the convergence of the
perturbation expansion for the
functional can be improved
by shifting more physics into the Kohn-Sham potential.
These developments are based on applying MBPT at low order,
so we next discuss how this is meaningful for nuclear interactions
if they are transformed to low-momentum potentials.
Finally, we review recent and ongoing efforts to apply the density
matrix expansion to make \abinitio\ energy calculations by expanding
functionals about the uniform system (asymmetric nuclear matter).  
This also provides examples of how to
deal with momentum space potentials, non-localities,
and three-body forces, which will not be found in the Coulomb literature.

\subsection{Goldstone many-body perturbation theory}
  \label{subsec:mbpt}

A direct method to construct an orbital-based energy
functional is to consider the Kohn-Sham potential $\vKS(\rvec)$ as defining
the single-particle potential (typically called $U(\rvec)$ in the nuclear
physics literature) that is used in Goldstone many-body perturbation
theory (MBPT).  
This will give us a diagrammatic expansion 
for the energy as a functional of $U$
that will depend on KS orbitals and eigenvalues. 
There are various alternative ways to formulate MBPT. 
For example, in the Coulomb DFT literature a formalism
for perturbation theory based
on coupling constant integration~\cite{DREIZLER90,Fiolhais:2003} 
is typically used.  
Here we use the Goldstone formalism that 
historically led from ``hard core'' NN potentials
to the hole-line expansion~\cite{Day:1967zz,Rajaraman:1967zza,Baldo99},
but the end result is basically the same.
Note that ``perturbation theory'' does not exclude infinite
summations of diagrams in our functional, 
although we will argue in Section~\ref{subsec:vlowk} that
second-order perturbation theory is a good starting point for
low-momentum nuclear interactions.

Our schematic discussion is compatible with the detailed,
pedagogical expositions of the Goldstone-diagram expansion given in 
Refs.~\cite{Day:1967zz,Day:1978aa,Negele:1988vy,Baldo99}.
We start with a division of the second-quantized,
time-independent Hamiltonian:
\beqn
  \Hhat = \Hhat_0 + \Hhat_1
  \;,
\eeqn
where the separation is our choice.
One might imagine taking
$\Hhat_0$ to be the
kinetic energy plus the fixed external potential and
$\Hhat_1$ to be the interaction potential.
Such a division will be invoked in Section~\ref{subsec:gf} 
when we discuss
the connection to the perturbation expansion using Green's functions.  
However, for our applications, we will require $\Hhat_0$ to 
be the sum of the kinetic energy%
 \footnote{For finite nuclei, we would subtract the center-of-mass
kinetic energy $\wh T_{\rm cm}$ from
$\wh T$.}
and a single-particle potential (including any $\Vextop$),
$\Hhat_0 = \wh T + \wh U$.
Then $\Hhat_1$ corrects for the new part of $\wh U$, 
$\Hhat_1 = \wh \VN - \wh U + \Vextop$.
With this restricted form, the many-body ground- and excited-states
of $\Hhat_0$ will be simple to construct.
The ground state of $\Hhat_0$, designated $\Phiket$, will be our
reference state and our Fock space will be built from particles
and holes with respect to it.
More precisely, the complete set of orbitals defined by the
single-particle
potential establishes the single-particle basis for the Fock space, and
second-quantized field operators are expanded in this basis.
So $\Phiket = \prod_{\varepsilon_i \leq \efermi} a_i^\dagger | 0
\rangle$ is a Slater determinant of orbitals where
$a_i(a_i^\dagger)$ is the destruction (creation) operator for the
single-particle state $i$ and $a_i | 0 \rangle = 0$
for all $i$ and $\efermi$ is the energy of the last filled level.
The idea will be to identify $\wh U$
with the Kohn-Sham potential
[i.e., $\wh U = \VKS = \int\!d\rvec\, \vKS(\rvec)\wh\rho(\rvec)$], which is to be
determined self-consistently in accordance with
Section~\ref{sec:dftoep}.  So the orbitals $\phi_i$ are then
KS orbitals.  

\begin{figure}[t]
\centering
 \includegraphics*[width=3.0in,angle=0.]{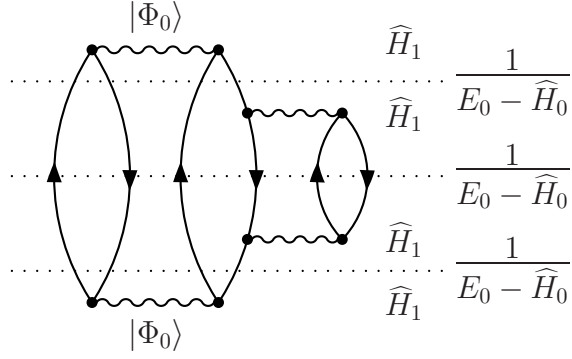}
 \caption{A Goldstone diagram contributing to Eq.~\eqref{eq:goldstone} at 
 fourth order in $\Hhat_1$ with a two-body potential only 
 (wiggle lines)~\cite{FETTER71}.  
 }
 \label{fig:gs2}       
\end{figure}

Goldstone's theorem expresses the energy of the ground state
as 
\beqn
  E = E_0 + \Phibra \Hhat_1 \sum_{n=0}^{\infty} 
    \left(
     \frac{1}{E_0 - \Hhat_0} \Hhat_1 
    \right)^n \Phiket_{\rm connected}
    \;,
    \label{eq:goldstone}
\eeqn 
where $E_0$ is the ground state energy of $\Hhat_0$;
that is, $\Hhat_0 \Phiket = E_0 \Phiket$.  (Note: These
operators are in the Schr\"odinger picture.)
The general idea of an $n^{\rm th}$ order contribution 
(see Fig.~\ref{fig:gs2})
is that we start with $\Phiket$
and apply $\Hhat_1$, which creates particles and holes.
For a two-body potential, this is two particles and two holes
(called ``doubles'' in quantum chemistry parlance), but in
our case $\Hhat_1$  in addition has
one-particle--one-hole excitations (``singles'')
as well as few-particle--few-hole excitations from few-body
interactions (``triples'' and beyond).
The factor $(E_0 - \Hhat_0)^{-1}$ \emph{propagates} the state and then
the next $\Hhat_1$ hits.  The proviso ``connected'' means that we
do not have $\Phiket$ as an intermediate state; we only get back
to it at the end after the final $\Hhat_1$.

\begin{figure}[t]
\centering
 \includegraphics*[width=4.5in,angle=0.]{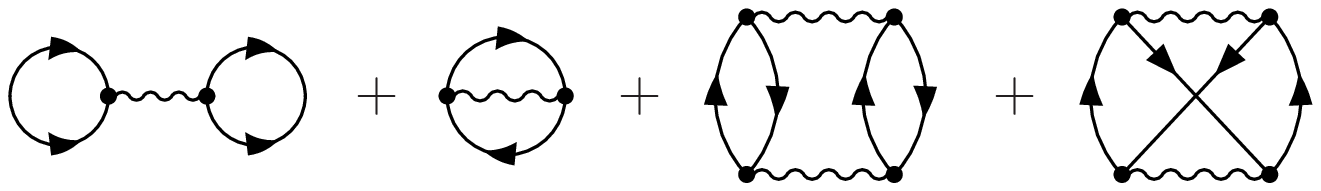}

 \vspace*{.1in}

 \includegraphics*[width=4.5in,angle=0.]{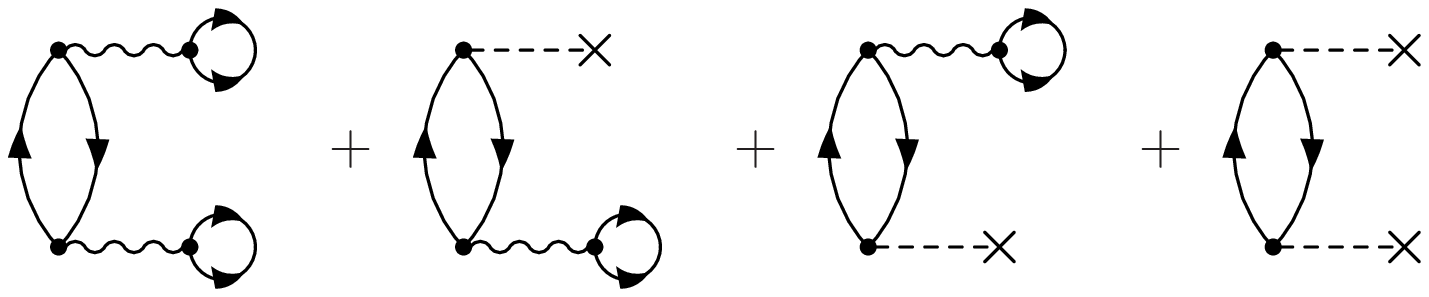}
 \caption{Goldstone diagrams at first and second order in $\Hhat_1$ with
 a single-particle potential $-\widehat U$ (dash lines) and
 a two-body potential $\widehat V_{\rm NN}$ only (wiggle lines).  
 }
 \label{fig:gs1}       
\end{figure}

Goldstone diagrams can be used to organize the contributions
from Eq.~\eqref{eq:goldstone}.
A set of rules is given in Ref.~\cite{Day:1967zz} for the general
case of interest with nonzero $\wh U$.
The diagrams that contribute in a finite system 
at first and second order in $\Hhat_1$
are shown in Fig.~\ref{fig:gs1}~\cite{Becker:1968aa} (those in the second row vanish in
uniform matter by momentum conservation).
A sample diagram contributing to Eq.~\eqref{eq:goldstone} is shown in 
Fig.~\ref{fig:gs2}.  Lines with upward arrows are particles and those
with downward arrows are holes; they will carry labels for the KS
orbitals.
By inserting complete sets of states, any diagram is reduced to products
of matrix elements of $\Hhat_1$ and energy denominators that are sums
and differences of KS eigenvalues.
For example, a general expression for the energy shift
$E - E_0$ with $\wh U=0$ is found to be (see Ref.~\cite{Negele:1988vy}
for more details):
\beqn
 E - E_0 = \sum_{\rm connected} \frac{(-1)^{n_L + n_h}}{2^{n_e}}
   \prod \frac{1}{-(\sum_a \epsilon_a - \sum_A\epsilon_A)}
  \prod \{ij | \wh V_{\rm NN} | kl \}
  \;,
\eeqn
where the matrix elements are antisymmetrized Hugenholtz matrix
elements (and only the two-body interaction is included; the
generalization to three-body and beyond is similar only with 
few-body matrix elements such as $\{ijk|\wh V_{\rm NNN}|lmn\}$).  The
energy denominator comes in between each successive interaction and
includes all particle lines (``$A$'') and hole lines 
(``$a$'') cut by the dotted lines as in Fig.~\ref{fig:gs2}.
The number of hole lines is $n_h$ and the number of equivalent pairs
is $n_e$, while $n_L$ is the number of closed loops that the diagram
with only direct matrix elements at each vertex would have.

A \emph{perturbative} calculation applying Goldstone's theorem (which is
just a restatement to all orders of time-independent perturbation
theory for the ground-state energy%
\footnote{A perturbative expansion with the same
$\Hhat_0$ (often called $\Hhat_s$ in the DFT literature) and $\Hhat_1$ 
can be derived~\cite{Fiolhais:2003} using
a coupling constant integration to adiabatically switch on
$\Hhat_1$ as in Refs.~\cite{FETTER71,Negele:1988vy}.
Thus $\Hhat = \Hhat_0 + \lambda\Hhat_1$ and
\beqn
  E - E_0 \equiv E_1 = \int_0^1\! d\lambda\,
   \langle \Psi_0(\lambda) | \Hhat_1 | \Psi_0(\lambda \rangle
   \;,
\eeqn
which is developed using the interaction picture time-evolution
operator.  Then $E_{\rm xc}$ is given by
\beqn
  E_{\rm xc} = E_1  + \int\!d\rvec\,
    \rho(\rvec)v_{\rm xc}
    \;.
\eeqn 
}) is organized in the number of
times $\Hhat_1$ is applied. 
Experience using nuclear potentials with strong short-range repulsion reveals
that such calculations do not converge.  This led to the development
of the Brueckner-Bethe-Goldstone approach and the hole-line expansion.
The latter is a prescription for infinite resummations of diagrams
(e.g., summing the potential first into a G-matrix and then including
all diagrams with a given number of independent hole
lines~\cite{Day:1967zz})
along with constraints on the background potential $\wh U$
in order to cancel certain diagrams.  These constraints are incompatible
with the choice $\wh U = \wh V_{\rm KS}$.  However, with low-momentum
interactions we are (apparently) free to choose $\wh U$ in this way
and even a simple {perturbative} truncation (at sufficient density)
is a reasonable approximation.

The matrix elements of the low-momentum potentials are typically
given in momentum representation or, more appropriately for the
present discussion, in a harmonic oscillator basis.
Then the KS orbitals are expanded in the same basis 
so that matrix elements such as $\{ij | V_{\rm NN} | kl \}$, 
which is in the orbital basis, can be evaluated.
(Note: the necessary
angular momentum recoupling is conventional;
see for example Ref.~\cite{Roth:2005ah}.)
To calculate $\vKS(\rvec)$ according to the OEP prescription, 
we need to compute the functional
derivatives of $E_{\rm Hxc}$ [see Eq.~\eqref{eq:vHplusvxc}] 
with respect to the orbitals and eigenvalues.
The eigenvalues appear explicitly in the Goldstone expansion 
and the derivatives of the potential
matrix elements can be carried out using the basis expansion without
having to evaluate the matrix elements in coordinate space.
For example, the two-body contribution from the Hartree-Fock terms is
\beqn
  \frac{\delta}{\delta\phi^\ast_i(\rvec)}
 \biggl\{ \frac12 \sum_{j,k=1}^A
  [\langle j k | \wh V_{\rm NN} | j k \rangle - \langle j k | \wh V_{\rm NN} | k j \rangle]
 \biggr\}
  = \sum_{j,k=1}^A \phi_k(\rvec)
  [\langle j k | \wh V_{\rm NN} | j i \rangle - \langle j k | \wh V_{\rm NN} | i j \rangle]  
 \;.
\eeqn
If the matrix element integrals are written out in coordinate space
for a local $V_{\rm NN}(\rvec,\rvec')$ and completeness
of the orbitals used to remove the sum over $k$, 
the corresponding part of Eq.~\eqref{eq:HFeq} is reproduced.

So we take~\cite{Sham:1985} 
\beqn
  \Hhat_0 = \Hhat_s = \wh T + \int\!d\rvec\, \wh\rho(\rvec)
    \vKS(\rvec)
    \;,
\eeqn
where $\vKS = \Vext + v_{\rm H} + v_{\rm xc} \equiv \Vext + v_{\rm Hxc}$.
The KS energy and density are
\bea
   E_0 &=& E_s = T_s + \int\!d\rvec\, \wh\rho(\rvec) \vKS(\rvec)
     = \sum_i n_i \varepsilon_i 
     \;, \\ 
   \rho(\rvec) &=& \Phibra \wh \rho(\rvec) \Phiket
     = \sum_i n_i |\phi_i(\rvec)|^2 \equiv \rho_{\rm KS}(\rvec)
     \;,  
\eea
which is the exact density according to the conditions (to be) imposed
on $\vKS$.
(Note: $\wh \rho(\rvec) = \wh\psi^\dagger(\rvec)\wh\psi(\rvec)$
where the field operators are
$\wh\psi^\dagger(\rvec) = \sum_i \phi_i^\ast(\rvec) a_i^\dagger$
and $\wh\psi(\rvec) = \sum_i \phi_i(\rvec) a_i^\dagger$.)
Then we have
\beqn
  \Hhat_1 = \wh \VN - \int\!d\rvec\, \wh\rho(\rvec) v_{\rm Hxc}(\rvec)
  \;.
\eeqn
At this point, we can carry out the Goldstone expansion to
any order or with whatever infinite summation we wish.

We note that because this expansion depends on $\vKS(\rvec)$, it is a highly
nonlinear functional equation, as $E_{\rm Hxc}$ depends on its
own functional derivative.  This is not in conflict with DFT because
$v_{\rm Hxc}$ is a density functional, so
\beqn
   E_{\rm Hxc} = 
   E_{\rm H} + E_{\rm xc} = E_1 + \int\!d\rvec\, [v_{\rm H}(\rvec)
         + v_{\rm xc}(\rvec)] \rho(\rvec)
\eeqn
is an implicit functional.
Furthermore, it is the same nonlinearity as in the OPM equation.
A perturbative approach to solving for the functional
derivative $v_{\rm Hxc}(\rvec) = \delta E_{\rm Hxc}[\rho]/\delta\rho(\rvec)$ 
leads to a linearization
of the OPM equation.  This will correspond to the inversion
method discussed in Section~\ref{sec:lt}.
The details of such a perturbative expansion and the nonperturbative
resummation of ring diagrams (the random phase approximation or RPA)
is described by Engel 
(with references to the original literature)~\cite{Fiolhais:2003}.

According to Engel the results for KS perturbation theory
are not satisfactory.
However, more recent developments have shown promise by
improving the perturbation theory.  
We have emphasized that with low-momentum
potentials, $\wh U$ can be chosen freely to allow the DFT constraints
to hold. 
But there is still freedom in defining it, which can be
used to significantly improve perturbative convergence in
the Coulomb problem.  We consider this in the next section.

\subsection{Improved perturbation theory}
\label{subsec:improvedpt}

In this section, we summarize the basic features of the approach
to \abinitio\ DFT under development by Bartlett and
collaborators~\cite{Bartlett:2005aa,Bartlett:2006aa}.
In Ref.~\cite{Bartlett:2006aa} they explain their strategy,
which is closely in line with the highly successful coupled
cluster approach to electronic systems.
A key part of the formulation as well as its numerical execution
is to use a finite basis set (which is almost always chosen to be gaussians
for finite systems in quantum chemistry so that all
integrals can be done analytically).
Above all, they emphasize that DFT should be formulated
so that it converges
in the limit of a full basis set,%
\footnote{The diagonalization of the Hamiltonian including all
possible particle-hole excitations in a given basis set is called
``full CI''.  That solution is an upper bound to the exact energy and
reproducing it using that same basis set is considered to be
an unambiguous measure of success in describing electron 
correlation~\cite{Bartlett:2007aa}.} 
as obtained
for wave function methods (cf.\ the partial hierarchy of 
CC approximations: $\mbox{MBPT(2)} < \mbox{CCSD} < \mbox{CCSD(T)}
< \mbox{full CI})$.
They contrast this precept with conventional DFT approaches such
as LDA or GGA or hybrid DFT (HDFT, which mixes semi-local GGA and
non-local HF exchange functionals), which do not consistently
predict better answers when applied in that order, 
even though they are claimed to
be better approximations.  That is, there is no definite hierarchy. 

The Bartlett construction follows most cleanly by requiring that corrections
to the density beyond the Kohn-Sham density, as calculated according
to MBPT, must vanish.  
As noted in Section~\ref{subsec:oepdensity},
this is an alternate
but equivalent prescription for orbital-based \abinitio\ DFT.
The Hamiltonian is split as in the last section into 
$\Hhat = \Hhat_0 + \Hhat_1$ using the Kohn-Sham potential
to define the reference state.
We will use $\PhiketKS$ instead of $\Phiket$ to make this clear.
The perturbative expansion for the energy 
is~\cite{Bartlett:2005aa,Bartlett:2006aa}
\beqn
  E = E_0 + E^{(1)} + E^{(2)} + \cdots
\eeqn
where (in quantum chemistry coupled cluster notation)
\bea
  E_0 &=& \PhibraKS \Hhat_0 \PhiketKS = \sum_i n_i \varepsilon_i \;,
  \\
  E^{(1)} &=& \PhibraKS \Hhat_1 \PhiketKS  \;,
  \\
  E^{(2)} &=& \PhibraKS \Hhat_1 \wh R_0 \Hhat_1 \PhiketKS  \;,
\eea
and so on.
Here the operator $\wh R_0$ is
\beqn
  \wh R_0 = \wh Q \frac{1}{E_0 - \Hhat_0} \wh Q \;,
\eeqn
with
\beqn
  \wh Q = \sum | \Phi_i^a\rangle \langle \Phi_i^a | +
  | \Phi_{ij}^{ab}\rangle \langle \Phi_{ij}^{ab} | +
  | \Phi_{ijk}^{abc}\rangle \langle \Phi_{ijk}^{abc} | + \cdots
  \;, 
\eeqn
which are singles, doubles, and higher excitations with respect
to the reference KS state (which is a determinant).
By convention, $a,b,c$ are unoccupied and $i,j,k$ are
occupied in the KS state.
We see that this agrees with the Goldstone expansion from before,
because $\wh Q$ ensures that only connected contributions are included
(i.e., it is a complete set of states excluding $\PhiketKS$).

Rather than use the functional derivative  to calculate $\vKS$,
the condition that the density be unchanged from the
Kohn-Sham density is used.  For calculating
the exchange-only KS potential $\wh V_{\rm x}$,
this implies the density be unchanged to first order in perturbation theory.
We expand the wave function and the density:
\bea
  | \Psi \rangle &=& \PhiketKS + | \Psi^{(1)}\rangle
        + | \Psi^{(2)}\rangle + \cdots  \;, \\
     \rho &=& \rho_{\rm KS} + \rho^{(1)} + \rho^{(2)} + \cdots  
     \;, 
\eea
where the density is computed from the expectation value of
 $\wh\rho(\rvec)$ (note that
Bartlett et al.\ write this using  $\wh\delta$, which is the
same as $\wh\rho$):
\beqn
   \rho(\rvec) = \frac{\langle \Psi | \wh\bdelta(\rvec) | \Psi \rangle}
                       {\langle \Psi | \Psi \rangle}
   \;.
\eeqn
So at leading order, we determine $v_{\rm x}(\rvec)$
such that $\rho^{(1)}(\rvec)=0$. 
Thus
\beqn
     \PhibraKS  \wh\bdelta(\rvec) | \Psi^{(1)} \rangle
       + \cc = 0
       \;,
\eeqn
where writing $\Hhat |\Psi\rangle = E |\psi\rangle$ to this order gives
\beqn
  (E_0 - \Hhat_0) | \Psi^{(1)} \rangle = (\Hhat_1 - E^{(1)}) \PhiketKS
\eeqn
and therefore
\beqn
   | \Psi^{(1)} \rangle = \wh R_0 (\Hhat_1 - E^{(1)}) \PhiketKS
   \;. 
\eeqn
Substituting for $| \Psi^{(1)} \rangle$,
\beqn
  \rho^{(1)} = 0 = \PhibraKS \wh\bdelta(\rvec) \wh R_0 \wh H_1
  \PhiketKS + \cc \;,
\eeqn
and noting that at this order only single excitations contribute,  we obtain
\bea
  0 &=& \sum_{j,i,a} \langle i | \wh\bdelta(\rvec) | a \rangle
      (\{ a j | \Hhat_1 | i j \}
       -\langle a | \wh V_{\rm Hx} | i \rangle) / (\varepsilon_i -
       \varepsilon_a) + \cc \nonumber
  \\
    &=& -\sum_{i,a} \phi_i^\ast(\rvec) \phi_a(\rvec)
      (\langle a | \wh K | i \rangle
       +\langle a | \wh V_{\rm x} | i \rangle) / (\varepsilon_i -
       \varepsilon_a) + \cc \;.
       \label{eq:bartoep}
\eea
where we have used 
$\langle \Phi_i^a | E_0 - \Hhat_0 | \Phi_i^a \rangle = \varepsilon_i -
\varepsilon_a$.
Here $\wh K$ is the conventional
exchange (Fock) energy operator $\wh \VN P_{12}$,
where $P_{12}$ is the particle-exchange operator.

Note that we cannot choose $\langle a | \wh K | i \rangle
= -\langle a | \wh V_{\rm x} | i \rangle$ because $\wh V_{\rm x}$ must
be local.
We can think of Eq.~\eqref{eq:bartoep} as the solution to a weighted least squares
problem to replace the non-local $\wh K$ by the local $\wh V_{\rm x}$
in the space spanned by $\{\phi_i^\ast,\phi_a\}$~\cite{Bartlett:2005aa}.  
This is a
point-wise identity that can be written
[with $\chi_s$ from Eq.~\eqref{eq:chisdef}]
\beqn  
  \int\! d\rvec'\, \chi_s(\rvec,\rvec') [v_{\rm x}(\rvec')
    +  K(\rvec')] = 0 \;,
    \label{eq:bartchis}
\eeqn
which agrees with the OEP equation 
in Section~\ref{sec:dftoep}; that is, we have
rederived the OEP equation with exchange only.
Here
\beqn
   K(\rvec)\phi_j(\rvec)
  = \int\! d\rvec'\, \rho(\rvec,\rvec') \VN(\rvec,\rvec')  
  \phi_j(\rvec')
\eeqn
for a local potential with
\beqn
   \rho(\rvec,\rvec') = \sum_i n_i
   \phi_i^\ast(\rvec')\phi_i(\rvec)
   \;.
\eeqn
Details on solving Eq.~\eqref{eq:bartchis} in matrix form are given in 
Refs.~\cite{Bartlett:2005aa,Bartlett:2006aa} with
pointers to the literature.

At second order, where the correlation potential first enters, 
one encounters issues with slow convergence.  
Here we simply sketch the problem and
proposed solution to give the flavor.  
Working to second-order MBPT, the second order
energy is straightforward to write but naturally more complex.
One insists that $\rho^{(1)} + \rho^{(2)} = 0$ and this defines
$\wh V_{\rm c}^{(2)}$ consistent to this order.  The problem is that the
perturbation $\Hhat_1^{(1)} - E^{(1)} = \wh \VN - (\wh V_{\rm H}
 + \wh V_{\rm x}^{(1)}) - E^{(1)}$ 
(note: no $\wh V_{\rm c}$ at this order) can have large, diagonal
contributions~\cite{Bartlett:2006aa}:
\bea
  && \PhibraKS  \wh\bdelta(\rvec) | \Phi_i^a \rangle
  \langle \Phi_i^a | (E_0 - \Hhat_0)^{-1} | \Phi_i^a \rangle
  \langle \Phi_i^a | \Hhat^{(1)} - E^{(1)} | \Phi_i^b\rangle
   \\
   && \hspace*{1cm}
   = - \Bigl[\langle a j | \wh \VN (1-P_{12}) | b i \rangle - \delta_{ij}
     \langle a | \wh K + \wh V_{\rm x} | b \rangle
     - \delta_{ab} \langle j | \wh K + \wh V_{\rm x} | i \rangle
       \Bigr] \frac{\phi_i(\rvec)\phi_a^\ast(\rvec)}
                   {\varepsilon_i - \varepsilon_a}
      \;.             
\eea
Because we can have $a=b$ and $i=j$, these are diagonal elements of
$\wh K + \wh V_{\rm x}$, which make $\Hhat^{(1)}$ a larger
perturbation than usual~\cite{Bartlett:2005aa,Bartlett:2006aa}.  
The remedy is to resum diagonal and near-diagonal
one-particle terms into $\Hhat'_0$ (this includes a rotation of the KS
orbits so that off-diagonal terms in $\Hhat'_0$ vanish).
Most of the effect from the resummation comes
with the replacement of KS denominator
energies $\varepsilon_p$ ($p$ is either a particle or hole) 
with diagonal Fock matrix elements $f_{pp}$,
where
\beqn
  f_{pp} = \langle p | f | p \rangle
    = \varepsilon_p - \langle p | \wh K + \wh V_{\rm xc} | p \rangle
    \;.
\eeqn
The details are beyond the scope of this review but can be found
in Ref.~\cite{Bartlett:2005bb} along with results.

In summary, the key issue is that the freedom in how we
 split the Hamiltonian
can be exploited to improve convergence of the perturbation series
without sacrificing the form of \abinitio\ DFT.  In particular,
rather than use the standard Kohn-Sham choice (called
G\"orling-Levy perturbation
theory~\cite{Gorling:1994aa,Gorling:1995aa}),
we shift as much physics as possible into $\Hhat_0$.
While far from a complete treatment, we hope we have conveyed
the flavor of the Bartlett approach.
As a side note, Bartlett et al.\ claim that observations made for
conventional DFT that orbital energies are meaningless (other than
the least bound) are mostly because functionals and potentials
were not accurate.


\subsection{Low-momentum interactions}
\label{subsec:vlowk}

The last two sections illustrate how exchange-correlation functions 
can be constructed for Coulomb systems
for which many-body perturbation theory is applicable.
This is irrelevant for nuclear physics unless we have similar control
of MBPT. 
Here we highlight recent results that show that modern RG methods can be used
to obtain a convergent expansion of nuclear matter properties by
evolving to low-momentum
interactions.
There are various methods to achieve these interactions starting
from phenomenological or EFT potentials (e.g., see
Refs.~\cite{Roth:2005pd,Bogner:2006ai,Bogner:2006pc}),
which for our purposes here are equivalent.
We simply cite two proof-of-principle results and refer the reader
to the literature.

First we consider the uniform system.
Despite a decades-old emphasis on infinite nuclear matter,
most advances in microscopic nuclear structure theory over the
last decade have been through 
expanding the reach of few-body calculations.
This has unambiguously established the quantitative
role of three-nucleon forces (3NF) for light nuclei 
($A \leqslant 12$)~\cite{Pieper:2001mp,Pieper:2004qh,Navratil:2007we,Hagen:2007ew}.
Until recently, few-body fits
have not sufficiently constrained 3NF contributions at
higher density such that nuclear matter calculations are predictive. 
Nuclear matter saturation is very delicate, with the binding energy
resulting from
cancellations of much larger potential and kinetic energy contributions.
When a quantitative reproduction of empirical saturation properties
has been obtained, it was
imposed by hand through adjusting
short-range three-body forces (see, for example, 
Refs.~\cite{Akmal:1998cf,Lejeune:2001bg}).

Progress for controlled nuclear matter calculations
has long been hindered by the difficulty
of the nuclear many-body problem when conventional
nuclear potentials are used.
But recent calculations~\cite{Bogner:2009un} 
overcome the hurdles by combining
controlled starting Hamiltonians based on chiral
effective field theory (EFT)~\cite{Entem:2003ft,Epelbaum:2004fk} with
renormalization group (RG) methods~\cite{Bogner:2003wn,Bogner:2006vp} to soften
the short-range repulsion and short-range tensor components of the
initial chiral interactions~\cite{Bogner:2006tw}. By doing so, 
the convergence of 
many-body calculations is vastly
accelerated~\cite{Bogner:2005sn,Bogner:2007rx,Bacca:2009yk}. 

\begin{figure}[t]
\begin{center}
\includegraphics[width=7in]{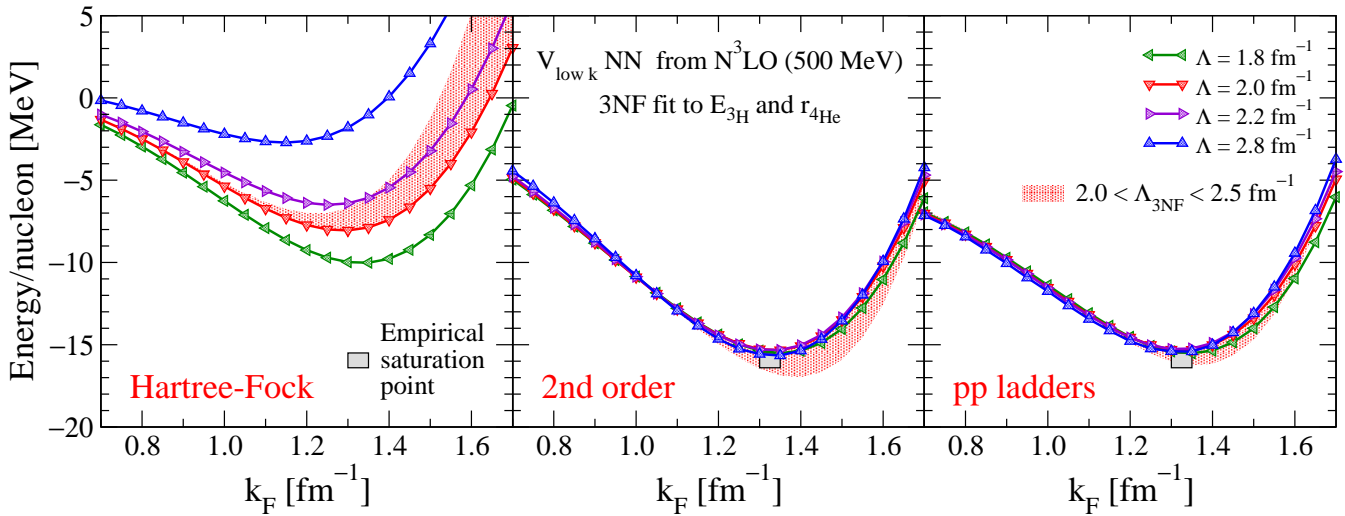}
\end{center}
\caption{Nuclear matter energy per particle as a
function of Fermi momentum $\kf$ at the Hartree-Fock level (left)
and including second-order (middle) and particle-particle-ladder 
contributions (right), based on evolved N$^3$LO NN potentials and
3NF fit to $E_{\rm^3H}$ and $r_{\rm^4He}$. Upper bounds to the theoretical 
uncertainties are estimated by the NN (lines) and 3N (band)
cutoff variations~\cite{Bogner:2009un}.}
\label{nm_all}
\end{figure}

Some key results are summarized in Fig.~\ref{nm_all},
which shows the energy per particle of symmetric matter
as a function of Fermi momentum $\kf$, or the density $\rho 
= 2 \kf^3/(3\pi^2)$.
A grey square representing the empirical saturation 
point is shown in each of the nuclear matter figures.
Its boundaries reflect the ranges of nuclear matter saturation
properties predicted by phenomenological Skyrme energy functionals that
most accurately reproduce properties of finite nuclei.
The calculations of Fig.~\ref{nm_all} start from the
N$^3$LO nucleon-nucleon (NN) potential (EM $500 \mev$) of 
Ref.~\cite{Entem:2003ft}. This NN potential is RG-evolved to
low-momentum interactions $\vlowk$ with a smooth 
$n_{\rm exp}=4$ regulator~\cite{Bogner:2006vp}. For each 
cutoff $\Lambda$, two couplings that determine the 
shorter-range parts of the ${\rm N^2LO}$ 
3NF~\cite{VanKolck:1994yi,Epelbaum:2002vt}
are fit to the $^3$H binding energy and the $^4$He matter
radius~\cite{Navratil:2007we} using exact Faddeev and Faddeev-Yakubovsky
methods as in Ref.~\cite{Nogga:2004ab}.

The Hartree-Fock results show that nuclear matter is bound even at
this simplest level.
A calculation without approximations should be independent of the
cutoffs, so the spread in Fig.~\ref{nm_all}
sets the scale for omitted many-body
contributions (more precisely, it sets a lower bound to omitted
contributions). The second-order results show a dramatic 
narrowing of this spread, with predicted saturation 
consistent with the empirical range. 
The narrowing happens across the full density range. This 
is strong evidence that these encouraging results are not 
fortuitous.
The particle-particle-ladder sum is little changed from 
second order except at the lowest densities shown. The latter
is not surprising because at very low density the presence of 
a two-body bound state necessitates a nonperturbative summation.
Similar results are obtained using 
flow equations to evolve
Hamiltonians, which is called the Similarity Renormalization
Group (SRG)~\cite{Glazek:1993rc,Wegner:1994aa,Bogner:2006pc,Roth:2005pd}
in nuclear physics.

The decrease in cutoff dependence in Fig.~\ref{nm_all} with more
complete approximations is necessary but not sufficient to conclude
that the calculations are under control.
Indeed, approximations that are independent of the cutoff will shift the answer
but not widen the error band from cutoff variation.
The theoretical errors 
arise from truncations in the initial chiral EFT Hamiltonian,
the approximation of the 3NF, and the many-body approximations.
The 3NF approximation is particularly uncertain because 
it involves long-range contributions independent of the cutoff.
Many-body corrections to the current approximations  
include higher-order terms
in the hole-line expansion and particle-hole corrections.
An approach such as coupled cluster theory that can
perform a high-level resummation including
long-range correlations will ultimately be necessary for a robust
validation.

These results, while not conclusive, open the door
to ab-initio density functional theory (DFT) both directly
(as in the last sections) but also
based on expanding about nuclear matter~\cite{Bogner:2008kj} (next
section).
This is analogous to the application of
DFT in quantum chemistry and condensed matter starting with the uniform
electron gas in local-density approximations and adding constrained derivative
corrections.  Phenomenological energy functionals (such as Skyrme) 
for nuclei have
impressive successes but lack a (quantitative) microscopic foundation
based on
nuclear forces and seem to have reached the limits of
improvement with the current form of 
functionals~\cite{Bertsch:2004us,Kortelainen:2008rp}.
On the other hand,
the theoretical errors of the calculations in Fig.~\ref{nm_all}, while impressively
small on the scale of the potential energy per particle,
are far too large to be quantitatively competitive with 
existing functionals. However, there is the possibility of fine 
tuning to heavy nuclei, of using EFT/RG to guide next-generation
functional forms, and of benchmarking with ab-initio methods 
for low-momentum interactions. 
Overall, these results are quite promising for a unified
description of all nuclei and nuclear matter but much work is
left to be done.  

Calculations of finite nuclei by another group using soft potentials
further support the use of MBPT.
Roth and collaborators have developed a method using designed
unitary transformations to remove short-range ``hard core'' and
tensor correlations~\cite{Roth:2006gx}.  
The approach is called the ``Unitary Correlation
Operator Method'' or UCOM.  
The result is a soft Hamiltonian that shares the favorable features of
the RG-based potentials.  They have also shown close parallels (and
some distinctions) of UCOM
to the SRG approach~\cite{Hergert:2007wp,Roth:2008km}.
Using a UCOM NN potential, which is phase equivalent to a given
initial Hamiltonian (in this case Argonne $v_{18}$~\cite{Wiringa:1994wb}), 
they have
calculated many nuclei (including heavy nuclei
such as $^{208}$Pb) in HF plus second- and third-order 
MBPT corrections.  The energies and radii obtained seem to be remarkably
converged at second order with good systematics~\cite{Roth:2005ah}.
Preliminary results from
supplementing the NN potential 
by a contact 3-body interaction with fitted strength are
very encouraging~\cite{Roth:2009private}.
They have also examined corrections from particle-hole states in the
RPA for closed-shell nuclei; these are found to be relatively small
corrections~\cite{Barbieri:2006vf}.

Thus MBPT appears to be a viable candidate for calculating a nuclear
exchange-correlation functional.  However, there
are three important caveats.
Detailed comparisons with a method that does high-order resummations
will be necessary before one can make robust conclusions.  In this regard,
converged
calculations using soft potentials with coupled cluster and 
no-core-shell-model 
techniques are feasible now in $^{16}$O including 3-body forces
and soon%
\footnote{These calculations may require an approximation
based on a normal-ordering truncation
to treat the 3-body force~\cite{Hagen:2007ew}, 
which has not been fully validated for these nuclei.} 
 in $^{40}$Ca
 (NN-only calculations for CC are possible
now even for an unevolved chiral EFT potential~\cite{Hagen:2008iw}).
Second, tests of MBPT have so far been largely limited to spherical nuclei
where pairing plays a small role.
The final point to emphasize again 
is that corrections are \emph{relatively} small,
but not absolutely small and are large compared to the accuracy
sought for functionals, so additional contributions must
eventually be considered.


\subsection{Density matrix expansion}
  \label{subsec:dme}

While formally the construction of an \abinitio\ DFT 
energy functional is well-defined based
on a many-body perturbation expansion (resummed as needed) 
about a Kohn-Sham reference state, in practice the feasibility
of using KS potentials from such a functional beyond Hartree-Fock
has only recently been demonstrated.  
This is an ongoing area of research in quantum
chemistry, with mixed results, and progress in nuclear physics
will require many further developments.  
A more immediate route to a DFT functional
based on microscopic nuclear interactions is to make a quasi-local
expansion of the energy in terms of various densities,
so that functional derivatives needed to define Kohn-Sham
potentials are immediate.
Of course, in doing so one sacrifices the full non-locality
present in an orbital-based functional.
An example of such an expansion is the density matrix expansion
(DME) introduced by Negele and
Vautherin~\cite{Negele:1972zp,Negele:1975zz}, which we
describe here in some detail. 

The strategy is to follow a path that will be compatible
with current nuclear DFT
technology but testable and systematically improvable.
In this regard, 
the phenomenological nuclear energy density functionals of
the Skyrme form are a good starting point to build on the
MBPT with low-momentum interactions.  
Modern Skyrme functionals have been applied over a very wide range
of nuclei, with quantitative success in reproducing properties
of nuclear ground states and low-lying 
excitations~\cite{Dobaczewski:2001ed,Stoitsov:2003pd,Bender:2003jk}.
Nevertheless, a significant reduction of the current global and local errors
is a major goal~\cite{unedfweb}.
One strategy is to improve the functional itself;
the form of the basic Skyrme functional in use is very restricted,
consisting of a sum of local powers of various nuclear
densities, e.g., for $N=Z$ nuclei~\cite{Ring:2005}:
\bea
  E_{\rm Skyrme}[\rho,\tau,\textbf{J}] 
    &=& \int\!d^3x\,
    \biggl\{ \frac{1}{2M}\tau + \frac{3}{8} t_0 \rho^2
  + \frac{1}{16} t_3 \rho^{2+\alpha}
 + \frac{1}{16}(3 t_1 + 5 t_2) \rho \tau  \nonumber
  \\ & & \hspace*{-.1in}\null
  + \frac{1}{64} (9t_1 - 5t_2) (\bm{\nabla} \rho)^2  
  - \frac{3}{4} W_0 \rho \bm{\nabla}\bfcdot\textbf{J}
  + \frac{1}{32}(t_1-t_2) \textbf{J}^2 \biggr\}  
  \;,
  \label{eq:ESHF}
\eea
where the density $\rho$, the kinetic density $\tau$, and the spin-orbit
density $\textbf{J}$ are expressed as sums 
over single-particle orbitals.
(Expressions for the Skyrme functional including isovector
and more general densities can be found in Ref.~\cite{Perlinska:2004wh}.)
Fits to measured nuclear data 
have given to date only limited constraints on possible
density and isospin dependencies and on the form of the spin-orbit interaction.
Even qualitative insight into these properties from realistic microscopic
calculations could be beneficial
in improving the effectiveness of the energy density functional.

A theoretical connection of the Skyrme functional to free-space NN
interactions was made long ago by Negele and Vautherin using the
density matrix expansion 
(DME)~\cite{Negele:1972zp,Negele:1975zz,Hofmann:1997zu}, 
but there have
been few subsequent microscopic developments.
The DME originated as an expansion of the Hartree-Fock energy
constructed using the nucleon-nucleon 
G matrix  \cite{Negele:1972zp,Negele:1975zz}, which was
treated in a local (i.e., diagonal in coordinate representation) approximation.
Recently the DME has been revisited for spin-saturated nuclei
using non-local low-momentum interactions in 
momentum representation~\cite{Bogner:2008kj}, 
for which G matrix summations are
not needed because of the softening of the interaction (see
Section~\ref{subsec:vlowk}).
When applied to a Hartree-Fock energy functional, the DME yields
an energy functional in
the form of a generalized Skyrme functional that is compatible
with existing codes, by replacing 
Skyrme coefficients with density-dependent functions.
As in the original application, a key feature of the DME is
that it is not a pure short-distance expansion but includes
resummations that treat long-range pion interactions correctly
in a uniform system.
However, we also caution that
the Negele-Vautherin DME involves prescriptions for the resummations without
a corresponding power counting to justify them.

\begin{figure}[t]
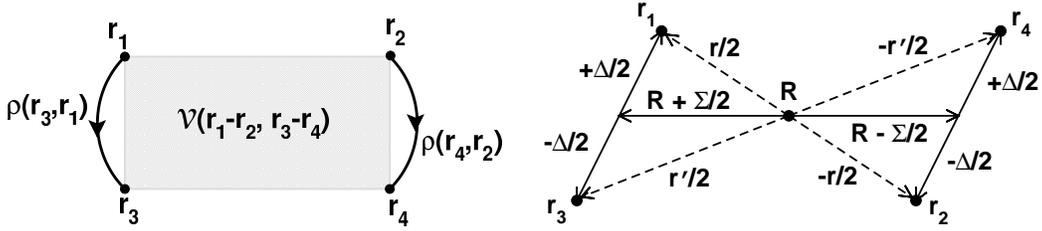

 \begin{center}
  \includegraphics*[width=2.6in]{fig_dme_Vblob}
  \hspace*{.1in}
  \includegraphics*[width=2.6in]{fig_dme_nonlocal2}
 \end{center}
 \caption{(a)~Schematic diagram for approximations
 to $\Wint$ that can be
 expanded using the DME. 
 (b)~Coordinates appropriate for the DME applied
 to the Hartree-Fock potential energy
 with a non-local potential.}
 \label{fig:dme_nonlocal}
\end{figure}

In essence, the DME
maps the orbital-dependent expressions for 
contributions to the interaction energy $\Wint$ of the type in 
Fig.~\ref{fig:dme_nonlocal}(a) into 
a semi-local form, with explicit dependence on the local 
densities $\rho(\Rvec)$, $\tau(\Rvec)$, $\nabla^2\rho(\Rvec)$, 
and so on.
This greatly simplifies the determination of the
Kohn-Sham potential because the functional
derivatives determining the KS potentials can be evaluated directly.
The density matrix expansion 
(DME) for spin-saturated nuclei has been formulated for low-momentum
interactions and applied to a Hartree-Fock energy functional including
both NN and NNN potentials in Ref.~\cite{Bogner:2008kj}.
The output is a set of functions of density that can replace 
density-independent
parameters in standard Skyrme Hartree-Fock codes. 
Furthermore, the upgrade from Skyrme energy functional to DME energy
functional can be carried out in stages.  For example, the spin-orbit
part and pairing can be kept in Skyrme form with the rest given by 
the DME. 
A further upgrade to orbital-based methods would only modify the same
part of the code, although the increased computational load will
be significant.

Here we outline how the density matrix
expansion is carried out for a microscopic DFT at HF order 
using low-momentum (and non-local) two-body potentials. 
The relevant object we need to
expand is $\Wint$, which is expressed in terms of 
the Kohn-Sham orbitals and eigenvalues that comprise the Kohn-Sham
single-particle propagators as in Section~\ref{subsec:mbpt}. 
For Hartree-Fock contributions
only the orbitals enter. Higher-order 
contributions such as the ladder diagrams in the particle-particle 
channel can also be put approximately into the required form by averaging
over the state dependence arising from 
the intermediate-state energy denominators (cf.\ the KLI
approximation). 
Therefore, results presented here for the 
Hartree-Fock contributions to the functional can be
generalized to approximately include selected higher-order contributions.
However, a framework for systematic improvement is yet to be
developed. 

Before considering the DME derivation and
its application to non-local low-momentum interactions, it
is useful to first derive in some detail the starting expression
for $\WHF$,
the Hartree-Fock contribution to $\Wint$. 
This introduces the basic notation
and highlights the differences between most existing DME 
studies, which are formulated with local interactions and in coordinate 
space throughout, and an approach formulated
in momentum space and geared towards non-local potentials.
For a local potential, the distinction between the direct (Hartree) 
and exchange (Fock)
contributions is significant, 
and is reflected in the conventional decomposition
of the DFT energy functional for Coulomb systems, 
which separates out the Hartree piece.
For a non-local potential, the distinction is blurred because
the Hartree contribution now involves the density matrix
(as opposed to the density)
and it is not useful to make this separation when the range
of the interaction is comparable to the non-locality.%
\footnote{However, it is useful to separate out the long-distance
part of the potential, which is local, and treat its direct 
(Hartree) contribution
exactly.}
Consequently, throughout this section we work instead 
with an antisymmetrized interaction. 

For a general (i.e., non-local)
free-space two-body potential $\wh V_{\rm NN}$, $\WHF$ is defined in terms
of Kohn-Sham states [Eq.~(\ref{eq:kseq})] labeled by $i$ and $j$, 
\bea
 \WHF &=&\frac{1}{2}
\sum_{i j}^{A} \langle i j | \widehat V_{\rm NN}(1-P_{12})|i j\rangle 
= \frac{1}{2}\sum_{i j}^{A} \langle i j | \widehat{ \mathcal{V}}|i
j\rangle  \;.
\eea
The summation is over the occupied states
 and the
antisymmetrized interaction $\widehat{\mathcal{V}}=\widehat V_{\rm NN} (1-P_{12})$  has
been introduced, with the exchange operator $P_{12}$ equal to
the product of operators for spin, isospin, and space exchange,
$P_{12} = P_{\sigma}P_{\tau}P_r$.  Note that the dependence
of $\WHF$ on the Kohn-Sham potential is implicit. 
By making repeated use of the completeness relation 
in space ({\bf r}), spin ($\sigma$), and isospin ($\tau$),
\beqn
 \openone =  \sum_{\sigma\tau}\int\!d{\bf r}|{\bf r}\sigma\tau\rangle\langle 
{\bf r}\sigma\tau|
  \;,
\eeqn
$\WHF$ can be written in
terms of the coordinate space Kohn-Sham orbitals as
\bea
\label{eq:HF_orbitals}
\WHF &=&\frac{1}{2}\sum_{i j} \sum_{\{\sigma \tau\}}
\!\int\! d\rone\! \int\! d\rtwo\! \int\! d\rthree\! \int\! d\rfour\,
\langle \rone\sigma_1\tau_1\rtwo\sigma_2\tau_2|\widehat{\mathcal{V}}
|\rthree\sigma_3\tau_3\rfour\sigma_4\tau_4\rangle
\nonumber
\\
&&\quad\null\times
\phi^*_{i}(\rone\sigma_1\tau_1)\phi_{i}(\rthree\sigma_3\tau_3)
\phi^*_{j}(\rtwo\sigma_2\tau_2)\phi_{j}(\rfour\sigma_4\tau_4) \;.
\eea
From the definition of the Kohn-Sham density matrix,
\beqn
  \label{eq:densitymat}
  \rho(\rthree\sigma_3\tau_3,\rone\sigma_1\tau_1)=
  \sum_{i}^{A}\phi^*_{i}(\rone\sigma_1\tau_1)\phi_{i}(\rthree\sigma_3\tau_3)
  \;,
\eeqn
so Eq.~(\ref{eq:HF_orbitals}) can be written as 
\bea
  \label{eq:HF_DM1}
  \WHF &=&\frac{1}{2} \sum_{\{\sigma \tau\}}
  \!\int\! d\rone\cdots \int\! d\rfour\,
  \langle \rone\sigma_1\tau_1\rtwo\sigma_2\tau_2|\widehat{\mathcal{V}}
  |\rthree\sigma_3\tau_3\rfour\sigma_4\tau_4\rangle
  \rho(\rthree\sigma_3\tau_3,\rone\sigma_1\tau_1)
  \rho(\rfour\sigma_4\tau_4,\rtwo\sigma_2\tau_4)\nonumber
  \\
  &=&\frac{1}{2}\tr_{1}\tr_{2}
  \!\int\! d\rone\cdots \int\! d\rfour\, \langle\rone\rtwo| \antisymV^{1\otimes 2} |\rthree\rfour\rangle
  \rhobold^{(1)}(\rthree,\rone)\rhobold^{(2)}(\rfour,\rtwo)
   \;,
\eea
where a matrix notation is used
in the second equation and the traces denote 
summations over the spin and isospin indices 
for ``particle 1" and ``particle 2''. Hereafter we drop the 
superscripts on  $\antisymV$ and $\rhobold$
that indicate
which space they act in as it will be clear from the context.  

Expanding the $\rhobold$ matrices on Pauli spin and isospin matrices we have
\beqn
\rhobold(\rovec_1,\rovec_2) = \frac{1}{4}[\rho_0(\rovec_1,\rovec_2) 
                            + \rho_1(\rovec_1,\rovec_2)\tau_z
                       + \vec{S}_0(\rovec_1,\rovec_2)\cdot\vec{\sigma}
           +\vec{S}_1(\rovec_1,\rovec_2)\cdot\vec{\sigma}\tau_z]
	   \;,
\eeqn
where we have assumed the absence of charge-mixing in the
single-particle states and the
components are obtained by taking the relevant traces of $\rhobold$.
From now on we consider only terms in the energy functional
arising from products of the scalar-isoscalar ($\rho_0$) density matrices in
Eq.~(\ref{eq:HF_DM1}), which 
are the relevant terms for spin-saturated systems with $N=Z$. 
Because there will be no confusion, 
we will drop the subscript ``$0$'' on the density matrices.

The first step is to switch to relative/center-of-mass (COM) coordinates
(see Fig.~\ref{fig:dme_nonlocal}).  The
free-space two-nucleon potential is diagonal in the COM coordinate, 
so the starting 
point for the DME of the two-body Hartree-Fock contribution 
from a non-local interaction is 
\beqn
\label{eq:hf_potential_contr}
\WHF =\frac{1}{32}
\int d{\bf R}\,d{\bf r}\,
d{\bf r'}\,\rho({\bf R}+\frac{\bf r'}{2},{\bf R}+\frac{\bf r}{2})
\rho({\bf R}-\frac{\bf r'}{2},{\bf R}-\frac{\bf r}{2}) 
 \tr_{\sigma\tau}[\langle\rovec|\antisymV|\rovec'\rangle] \;,
\eeqn
where $\antisymV$ denotes the antisymmetrized interaction
and the trace is defined as
\beqn
\tr_{\sigma\tau}[\langle\rovec|\antisymV|\rovec'\rangle] \equiv
\sum_{\{\sigma\tau\}} \langle \rovec\sigma_1\tau_1\sigma_2\tau_2|
\widehat{V}(1-P_{12})|\rovec'\sigma_1\tau_1\sigma_2\tau_2\rangle
\;.
\eeqn
The DME derivation of Negele and
Vautherin (NV)~\cite{Negele:1972zp} 
focuses
on applications to local potentials,
which satisfy 
$\langle\rovec |\widehat{V}
|\rovec'\rangle = \delta(\rovec-\rovec')\langle
\rovec|\widehat{V}|\rovec'\rangle$. 
While the original NV work included 
coordinate-space formulas applicable for non-local interactions, for
low-momentum potentials it is advantageous to revisit
and extend the original derivation to a momentum-space formulation.
(Note that Kaiser et al.\ have shown how to
use medium-insertions in momentum space in their application of
the DME to chiral perturbation theory at finite 
density~\cite{Kaiser:2002jz,Kaiser:2003uh,Fritsch:2004nx}.)

For the momentum space formulation, we first 
rewrite the density matrices in
Eq.~(\ref{eq:hf_potential_contr}) as
\bea
  \rho({\bf R}\pm{\bf r'}/{2},{\bf R}\pm{\bf r}/{2})=
  \rho({\bf R}^\pm\pm{\bDel }/{2},{\bf R}^\pm\mp{\bDel}/{2})\;,
\eea
where the vectors appearing on the right-hand side are defined by  
(see Fig.~\ref{fig:dme_nonlocal})
\bea
  {\bf R}^\pm={\bf R}\pm\frac{1}{2}\bSig \;, \qquad
\bSig=\frac{1}{2}({\bf r'}+{\bf r})\;, \qquad
\Delta=\frac{1}{2}({\bf r'}-{\bf r}) \;.
\eea
Introducing the Fourier transform
of $\antisymV$ in the momentum transfers conjugate to
$\bSig$ and $\Delta$, 
\beqn
 \qvec=\kvec-\kpvec \;, \qquad \pvec = \kvec+\kpvec \;,
\eeqn
(where $\kpvec$, $\kvec$ correspond to 
relative momenta) gives
\bea
\label{eq:HF_kspace}
\WHF = \frac{1}{32}\int d\Rvec\int 
        \frac{d\qvec\,d\pvec}{(2\pi)^6}\,F(\Rvec,\qvec,\pvec)
   \, \tr_{\sigma\tau}[\widetilde{\antisymV}(\qvec,\pvec)]
  \;,\nonumber
\eea
where we have defined
\beqn
\label{eq:Fqp}
F(\Rvec,\qvec,\pvec) \equiv \int d\bSig\,d\bDel\, 
  e^{i\qvec\cdot\bSig}\,e^{i\pvec\cdot\bDel}
  \,\rho(\Rvec^+ -\bDel/2,\Rvec^+ +\bDel/2)
\rho(\Rvec^- +\bDel/2,\Rvec^- -\bDel/2)\;, 
\eeqn
and
\beqn
\label{eq:Vqp}
\widetilde{\antisymV}(\qvec,\pvec)
 \equiv 8\int d\bSig\,d\bDel\, e^{-i\qvec\cdot\bSig}\,
e^{-i\pvec\cdot\bDel}\,\langle\bSig-\bDel|\antisymV|\bSig+\bDel\rangle\;.
\eeqn
The momenta $\qvec$ and $\pvec$ correspond to the momentum transfers for a
local interaction in the direct and exchange channels. That is, the direct
matrix element is a function of $\qvec$ and the exchange is a function of
$\pvec$. In contrast, for a non-local interaction the direct and exchange matrix
elements depend on both   $\qvec$ and $\pvec$. This is the reason why we do not
attempt to separate out the Hartree (direct) and Fock (exchange) contributions
to $\WHF$, as is commonly done for local interactions.

The trace of Eq.~(\ref{eq:Vqp}) can be written in a more convenient form for
the application of low momentum NN interactions
as a sum over partial wave matrix elements (see Ref.~\cite{Bogner:2008kj}
for more complete definitions),
\bea
 \label{eq:Vqplsjt}
 \tr_{\sigma\tau}[\widetilde{\antisymV}(\qvec,\pvec)] =
 8\pi \sum_{lsj}\,^{'} (2j+1)(2t+1)\,P_{l}(\widehat{\kvec}
 \cdot\widehat{\kpvec})\langle klsjt|V|k'lsjt\rangle\;,
\eea
 where the primed summation 
 means that it is restricted to values where $l+s+t$ is odd,
 with $\kvec = \frac{1}{2}(\pvec+\qvec)$ and $\kpvec=\frac{1}{2}
 (\pvec-\qvec)$. For simplicity we assume a 
 charge-independent two-nucleon interaction, although charge-dependence
 can easily be included. 

The expression Eq.~(\ref{eq:HF_kspace}) for $\WHF$ 
is written in terms of off-diagonal density 
matrices constructed from the Kohn-Sham orbitals and so 
is an \emph{implicit} functional of the density. 
To circumvent the application of the chain rule for the KS potentials,
we apply Negele and Vautherin's DME to $\WHF$, resulting in an expression
of the form
\beqn
   E_{\rm DME}[\rho,\tau,\textbf{J}] 
     = \int\!d\Rvec\, \mathcal{E}_{\rm DME}(\rho(\Rvec),\tau(\Rvec),
                \textbf{J}(\Rvec))
     \;,
     \label{eq:EDME}
\eeqn
with \emph{explicit} dependence on the local quantities
$\rho(\Rvec)$,  $\tau(\Rvec)$, and $|\nabla \rho(\Rvec)|^2$
that we write for $\WHF$ as
\beqn
  \WHF = \int\! d\Rvec\,( A[\rho] + B[\rho]\tau +
    C[\rho](\nabla\rho)^2 + \cdots)
  \;.
  \label{eq:WHFdme}
\eeqn
(We have suppressed
terms that go beyond the present limited discussion; 
also, when $N\neq Z$ these are functions of the isovector densities
as well.) 
The goal is to find the coefficient functions in Eq.~\eqref{eq:WHFdme}.
The starting point of the DME is the 
formal identity~\cite{Negele:1972zp}
\beqn
  \rho({\bf R}+{\bf s}/{2},{\bf R}-{\bf s}/{2})  =
  \sum_a \phi^{*}({\bf R}+{\bf s}/{2})\phi({\bf R}-{\bf s}/{2})
  = \bigl[e^{{\bf s}\bfcdot(\nabla_1-\nabla_2)/2}\sum_a
  \phi^{*}({\bf R}_1)\phi({\bf R}_2)\bigr]_{{\bf R}_1={\bf R}_2={\bf R}}\;,
\eeqn 
where $\nabla_1$ and $\nabla_2$ act on ${\bf R}_1$ and ${\bf R}_1$,
respectively, and the result is evaluated at 
${\bf R}_1={\bf R}_1={\bf R}$. 
We assume here that
time-reversed orbitals are filled pairwise, so that the linear term of the
exponential expansion vanishes. 
After applying a 
Bessel-function expansion
(which is simply the usual plane-wave expansion with real arguments),
the angle-averaged density matrix takes the form
\beqn
\label{eq:canonicalDME}
  \hat{\rho}({\bf R}+{\bf s}/{2},{\bf R}-{\bf s}/{2})
   =
   \frac{1}{s \kf (\Rvec)}\Biggl[\sum_{n=0}^\infty(4n+3)j_{2n+1}(s \kf (\Rvec))
  \mathcal{Q}_n\biggl(\biggl(\frac{\nabla_1-\nabla_2}{2\kf (\Rvec)}\biggr)^2\biggr)\Biggr]
  \rho({\bf R}_1,{\bf R}_2)\;,
\eeqn
where $\mathcal{Q}$ is related to the usual Legendre polynomial by
$\mathcal{Q}(z^2)={P_{2k+1}(iz)}/(iz)$ and
an arbitrary momentum scale $\kf (\Rvec)$ has been introduced. 
Equation~(\ref{eq:canonicalDME}) is 
independent of $\kf $ if all terms are kept, but any truncation will give
results depending on the particular choice for $\kf $. The 
standard LDA choice of Negele and Vautherin,
\beqn
 \kf (\Rvec)=(3\pi^2 \rho({\bf R})/2)^{1/3}  \;, 
 \label{eq:standardLDA}
\eeqn
is used here.
Alternative  choices for $\kf (\Rvec)$ may better optimize the
convergence of truncated expansions of Eq.~(\ref{eq:canonicalDME}) 
and lead to a systematic power counting. 

Following Negele and Vautherin, Eq.~(\ref{eq:canonicalDME}) is 
truncated to terms with $n\leqslant 1$, which yields the fundamental 
equation of the DME,
\beqn
\hat{\rho}({\bf R}+\frac{\bf s}{2},{\bf R}-\frac{\bf s}{2})
    \approx  \rhoSL(\kf (\Rvec)s)\,\rho(\Rvec) 
 + s^2 g(\kf (\Rvec)s)\bigl[\frac{1}{4}\nabla^2\rho(\Rvec) - \tau(\Rvec) 
+ \frac{3}{5}\kf (\Rvec)^2\rho(\Rvec)\bigr]
\;,
\label{eq:DMEmastereqn}
\eeqn
where
\beqn 
     \rhoSL(x) \equiv 3j_1(x)/x \;, \qquad 
           g(x) \equiv 35 j_3(x)/2x^3 \;,
\eeqn
and the  kinetic energy density is $\tau(\Rvec) = \sum_{i}|\nabla
\phi_i(\Rvec)|^2$. 
If a short-range interaction is folded with the density matrix,
then a truncated Taylor series expansion of Eq.~(\ref{eq:DMEmastereqn})
in powers of $s$
would be justified and would produce a quasi-local functional.
But the local $\kf $ in the interior of a nucleus
is typically greater than the pion mass $m_{\pi}$,
so such an expansion would give a poor representation of the
physics of the long-range pion exchange interaction.

Instead, the DME is constructed as an expansion about the exact
nuclear matter density matrix.
Thus,
Eq.~(\ref{eq:DMEmastereqn}) has the important feature
that it reduces to the density matrix in the homogeneous nuclear matter
limit, $\rhoNM(\Rvec + {\bf s}/2, \Rvec - {\bf s}/2) = \rhoSL(\kf s)\,\rho$. 
As a result, the resummed expansion in
Eq.~(\ref{eq:DMEmastereqn}) 
 does not distort the finite range physics, as the
long-range one-pion-exchange 
contribution to nuclear matter is exactly reproduced and the
finite-range physics is encoded as non-trivial (e.g., non-monomial) density
dependence in the resulting functional. 
The small parameters justifying this expansion emerge in the functionals
as integrals over the inhomogeneities of the density.
(See Section~\ref{subsec:eft} and
Ref.~\cite{Bhattacharyya:2004qm} for examples
of estimated contributions to a functional for a model problem.)

In the case of a local interaction, 
the  Fock term is schematically given by $W_{\rm F} \sim \int
d\Rvec\,d{\bf s}\, \rho^2(\Rvec+\svec/2,\Rvec - \svec/2)V(\svec)$,
so a single application of
Eq.~(\ref{eq:DMEmastereqn}) is sufficient to cast $\WHF$ into the desired form.
For a
non-local interaction the calculation is more involved as two applications of
the DME are required. 
There is arbitrariness in what is kept in higher orders of $\Sigma^2$,
which is used (following Negele and Vautherin) 
to ``reverse engineer'' the expansion
so that the exact nuclear matter limit is always exactly reproduced by
the leading term~\cite{Negele:1972zp}.
We emphasize that this is a prescription without established power
counting or error estimates; as shown in Ref.~\cite{Bogner:2008kj}, 
different prescriptions can lead
to significant changes in nuclear observables.
The end result for the product of two density
matrices is~\cite{Bogner:2008kj}
\bea
 &&
 \rho({\bf R}+\frac{\bf r'}{2},{\bf R}+\frac{\bf r}{2})
 \rho({\bf R}-\frac{\bf r'}{2},{\bf R}-\frac{\bf r}{2}) 
   \approx \rhoSL^2(\kf \Delta)\rho^2
    +\frac{1}{2} \Sigma^2 g(\kf \Sigma)
   \Bigl(\rho\nabla^2\rho\, \rhoSL(\kf \Delta) j_0(\kf \Delta)    
   \nonumber \\ && \quad\qquad \null
    -|\nabla\rho|^2 
   [j_0^2(\kf \Delta)+j_1^2(\kf \Delta)]
  \Bigr)
   + 2\Delta^2 g(\kf \Delta)\rhoSL(\kf \Delta)
  \Bigl(\frac{1}{4} \rho\nabla^2\rho
     -\rho\,\tau  +\frac{3}{5}\kf ^2\rho^2
  \Bigr) \;. 
  \label{eq:expandedDMs_non-local}
\eea
In the momentum space expression
for $\WHF$, it remains to 
evaluate the Fourier transforms defined in Eq.~(\ref{eq:Fqp})
for the expanded density matrices in Eq.~(\ref{eq:expandedDMs_non-local}). 
Identifying the terms in Eq.~(\ref{eq:WHFdme}) that give
the DME functionals $A[\rho]$, $B[\rho]$, and $C[\rho]$, we have
\bea
  \label{eq:AB_NN}
  A[\rho] &=& \frac{\rho^2}{16\pi \kf ^3}\int_{0}^{2\kf }\!p^2dp\,  
  \tr_{\sigma\tau}
  [\widetilde{\antisymV}(0,\pvec)] 
     \, (I_1(\pb) + \frac{6}{5}I_2(\pb) ) \;,  \\
  B[\rho] &=& -\frac{\rho}{8\pi \kf ^5}\int_{0}^{2\kf }\!p^2dp\,  
  \tr_{\sigma\tau}
  [\widetilde{\antisymV}(0,\pvec)]\, I_2(\pb)\;,
\eea
where $\tr_{\sigma\tau}[\widetilde{\antisymV}(0,\pvec)]$ 
is given by a
simple sum of diagonal matrix elements in the different partial waves, 
\beqn
  \tr_{\sigma\tau}[\widetilde{\antisymV}(0,\pvec)] =
  8\pi  \sum_{lsj}\,^{'} (2j+1)(2t+1)\,\langle 
  \frac{p}{2}lsjt|V|\frac{p}{2}lsjt\rangle \;.
\eeqn
The primed sum is over all channels for which $l+s+t$ is odd.

In these expressions, the functions $I_{j}(\bar{p})$ and $I_j(\bar{q})$ are
simple polynomials (and theta functions)
in the scaled momenta $\bar{p} \equiv p/\kf$  and $\bar{q} \equiv q/\kf$: 
\bea
  \label{eq:DMEintegrals_NN1}
  I_1(\bar{p}) &\equiv& \int x^2\,dx\, j_0(\bar{p}x)\,\rhoSL^2(x) 
  = \frac{3\pi}{32} (16 - 12\pb + \pb^3)\,\theta(2-\pb) \;, \\
  I_2(\bar{p}) &\equiv& \int x^4\,dx\, j_0(\bar{p}x)\, \rhoSL(x) \,g(x) 
  =-\frac{35\pi}{128} (\pb^5-18\pb^3 
     +40\pb^2-24\pb) \, \theta(2-\pb) \;,
  \label{eq:DMEintegrals_NN2}
  \\
  \label{eq:DMEintegrals_NN3}
  I_3(\bar{q}) &\equiv& \int x^4\, dx\, j_0(\bar{q}x) \, g(x) =
  -\frac{35\pi}{8} (5\qb^2-3) \, \theta(1-\qb) \;, \\
  \label{eq:DMEintegrals_NN4}
  I_4(\bar{p}) &\equiv& \int x^2 \,dx\, j_0(\bar{p}x)\, j_0(x)\, \rhoSL(x) =
  \frac{3\pi}{8} (2-\pb) \, \theta(2-\pb) \;, \\
  \label{eq:DMEintegrals_NN5}
  I_5(\bar{p}) &\equiv&\int x^2 \, dx\, j_0(\bar{px})
            [j_0^2(x) + j_1^2(x)]
  = \frac{\pi}{8\pb}  (4-\pb^2) \, \theta(2-\pb)\;.
\eea
Note that the trivial angular dependence of 
Eqs.~(\ref{eq:DMEintegrals_NN1})--(\ref{eq:DMEintegrals_NN5})
is a consequence of the angle averaging that is implicit with each application 
of the DME.

The contributions to $\WHF$ that have gradients of the local density
take the form
\beqn
  \WHF\bigr|_{|\nabla\rho|^2} = \int d\Rvec\,
    \bigl( C_{\nabla^2\rho}\nabla^2\rho(\Rvec)
  + C_{|\nabla\rho|^2}|\nabla\rho(\Rvec)|^2\bigr) 
   \;.
\eeqn
We can perform a partial integration on the $\nabla^2\rho$ terms to cast
them into the canonical form proportional to only $|\nabla \rho|^2$;
that is, 
\beqn
  \WHF\bigr|_{|\nabla\rho|^2} 
  =  \int d\Rvec\,|\nabla\rho(\Rvec)|^2\,\bigl[ C_{|\nabla\rho|^2} 
  -\frac{d}{d\rho}C_{\nabla^2\rho}\bigr] \;,
\eeqn
so that
\beqn
  C[\rho] = C_{|\nabla\rho|^2} - \frac{d}{d\rho}C_{\nabla^2\rho}\;.
  \label{eq:deriv}
\eeqn
In practice it is efficient and accurate to calculate the derivative
in Eq.~\eqref{eq:deriv} numerically rather than analytically. 
The expressions for $C_{|\nabla\rho|^2}$ and $C_{\nabla^2\rho}$ are 
\bea
  C_{|\nabla\rho|^2} &=& 
    \frac{1}{32}\int\frac{d\qvec\, d\pvec}{(2\pi)^6}\, 
  \bigl( -\frac{8\pi^2}{\kf ^8}I_3(\qb)\, I_5(\pb) \bigr)  \,
    \tr_{\sigma\tau}
  [\widetilde{\antisymV}(\qvec,\pvec)]
  \\
  &=& -\frac{1}{16\pi^2\kf ^8}\int_0^{\kf}\! q^2dq \int_0^{2\kf}\! p^2dp\,  
  I_3(\qb)\, I_5(\pb) \,
  \widetilde{\antisymV}_{av}(q,p)
  \;,
\eea
\bea
  C_{\nabla^2\rho} &=& 
      \frac{\rho}{32}\int\frac{d\qvec\, d\pvec}{(2\pi)^6}\, 
    \bigl(\frac{1}{\kf ^5}(2\pi)^4 \delta^3(\qvec)\,I_2(\pb) 
    +\frac{8\pi^2}{\kf ^8}I_3(\qb)\, I_4(\pb)\bigr) \, 
      \tr_{\sigma\tau}
    [\widetilde{\antisymV}(\qvec,\pvec)]
  \nonumber\\
  &=& \frac{\rho}{32\pi \kf ^5}
    \int_0^{2\kf}\! p^2dp \,I_2(\pb)\, 
   \tr_{\sigma\tau}[\widetilde{\antisymV}(0,\pvec)]
  \nonumber\\
  &&\qquad\quad \null + 
    \frac{\rho}{16\pi^2 \kf ^8}\int_0^{\kf}\! q^2dq \int_0^{2\kf}\! p^2dp\,
    I_3(\qb)\,I_4(\pb)\,
   \widetilde{\antisymV}_{\rm av}(q,p)
   \;,
\eea
where $\widetilde{\antisymV}_{\rm av}(q,p)$ is the angle-averaged interaction,
\beqn
  \widetilde{\antisymV}_{av}(q,p) \equiv \frac{1}{2}\int d(\cos{\theta})\,
   \tr_{\sigma\tau} [\widetilde{\antisymV}(\qvec,\pvec)]  \;,
\eeqn
and $\widetilde{\antisymV}(\qvec,\pvec)$ is given 
by Eq.~(\ref{eq:Vqplsjt}). 
Note that care must be taken in the evaluation of  
$d C_{\nabla^2\rho}/d\rho$ if the vertex  
$\widetilde{\antisymV}(\qvec,\pvec)$ is density-dependent 
or if the local Fermi momentum is not taken to be
$\kf  = (3\pi^2\rho/2 )^{1/3}$. 

\begin{figure}[tp]
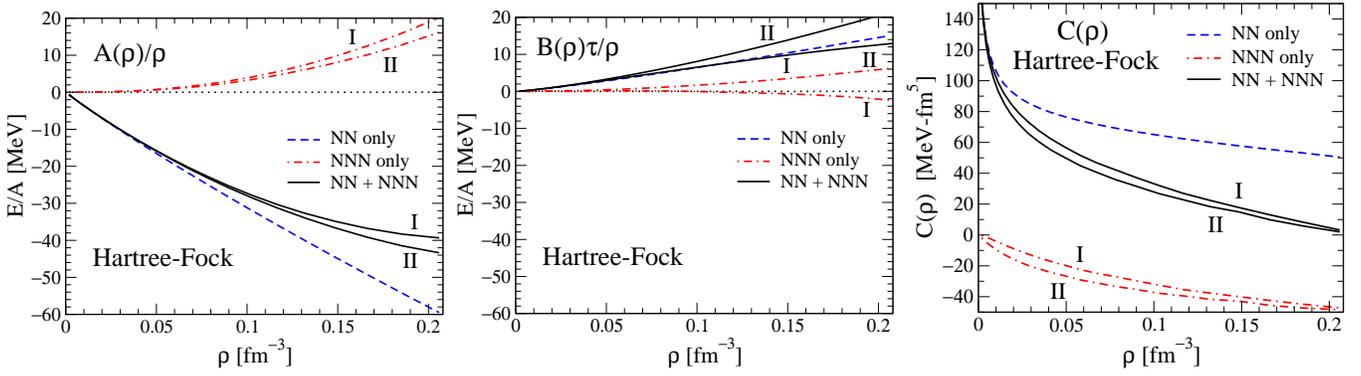

 \begin{center}
  \includegraphics*[width=2.3in]{Aterm-9-2008_rho}
  \includegraphics*[width=2.3in]{Bterm-9-2008_rho}
  \includegraphics*[width=2.3in]{Cterm-9-2008_rho}
 \end{center}
 \vspace*{-.1in}
 \caption{Contribution to the energy per particle in nuclear
 matter from the isoscalar coefficient functions $A(\rho)$,
 $B(\rho)$, and $C(\rho)$ 
 as a function of the density from the DME applied to the Hartree-Fock
 energy calculated using $\vlowk$ with $\Lambda = 2.1\fmi$.
 The result including the NN interaction alone is compared
 to NN plus NNN interactions for two DME expansions (I and II, see
 Ref.~\cite{Bogner:2008kj}).}
 \label{fig:ABC_coeffs}
\end{figure}

These calculations are extended to (local) three-body forces with
corresponding contributions to $A$, $B$, and $C$ in
Ref.~\cite{Bogner:2008kj} using the same type of expansions.  
Some representative numerical results given there
are reproduced in Fig.~\ref{fig:ABC_coeffs}, which illustrate the
relative size of NN and NNN contributions and the truncation uncertainty
introduced from different DME prescriptions. 
It is clear that the decreasing 
hierarchy of many-body contributions is maintained in the individual
coefficent functions, but cancellations magnify the sensitivity 
to the three-body part and how the truncation is carried out. 

These results
are limited and do not yet touch on many of the most
interesting aspects of microscopic DFT from low-momentum potentials.
Topics that should be explored in the future include:
\begin{itemize}
  \item Examine the resolution or scale dependence of the energy functional
  by evolving the input low-momentum potential.  There will be
  dependence on the RG cutoff or flow parameter
  both from omitted physics
  and from intrinsic scale dependence.  Calculations 
  at least to second order are needed to separate these dependencies.
  \item Examine the isovector part of the functional. Contributions
  from the more interesting long-range (pion) parts of the free-space 
  interactions can be isolated, allowing the derivation of
  analytic expressions for the dominant density dependence of the isovector DME coupling functions.   
    \item Study the dependence of spin-orbit contributions on NN
  vs.\ NNN interactions.  This includes the isospin dependence as
  well as overall magnitudes. The NN spin-orbit contributions arise from 
  short-range interactions, whereas NNN contributions arise from 
  the long-range two-pion exchange interaction. Therefore, we expect to find a rather
  different density dependence for the two types of spin-orbit contributions.   
  \item Explore the contribution of tensor contributions, which
  have recently been reconsidered
  phenomenologically~\cite{Lesinski:2007zz,Brink:2007it}.
  \item Understand the scaling of contributions from many-body forces.
  In particular, how does the four-body force (which is
  known at N$^3$LO
  in chiral EFT with conventional Weinberg counting) contribution
  at Hartree-Fock level impact the energy functional?  
  \end{itemize}

\noindent
There are both refinements possible
within the DME framework and generalizations that
test its applicability and accuracy. 
Most immediately,
the DME at the Hartree-Fock level
can be directly extended to approximately include second-order (or full
particle-particle ladder) contributions
by using averaged energies for the energy denominators.

In extending the first calculations from Ref.~\cite{Bogner:2008kj},
the standard DME formalism from Negele and Vautherin~\cite{Negele:1972zp} will
also need to be modified.
This formalism has problems even beyond the truncation errors
from different DME prescriptions seen in Fig.~\ref{fig:ABC_coeffs},
the most severe being that
it provides an extremely poor description of the 
vector part of the density
matrix. 
While the standard DME is reasonable at reproducing the scalar
density matrices, even here the errors are sufficiently large that the
disagreement with a full finite-range Hartree-Fock calculations can 
reach the MeV per particle level. 
Gebremariam and collaborators have
traced both of these problems to an
inadequate phase space averaging (PSA) used in the previous DME
approaches~\cite{Biruk}. In the derivation of the DME, one incorporates average
information about the local momentum distribution into the approximation.
The Negele-Vautherin DME uses the phase space of infinite nuclear
matter to perform this averaging for the scalar part
(and do not even average the vector part). 
However, the local momentum distribution
in finite Fermi systems exhibits two striking differences from that of
infinite homogeneous matter. First, mean-field calculations of nuclei show
that the local momentum distribution exhibits a diffuse Fermi surface that
is especially pronounced in the nuclear surface~\cite{Durand:1982aa}. 
Second, the local
momentum distribution is found to be anisotropic, with the deformation
accentuated in the surface region of the finite Fermi 
system~\cite{Bulgac:1996aa}.

Motivated by previous studies of the Wigner distribution function
in nuclei, Gebremariam et al.\ calculate the quadrupole
deformation of the local momentum distribution using wave
functions for that nucleus~\cite{Biruk}.  There are no free parameters.
The improvements are substantial for the
vector density matrices, typically reducing
relative errors in integrated quantities by as much as an order of
magnitude across many
different isotope chains~\cite{Biruk}.  

Future tests of the DME will include benchmarks
against \textit{ab initio} methods in the overlap region
of light-to-medium nuclei.
Additional information is obtained from putting 
the nuclei in external fields,
which can be added directly to the DFT/DME functional. 
Work is in progress on comparisons to both 
coupled cluster and full configuration interaction calculations
 as part of the UNEDF project.
A key feature is that the same Hamiltonian will be used for the microscopic
calculation and the DME approximation to the DFT.
The freedom to adjust (or turn off) external fields as well as to vary
other parameters in the Hamiltonian permits detailed evaluations of
the approximate functionals. 
In parallel there will be
refined nuclear matter calculations;
power counting arguments from re-examining the Brueckner-Bethe-Goldstone
approach in light of low-momentum potentials
will provide a framework for organizing higher-order contributions.
These investigations should provide insight into how
the energy density functional can be fine tuned for greater
accuracy in a manner
consistent with power counting and EFT principles.


\section{DFT as Legendre transform}
  \label{sec:lt}

In this section, we turn from MBPT in second quantization formalism
to the formulation of DFT using path integrals for effective actions
of composite operators.  This can be an intimidating formalism if
unfamiliar but offers complementary advantages.  For example:
\begin{itemize}
  \item Effective actions are the natural theoretical
  framework for Legendre
  transforms~\cite{Peskin:1995ev,Weinberg:1996II,Zinnjustin:2002}, 
  which is the underlying basis for DFT.  These aspects
  tend to be hidden in the framework of Section~\ref{sec:abinitio}.
  \item The path integral construction of DFT is transparent, such as
  the role and usefulness
  of additional densities/sources.  (As for implementation, we
  rapidly discover the same necessity to evaluate functional derivatives
  that was treated in Section~\ref{sec:dftoep}.)
  \item Path integral effective actions are particularly 
  suited for symmetry breaking, such as encountered with
  pairing.  The renormalization issues in the latter case
  are manifest rather than hidden.
  \item Connections to effective field theory (EFT) and power
  counting can be more accessible.
  \item The path integral formulation puts the DFT construction
  in a broader perspective, which can suggest connections and
  generalizations not apparent otherwise.  For example, there
  are alternative effective actions using auxiliary fields or
  with a two-particle-irreducible nature.  
  The latter may be related to more general EDF constructions
  as proposed in Ref.~\cite{Duguet:2009gc}.
  \item
  The quantization of
  gauge theories was greatly facilitated by Faddeev-Popov 
  and BRST methods
  using path integrals; the same techniques offer alternative
  possibilities for implementing collective coordinates to
  restore symmetries broken by the mean-field organization 
  of DFT~\cite{Bes:1990,Calzetta:2005}.
  \item The path integral formulation can suggest different
  types of nonperturbative approximations, such as $1/N$
  expansions, that might be needed for extensions beyond low-order
  nuclear MBPT (e.g., to handle long-range correlations).
\end{itemize}  
Space does not permit a detailed exposition of the path integral
formulation of DFT.
Furthermore,
in the short term the construction for nuclear DFT
based on MBPT and the DME is probably cleanest in the formalism 
of Section~\ref{sec:abinitio}.
Therefore we focus on presenting the main ideas through analogy
and schematic versions of the path integrals,
and concentrate the EFT discussion on general issues such as
power counting for functionals.

As stressed in Section~\ref{sec:intro}, 
microscopic DFT follows naturally
from calculating the response of a many-body system
to external sources, as in Green's function methods, only with local, static
sources that couple to densities rather than fundamental fields.  
(Time-dependent sources can be used for certain excited states.)
It is profitable to think in terms of a thermodynamic formulation of
DFT, which uses the effective action formalism \cite{Negele:1988vy}
applied to composite operators 
to construct energy
density functionals \cite{Fukuda:1994pq,Valiev:1997bb,Polonyi2001}. 
The basic plan is to consider the zero temperature limit
of the partition function $\mathcal{Z}$ for
the (finite) system of interest in
the presence of external sources coupled to various quantities of
interest (such as the fermion density).
We derive energy functionals of these quantities by Legendre
transformations with respect to the sources \cite{Kutzelnigg:2006aa}. 
These sources probe, in a
variational sense, configurations near the ground state. 

The work by Lieb~\cite{Lieb:1983} on the Hohenberg-Kohn 
theorem~\cite{Hohenberg:1964zz} establishes that the real issue 
for DFT is
the existence of the Legendre transform $F[\rho]$ of the
ground state energy as a functional $E[v]$ of the potential.
The details involve sophisticated
mathematics (e.g., convex-functional analysis) 
that is not readily accessible;
we  recommend Ref.~\cite{Kutzelnigg:2006aa} by Kutzelnigg
as a gateway to the mathematically rigorous literature behind
DFT in terms of Legendre transformations.%
\footnote{There are important formal details~\cite{Eschrig:2003}, 
such as that we need $E[v]$ to
be concave to carry out the transform.}   
Fortunately, for our purposes the familiarity of physicists
with Legendre transforms in the context of thermodynamics is
all we need.
We highly recommend Ref.~\cite{Argaman:2000xx} as an elementary
introduction to DFT that carries the
analogy to thermodynamics through various simple examples.

Effective actions are a natural framework to implement Legendre
transformations, motivate approximations not obvious in MBPT,
and consider generalizations.
One limitation of DFT is the exclusive
role of local potentials (sources)
and densities, where locality is in reference to coordinate
space.
Kutzelnigg points out that this is in contrast to many-body
methods that introduce a finite basis in which operators are expanded,
for which local operators have no privileged place.  In this sense,
density matrix functional theory, as proposed for nuclei
in Ref.~\cite{Duguet:2009gc}, seems more
natural~\cite{Kutzelnigg:2006aa}.  
By looking at effective actions as a broader context, the limitations
and problems of local sources are apparent, but also the opportunities for 
generalizations.

\subsection{Analogy to Legendre transform in thermodynamics}
\label{subsec:analogy}

We first preview  DFT as an effective
action~\cite{Polonyi2001} by
recalling ordinary thermodynamics with $N$ particles as
temperature $T \rightarrow 0$.
The thermodynamic potential is related to the grand canonical partition
function, with the chemical potential $\mu$ acting as a  
source to change $N = \langle \wh N\rangle$,
 \beqn
    \Omega(\mu) = -kT \ln Z(\mu)
    \qquad
    \mbox{and}
    \qquad
    N = -\left(\frac{\partial\Omega}{\partial\mu}
	 \right)_{TV} \; .
 \eeqn
Because $\Omega$ is convex, $N$ is a monotonically increasing
function of $\mu$ and we can
\emph{invert} to find $\mu(N)$ and apply a Legendre transform to obtain
 \beqn
   F(N) = \Omega(\mu(N)) + \mu(N) N \; .
 \eeqn
This is our (free) energy function of the particle number, which is 
analogous to the DFT energy functional of the density.%
\footnote{Because $\Vext$ is typically given rather than eliminated, for
a closer analogy we would also define $\Omega_\mu(N) \equiv F(N) - \mu N$,
which depends explicitly on both $N$ and $\mu$.  This gives the 
grand potential when minimized with respect to
$N$~\cite{Argaman:2000xx}.}
Indeed, if we
generalize to a spatially dependent chemical potential $J(\xvec)$, then
  \beqn
    Z(\mu) \longrightarrow Z[J(\xvec)]
    \qquad
    \mbox{and}
    \qquad
    \mu N = \mu\int\psi^\dagger\psi
      \longrightarrow
      \int J(\xvec)\psi^\dagger\psi(\xvec) \; .
  \eeqn
Now Legendre transform from $\ln Z[J(\xvec)]$ to 
$\Gamma[\rho(\xvec)]$, where $\rho = \langle\psi^\dagger\psi\rangle_J$,
and we have DFT with $\Gamma$ simply proportional to 
the energy functional! [Note that $J(\xvec) \rightarrow
\Vext(\xvec)$ to match our previous notation.]

The functional $\Gamma$ is a type of effective action~\cite{Polonyi2001}.
An effective action is generically the
Legendre transform of a generating functional
with an external source (or sources).  For DFT, we use a source to adjust
the density (cf.\ using an external applied magnetic field to
adjust the magnetization in a spin system).
Consider first the simplest case of a single 
external source $J(\bfx)$ coupled to the density operator 
$\wh \rho(x) \equiv \psi^\dagger(x)\psi(x)$ in the partition
function
\beqn
    \mathcal{Z}[J] = 
    e^{-W[J]} \sim {\rm Tr\,} 
      e^{-\beta (\wh H + J\,\wh \rho) }
    \sim \int\!\mathcal{D}[\psi^\dagger]\mathcal{D}[\psi]
    \,e^{-\int\! [\mathcal{L} + J\,\psi^\dagger\psi]} 
    \;,
    \label{eq:ZofJ}
\eeqn
for which we can construct a (Euclidean) path integral representation
with Lagrangian $\mathcal{L}$ \cite{Negele:1988vy}.
(Note: because our treatment is schematic, for convenience
we neglect normalization factors and take the inverse temperature
$\beta$ and the volume $\Omega$ equal to unity in the sequel.)
The static density $\rho(\bfx)$ in the presence of $J(\bfx)$ is
\beqn
  \rho(\bfx) \equiv \langle \wh \rho(\bfx) \rangle_{J}
   = \frac{\delta W[J]}{\delta J(\bfx)}
   \;,
\eeqn  
which we invert to find $J[\rho]$ and then Legendre transform from $J$ to
$\rho$:
\beqn
   \Gamma[\rho] = - W[J] + \int\!d{\bfx}\, J(\bfx) \rho(\bfx) \;,
   \label{eq:gammarho}
\eeqn
with
\beqn
   J(\bfx) = \frac{\delta \Gamma[\rho]}{\delta \rho(\bfx)}
   \longrightarrow 
   \left.
   \frac{\delta \Gamma[\rho]}{\delta \rho(\bfx)}\right|_{\rho_{\rm gs}(\bfx)
   } =0
   \;.
   \label{eq:Jofx}
\eeqn 
For static $\rho(\bfx)$, $\Gamma[\rho]$ is proportional to 
the conventional Hohenberg-Kohn energy functional, which
by Eq.~(\ref{eq:Jofx}) is extremized at the ground state density
$\rho_{\rm gs}(\bfx)$ (and thermodynamic arguments establish that it is
a minimum \cite{Valiev:1997bb}).\footnote{A Minkowski-space formulation of
the effective action with time-dependent sources
leads naturally to an RPA-like generalization
of DFT that can be used to calculate properties of collective excitations.}

Consider the partition function in the zero-temperature limit of
a Hamiltonian with time-independent source $J({\bf x})$
\cite{Zinnjustin:2002}:
 \beqn
   { \Hhat}(J) = {\Hhat}  + \int\!  J\, \psi^\dagger\psi \; .
 \eeqn
\emph{If} the ground state is isolated (and bounded from below),
 \beqn
   e^{-\beta \Hhat} = e^{-\beta E_0}
     \left[
       | 0 \rangle \langle 0 |
	+ {\cal O}\bigl(e^{-\beta (E_1 - E_0)}\bigr)
     \right]
     \; .
 \eeqn
 As  $\beta \rightarrow \infty$, ${\cal Z}[J]$ yields the
ground state of ${\Hhat}(J)$ with energy $E_0(J)$:
  \beqn    
  { E_0(J)} = \lim_{\beta\rightarrow \infty} -\frac{1}{\beta} \log 
  {\cal Z}[J]
    = \frac{1}{\beta}W[J] \; .
 \eeqn
Substitute and separate out the pieces:
 \beqn
 E_0(J) = \langle {\Hhat }(J) \rangle_J 
      = \langle \Hhat \rangle_J 
	 + \int\! J \langle \psi^\dagger\psi \rangle_J
      = \langle {\Hhat} \rangle_J + \int\! J\, \rho(J)
      \; .
 \eeqn
Rearranging,
the expectation value of ${\Hhat}$ in the ground state
generated by $J[\rho]$ is%
\footnote{The    
functionals will change with resolution or field redefinitions; 
only stationary points are observables.
This can be seen from Eq.~\eqref{eq:expH}, where $\Gamma[\rho]$
is not the expectation value of $\Hhat$ in an eigenstate 
unless $J = J[\rho_{\rm gs}]$.}
 \beqn
   { \langle {\Hhat} \rangle_J} =  E_0(J) - \int J\, \rho
    = \frac{1}{\beta}\Gamma[\rho]
    \; .
      \label{eq:expH}
 \eeqn
Now put it all together:
   \beqn
      \frac{1}{\beta}\Gamma[\rho] = \langle {\Hhat} \rangle_J 
     \stackrel{J\rightarrow 0}{\longrightarrow}
       E_0
     \quad \mbox{and} \quad
    J(x) = -\frac{\delta \Gamma[\rho]}{\delta \rho(x)}
     \stackrel{J\rightarrow 0}{\longrightarrow}
    {
    \left.\frac{\delta \Gamma[\rho]}{\delta \rho(x)}
            \right|_{\rho_{\rm gs}(\bfx)} =0 }
	    \; .
   \eeqn
So for static $\rho(\bfx)$, $\Gamma[\rho]$ is proportional to 
the DFT energy functional $F_{\rm HK}$.  
Furthermore, the true ground state (with $J=0$) is a  variational
minimum,
\emph{so additional sources should be better than just
one source coupled to the density} (as we will consider below).%
\footnote{For the Minkowski-space version of this discussion,
	  see Ref.~\cite{Weinberg:1996II}.}
The simple, universal dependence on a non-zero
external potential $v$ follows
directly in this formalism:
    \beqn
      \Gamma_{v}[\rho] = W_{v}[J] - \int\! J\,\rho
       = W_{v=0}[J+v] - \int\! [(J+v)-v]\, \rho
       = \Gamma_{v=0}[\rho] + \int\! v\,\rho
       \; .
    \eeqn
Thus allowing for non-zero $\Vext$ is a trivial modification
to $\Gamma[\rho]$.    

There are a various paths to a DFT effective action:
  \begin{enumerate}
    \item The auxiliary field (Hubbard-Stratonovich)
    method~\cite{Nagaosa:1999,Stone:2000}: 
      Couple $\psi^\dagger\psi$ to an auxiliary field $\varphi$,
      and eliminate all or part of $(\psi^\dagger\psi)^2$.
      Add a source term
       $J\varphi$ and perform a 
       loop expansion about the expectation value $\phi
        =\langle\varphi\rangle$.
      A Kohn-Sham version uses 
	the freedom of the expansion to require the density 
	be unchanged at each order.
    \item The inversion
    method~\cite{Fukuda:1995im,Valiev:1997bb,Valiev:1997aa,Rasamny:1998zz}
      yields a systematic Kohn-Sham DFT, based on
      an order-by-order expansion.  For example, we can
      apply the EFT power
         counting for a dilute system.
    \item Derive the functional with an RG approach~\cite{Schwenk:2004hm}.
    This is briefly discussed in Section~\ref{sec:summary}. 
  \end{enumerate}

\noindent
We will discuss here the inversion method, which connects most
closely to the developments in Section~\ref{sec:abinitio}.

\subsection{Effective actions for composite operators}
  \label{subsec:effact}

A formal constructive framework for Kohn-Sham
DFT based on effective
actions of composite operators can be carried out using the inversion 
method \cite{Fukuda:1994pq,Valiev:1997bb,Valiev:1997aa,Polonyi2001,Puglia:2002vk,%
Bhattacharyya:2004qm,Bhattacharyya:2004aw,Furnstahl:2004xn}.
This is an organization of the many-body problem
that is based on calculating the response of a finite 
system to external, static
sources rather than seeking the many-body wave function.
It requires a tractable expansion (such as an EFT momentum expansion
or many-body perturbation theory) that is controllable in the
presence of inhomogeneous
sources, which act as single-particle potentials.
As already noted in Section~\ref{subsec:vlowk},
this is problematic for conventional internucleon interactions,
for which the single-particle potential needs to be tuned
to enhance the convergence of the hole-line expansion~\cite{Day:1967zz,Baldo99},
but is ideally suited for low-momentum interactions.
Given an expansion, one can construct a free-energy
functional in the presence of the sources
and then Legendre transform order-by-order
to the desired functional of the densities.
Because these are complicated, non-local
functionals and we require functional
derivatives with respect to the densities, whose dependencies are
usually only implicit, we will ultimately apply the methods of
Section~\ref{sec:dftoep} to derive OEP equations.

The essential features of the inversion method can be illustrated
without the involved path integral formalism by considering 
the Kohn-Luttinger-Ward (KLW) theorem
\cite{Kohn:1960zz,Luttinger:1960ua}, which relates
the perturbative calculation of diagrams using the finite-temperature
Matsubara formalism in the zero-temperature limit
to the calculation of
diagrams using zero-temperature perturbation theory.
A demonstration of the KLW theorem using an inversion method
for the case of an electron gas presented
in Ref.~\cite{FETTER71} can
be adapted to any hierarchical expansion.
Besides perturbation theory in the interaction, this includes
an EFT expansion relevant to a natural
short-distance interaction (which is an expansion in the Fermi momentum
$\kf$ times the effective range parameters $a_s$, $r_s$, 
and so on~\cite{Hammer:2000xg}) and
nonperturbative (in diagrams) $1/N$ expansions~\cite{Furnstahl:2002gt}.

The basic plan is to carry out order-by-order the conventional thermodynamic
Legendre transformation already considered:
\beq
   F(N) = \Omega(\mu) + \mu N \ ,
   \label{eq:LT1}
\eeq
with $\mu(N)$ obtained by inverting 
$N(\mu)=-(\partial\Omega/\partial\mu)_{TV}$.
We expand each of the quantities in Eq.~(\ref{eq:LT1}) about the
non-interacting system:
\bea
       \Omega(\mu) &=& \Omega_0(\mu) + \Omega_1(\mu) + \Omega_2(\mu) +
        \cdots \ , \label{eq:Omegaexp} \\
        \mu &=& \mu_0 + \mu_1 + \mu_2 + \cdots \ , 
           \label{eq:muexp} \\
        F(N) &=& F_0(N) + F_1(N) + F_2(N) + \cdots \ ,
   \label{eq:Fexp}
\eea
where the subscript indicates the order of the expansion.
(Note that the subscript is just a counting parameter that
does not have to correspond to a \emph{power series} in the expansion
parameter; e.g., in the Coulomb case the
expansion parameter is $e^2$ but $\Omega_2$ has both an $e^4$ term and the
correlation energy of order $e^4\ln e$.)
The non-interacting system refers to a system of zeroth order in the EFT
expansion parameter.  This means the zeroth-order 
system has no \emph{internal} interactions among the particles, but it
can include external sources  (we exploit this freedom below).

The number of particles is
\beq
  N(\mu,T,V) = -\left(\frac{\partial\Omega}{\partial\mu}\right)_{TV}
    = - \frac{\partial\Omega_0}{\partial\mu}
    - \frac{\partial\Omega_1}{\partial\mu}
    - \frac{\partial\Omega_2}{\partial\mu}
    + \cdots
    \label{eq:Nexp}
\eeq
Note that we could simply use the \emph{unexpanded} first equality in
Eq.~(\ref{eq:Nexp}) together with the series in Eq.~(\ref{eq:Omegaexp}),
because they define a parametric relation between $N$ and $\Omega$ in
terms of $\mu$ \cite{FETTER71}.
Instead, to mimic DFT
we perform the inversion in Eq.~(\ref{eq:Nexp}) order-by-order, 
treating $N$ as
order zero in the expansion.
(That is, we ensure there are no corrections to $N$ at higher order.)
This means that the zeroth order equation,
\beq
   N = - \left[ \frac{\partial\Omega_0}{\partial\mu} \right]_{\mu=\mu_0}
   \ ,
   \label{eq:Neq}
\eeq
is the \emph{only} equation to which $N$ contributes (by construction).  
Thus the
``exact'' $N$ is obtained from the non-interacting thermodynamic
potential.  This might not sound impressive, but the analogous situation
holds when we generalize $\mu$ to be position dependent  
or coupled to the pair density in a finite system.
In these cases, it is the exact, 
spatially dependent fermion or pair density 
(with appropriate renormalization conditions, see below)
that is obtained
from a non-interacting system with a single-particle potential that
is the generalization of $\mu_0$.
This is precisely the description of
the Kohn-Sham system (see Refs.~\cite{Puglia:2002vk} and \cite{Bhattacharyya:2004qm} 
for details on carrying
out the DFT case without pairing). 

Equation (\ref{eq:Neq}) determines $N(\mu_0)$ at any temperature, from
which we can find $\mu_0(N)$ 
for any
system for which we can identify $\Omega_0$
(the inversion is unique because $\mu_0$ is a monotonic function of $N$
\cite{FETTER71}).
If we have a uniform system with no external sources,
$\mu_0$ is the chemical potential of a noninteracting 
Fermi gas at temperature $T$ with density $N/V$.  
In particular, at $T=0$
with no external potential and spin-isospin
degeneracy $\nu$,
 \beq
    \mu_0(N) = (6\pi^2 N/\nu V)^{2/3} \equiv \kf^2/2M \equiv
            \efermi^0 \ .
       \label{eq:mu0}
 \eeq
The first-order equation extracted from Eq.~(\ref{eq:Nexp})
has two terms, which lets us solve for
$\mu_1$ in terms of known (from diagrams) functions of $\mu_0$:
\beq
   0 = \left[ \frac{\partial\Omega_1}{\partial\mu} \right]_{\mu=\mu_0}
   + \mu_1 \left[ \frac{\partial^2\Omega_0}{\partial\mu^2} \right]_{\mu=\mu_0}
        \ \Longrightarrow\
         \mu_1 = -\frac{[\partial\Omega_1/\partial\mu]_{\mu=\mu_0}}
          {[\partial^2\Omega_0/\partial\mu^2]_{\mu=\mu_0}}
       \ .
       \label{eq:mu1eq}
\eeq
At second order, we can isolate and solve for $\mu_2$, eliminating $\mu_1$
using (\ref{eq:mu1eq}).
This pattern continues to all orders: $\mu_i$ is determined by functions
of $\mu_0$ only.

Now we apply the inversion to $F = \Omega + \mu N$:
\bea
  F(N) &=& \underbrace{\Omega_0(\mu_0) + \mu_0 N}_{F_0}
  \null + \null \underbrace{
      \mu_1 \left[\frac{\partial\Omega_0}{\partial\mu}\right]_{\mu=\mu_0}
      + \Omega_1(\mu_0) + \mu_1 N
      }_{F_1}
   \nonumber \\ & & 
  \null + \underbrace{
     \mu_2 \left[\frac{\partial\Omega_0}{\partial\mu}\right]_{\mu=\mu_0}
     + \mu_1 \left[\frac{\partial\Omega_1}{\partial\mu}\right]_{\mu=\mu_0}
      + \frac{1}{2}\mu_1^2 
      \left[\frac{\partial^2\Omega_0}{\partial\mu^2}\right]_{\mu=\mu_0}
      + \Omega_2(\mu_0) + \mu_2 N       
  }_{F_2} 
  \null + \cdots\ \
  \label{eq:Fexp2}
\eea    
The $\mu_i N$ term always cancels with $\mu_i
[\partial\Omega_0/\partial\mu]_{\mu=\mu_0}$ in $F_i$ 
 for $i\ge 1$
because of Eq.~(\ref{eq:Neq}), leaving
\beq
 F(N) = F_0(N) + \underbrace{\Omega_1(\mu_0)}_{F_1} +
   \underbrace{
   \Omega_2(\mu_0) - \frac{1}{2}
   \frac{[\partial\Omega_1/\partial\mu]^2_{\mu=\mu_0}}
        {[\partial^2\Omega_0/\partial\mu^2]_{\mu=\mu_0}}
   }_{F_2}
   + \cdots \ ,
   \label{eq:Fexp3}
\eeq
where we have also used Eq.~(\ref{eq:mu1eq}) to simplify $F_2$.
The expansion for $F$ can be extended systematically, but this is all we need
here.
(Higher orders can be found by following the prescription in
Refs.~\cite{Okumura:1996,Yokojima:1995hy}.)

\begin{figure}[t]
\begin{center}
  \includegraphics*[width=12.cm,angle=0]{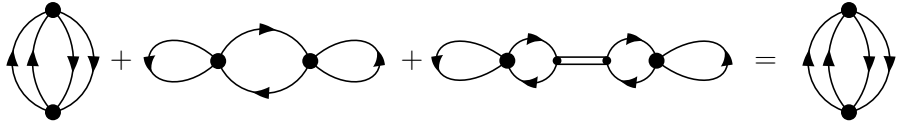}
\end{center}
\vspace*{-.2in}
\caption{Cancellation of the anomalous diagram at NLO. The double
lines represents the inverse of 
$[\partial^2\Omega_0/\partial\mu^2]_{\mu=\mu_0}$ in Eq.~(\ref{eq:Fexp3}).}
\label{fig_F2}
\end{figure}

This construction is rather general.
The Kohn-Luttinger-Ward theorem explores a particular case, the
$T\rightarrow 0$ limit, in which
the second term in $F_2$ in Eq.~(\ref{eq:Fexp3}) cancels precisely
against the anomalous diagram in $\Omega_2$, as illustrated
in Fig.~\ref{fig_F2}.
This cancellation of derivative terms and anomalous diagrams occurs to
all orders in the expansion.
An analogous cancellation occurs in the Kohn-Sham DFT 
inversion~\cite{Puglia:2002vk}.
The end result is an expression for the free-energy $F(N)$ in terms of
the diagrams used for $\Omega_i(\mu)$, only evaluated with
$\mu=\mu_0$ and excluding the anomalous diagrams (both of which simplify
the evaluation of $F(N)$!).
This is precisely the formalism used in Ref.~\cite{Hammer:2000xg} for a
uniform
low-density Fermi gas at zero temperature, where
$\mu_0$ appeared as the Fermi energy of Eq.~(\ref{eq:mu0}).

This procedure 
can be generalized rather directly~\cite{Fukuda:1994pq,Valiev:1997bb} to
carry out the inversion from $\rho[J]$ to $J[\rho]$
needed for Eqs.~\eqref{eq:ZofJ}--\eqref{eq:Jofx}.
The idea again is to expand the relevant quantities in a hierarchy,
here labeled by a counting parameter $\lambda$,
\bea
   W[J,\lambda] &\!=\!& W_0[J] + \lambda W_1[J] + \lambda^2 W_2[J] + \cdots 
    \;, \\
   J[\rho,\lambda] &\!=\!& J_0[\rho] + \lambda J_1[\rho] + \lambda^2 J_2[\rho] 
      + \cdots \;, \\
   \Gamma[\rho,\lambda] &\!=\!& \Gamma_0[\rho] 
            + \lambda \Gamma_1[\rho] + \lambda^2 \Gamma_2[\rho] + \cdots
            \;, 
\eea
treating $\rho$ as order unity (which is the same as requiring
that there are no corrections to the zero-order density),
and match order by order in $\lambda$ to determine the
$J_i$'s and $\Gamma_i$'s.  
Zeroth order is a noninteracting system with potential $J_0(x)$:
\beqn
  \Gamma_0[\rho] = -W_0[J_0] + \int\!d{\bfx}\, J_0(\bfx)\rho(\bfx) 
\eeqn
and
\beqn  
  \rho(\bfx) = \frac{\delta W_0[J_0]}{\delta J_0(\bfx)} 
  \;.   
  \label{eq:zeroth}
\eeqn    
Because $\rho$ appears only at zeroth order, 
it is always specified
from the non-interacting system according 
to Eq.~(\ref{eq:zeroth}); there are no corrections at
higher order.
\emph{This is the Kohn-Sham system with the same density as the
fully interacting system.}

What we have done is use freedom to
split $J$ into $J_0$ and $J - J_0$, which is the analog
of introducing a single-particle potential $\wh U$
and splitting the Hamiltonian according to 
$\wh H = (\wh T + \wh U) + (\wh \VN - \wh U)$,
as discussed in Section~\ref{sec:abinitio}.  Typically 
$\wh U$ is chosen to accelerate
(or even allow) convergence of a many-body expansion
(e.g., the Bethe-Brueckner-Goldstone theory \cite{Day:1967zz,Rajaraman:1967zza,Baldo99}).
For DFT,
we need to choose it 
to ensure that the \emph{density} is unchanged, order by order.
Thus, we need the flexibility in the many-body expansion
to choose $\wh U$ without seriously degrading the convergence; 
such freedom is
characteristic of low-momentum interactions.
(Note: If there is a non-zero external potential, it is simply
included with $J_0$.)

The path integral defined by $W_0$ (or $Z_0$ actually) 
is a gaussian, which is equal to
a functional determinant that is evaluated
by  diagonalizing $W_0[J_0]$.
To do so means to introduce Kohn-Sham orbitals $\phi_i$
and eigenvalues $\varepsilon_i$,
\beqn
  [-\bm{\nabla}^2/2m - J_0(\bfx)]\phi_i
  = {\varepsilon_i}\phi_i
  \label{eq:ksequation}
\eeqn
so that
\beqn   
  \rho(\bfx) = \sum_{i=1}^A |\phi_i(\bfx)|^2
   \;.
   \label{eq:ksdensity}
\eeqn
Then $Z_0$ is the product and $W_0$ is 
the sum of the $\varepsilon_i$'s.
Thus in the path integral formalism, the Kohn-Sham system
arises naturally as the zeroth-order approximation
to the problem.
The organization is based on a saddlepoint evaluation
about the system defined by $J_0$ (which still must be
specified) and subsequent corrections.

The orbitals and eigenvalues are used to construct the Kohn-Sham
Green's functions, which are used as the propagator lines in
calculations the diagrams generated by $W_i[J_0]$.
Finally, we find
$J_0$ for the ground state by truncating the
chain at $\Gamma_{\imax}$,
\beqn
  {J_0} \rightarrow W_1 \rightarrow \Gamma_1 \rightarrow J_1
   \rightarrow W_2 \rightarrow \Gamma_2 \rightarrow \cdots
   \rightarrow W_{\imax} \rightarrow \Gamma_{\imax}
\eeqn
and completing the self-consistency loop that enforces $J(\bfx)=0$:
\beqn
     {J_0(\bfx) = 
     -\sum_{i>0}^{\imax} J_i(\bfx) =
     \sum_{i>0}^{\imax} \frac{\delta\Gamma_{i}[\rho]}{\delta\rho(\bfx)}} 
     \equiv \frac{\delta\Gamma_{\rm int}[\rho]}{\delta\rho(\bfx)}
   \;.   
   \label{eq:loop}
\eeqn
Note that the sum of the $\Gamma_i$'s is directly
proportional to the desired energy functional.
When transforming from $W_i$ to $\Gamma_i$,
there are additional diagrams that take into account the adjustment of the
source to maintain the same density and also so-called anomalous diagrams
(these are two-particle reducible). 
This is the most complicated part but corresponds directly to the
extra terms in the KLW inversion [see
Eqs.~\eqref{eq:Fexp2},\eqref{eq:Fexp3} and
Fig.~\ref{fig_F2}]. 
A general discussion and Feynman rules for these diagrams are given in
Refs.~\cite{Valiev:1997bb,Valiev:1997aa,Rasamny:1998zz,Puglia:2002vk}.

We emphasize that even though solving for Kohn-Sham orbitals makes the approach
look like a mean-field Hartree calculation, 
the approximation to the energy and density is 
\emph{only} in the truncation of Eq.~(\ref{eq:loop}).
It is a mean-field formalism in the 
sense of a conventional
loop expansion, which is nonperturbative only in the
background field while including further correlations perturbatively
order-by-order in loops.  
The special feature of DFT is
that the saddlepoint evaluation applies the condition that there are no
corrections to the density.
Once again, this is not ordinarily an appropriate expansion
for internucleon interactions; it is the special features of
low-momentum interactions that make them suitable.

To generalize the energy functional to accommodate additional
densities such as $\tau$ and ${\bf J}$ (which appear in the
density matrix expansion for nuclei), 
we simply introduce an additional source coupled to each density. 
Thus,
to generate a DFT functional of the kinetic-energy density
as well as the density, add
$\eta({\bf x})\,\bm{\nabla}\psi^\dagger\bm{\cdot}\bm{\nabla}\psi$ 
to the Lagrangian and
Legendre transform to an effective action of $\rho$ and 
$\tau$~\cite{Bhattacharyya:2004aw}:
\beqn
  \Gamma[\rho,\tau] = W[J,\eta]
   - \int\! d{\bfx}\, J(\bfx)\rho(\bfx) - \int\! d{\bfx}\, \eta(\bfx)\tau(\bfx) 
  \;.
\eeqn
The inversion method results in two Kohn-Sham potentials,
\beqn
     J_0({\bf x}) = 
     \left. \frac{\delta \Gamma_{\rm int}[\rho,\tau]}{\delta
     \rho({\bf x})}\right|_{\tau}
      \quad \mbox{and} \quad 
     \left. \eta_0({\bf x}) = \frac{\delta \Gamma_{\rm int}[\rho,\tau]}{\delta
     \tau({\bf x})}\right|_{\rho} \;,     
\eeqn
where $\Gamma_{\rm int} \equiv \Gamma - \Gamma_0$.
The Kohn-Sham equation is now~\cite{Bhattacharyya:2004aw}
\beqn
   \bigl[ 
   -\bm{\nabla}{\frac{1}{\Mstar({\bf{x}})}}\bm{\nabla}
     - J_0(\bfx)
   \bigr]\, \phi_i = \varepsilon_i \phi_i
   \;,
\eeqn         
with an effective mass
$1/2\Mstar(\bfx) \equiv 1/2M - \eta_0(\bfx)$,
just like in Skyrme HF (see also Ref.~\cite{Bulgac:1995aa} for an
early application to Coulomb systems). 
Generalizing to the spin-orbit or other densities (including
pairing \cite{Furnstahl:2006pa}, see Section~\ref{subsec:pairing}) 
proceeds analogously.
We note that the variational nature of the effective action 
implies that adding sources will
always improve the effectiveness of the energy functional.

\begin{figure}[t]
 \begin{center}
  \includegraphics*[width=3.8in]{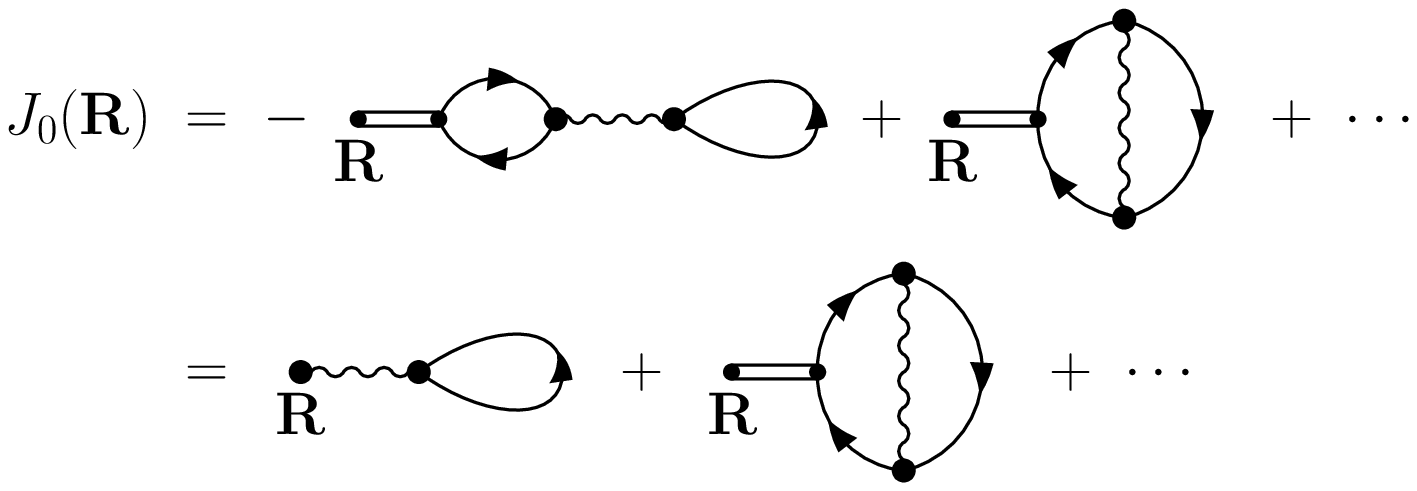}
 \end{center}
 \caption{Schematic representation of Eq.~(\ref{eq:J0chain}) for
 a local potential, where
   the double-line symbol denotes the $(\delta\rho/\delta J_0)^{-1}$ term. }
 \label{fig:J0}
\end{figure}

The Feynman diagrams for $W_i$ will in general include multiple
vertex points over which to integrate.  Further, the dependence on the
densities
will not be explicit except when we have Hartree terms with a 
local potential (that is, a potential diagonal in coordinate
representation).
One way to proceed is to calculate 
the Kohn-Sham potentials
using a functional chain rule, e.g.,
\beqn
 J_0(\Rvec)
  = \frac{\delta\Gamint[\rho]}{\delta\rho(\Rvec)}
  = \int\!d{\bf y}\, \left(
     \frac{\delta\rho(\Rvec)}{\delta J_0({\bf y})}
          \right)^{-1}
   \frac{\delta\Gamma_{\rm int}[\rho]}{\delta J_0({\bf y})}
   \;,
   \label{eq:J0chain}
\eeqn
and steepest descent~\cite{Valiev:1997bb}.
This is illustrated schematically for a local interaction
in Fig.~\ref{fig:J0}.
We see that the Kohn-Sham potential is always just a function
of $\Rvec$ but that the functional is generally very non-local.
If zero-range interactions are used, these diagrams collapse
into an expression for $J_0(\Rvec)$ that has no internal
vertices, but this is
no longer true for diagrams with more than one interaction.
The orbital-based methods from Section~\ref{sec:dftoep} take the chain rule 
in Eq.~(\ref{eq:J0chain}) one step further, adding
a functional derivative of the sources 
with respect to the $\phi_i$'s (and
$\varepsilon_i$'s)~\cite{Fiolhais:2003,Bartlett:2005aa,Gorling:2005aa,Baerends:2005aa},
which can be applied directly to the functionals.

\subsection{EFT and power counting for functionals}
\label{subsec:eft}

The expansion suggested by low-momentum interactions is perturbation
theory in powers of the softened interaction $\wh \VN$.
But is is also instructive to consider the simplest EFT example,
a dilute system of fermions in a harmonic trap, interacting via
natural-sized contact 
interactions~\cite{Puglia:2002vk,Bhattacharyya:2004qm,Bhattacharyya:2004aw}.
(Natural-sized means that the scattering length is not fine-tuned
to a large value compared to the effective range.)
We can construct the effective action as a path integral
by finding $W[J]$ order-by-order in an EFT expansion.
For a dilute short-range
system, 
the EFT Lagrangian
for a nonrelativistic spin-1/2 fermion
field with spin-independent interactions
in Euclidean form is ($\tauE$ is the Euclidean time)
\bea
  \Leuclid  &=&
       \psi^\dagger \biggl[\frac{\partial}{\partial\tauE}  
          - \frac{\nab^{\,2}}{2M}\biggr]
                 \psi + \frac{C_0}{2}(\psi^\dagger \psi)^2
            - \frac{C_2}{16}\Bigl[ (\psi\psi)^\dagger
                                  (\psi\galnab^2\psi)+\mbox{ h.c.}
                             \Bigr]
   -
         \frac{C_2'}{8} (\psi \galnab \psi)^\dagger \cdot
             (\psi\galnab \psi)
+  \ldots
  \nonumber \\
   &\equiv& 
    \psi^\dagger \biggl[\frac{\partial}{\partial\tauE}  
          - \frac{\nab^{\,2}}{2M}\biggr] \psi
	  + \Leuclid^{\rm int}(\psi^\dagger,\psi) \ ,
  \label{eq:lag}                                                   
\eea
where $\galnab=\overleftarrow{\nabla}-\nab$ is the Galilean invariant
derivative and h.c.\ denotes the Hermitian conjugate.%
\footnote{We use units with $\hbar = 1$.}
The terms proportional to $C_2$ and $C_2'$ contribute to $s$-wave and
$p$-wave scattering, respectively, 
while the dots represent terms with more derivatives and/or more
fields, as well as renormalization counterterms.

The Euclidean generating functional with chemical
potential $\mu$ and external sources
$\eta(x)$ and $\etadag(x)$ \cite{Negele:1988vy,Stone:2000}: 
\beq
  Z[\eta,\etadag; \mu] \equiv e^{-W[\eta,\etadag; \mu]}
      = \int\! D\psi_\alpha D\psi_\alpha^\dag 
      \ e^{-\int\! d^4x\, [\Leuclid\,-\,   \mu\,
    \psi_\alpha^\dag(x) \psi_\alpha(x)
    + \etadag_\alpha(x)\psi_\alpha(x) +
    \psi_\alpha^\dag(x)\eta_\alpha(x)]}
    \ ,
   \label{partfunc1}
\eeq
where $\int d^4x$ includes a $d\tauE$ integration that
runs from $-\beta/2$ to $\beta/2$ (to facilitate the
$\beta\rightarrow\infty$ limit) and
anti-periodic boundary conditions are imposed.
A conventional perturbative expansion is realized by removing the
interaction terms from the path integral in (\ref{eq:lag}) in favor of
functional derivatives with respect to $\eta$ and $\etadag$ and
performing the remaining Gaussian integration over $\psi$
and $\psi^\dagger$ \cite{Negele:1988vy,Stone:2000}:
\beq
  Z[\eta,\etadag; \mu] = Z_0\,
    e^{-\int\! d^4x\, \Leuclid^{\rm int}[\delta/\delta\eta(x),
      -\delta/\delta\eta^\dagger(x)]}
      \,
    e^{\int\! d^4y\, d^4y'\, \eta^\dagger(y)
    \Geucl(y,y') \eta(y')} \ ,       
\eeq
where the spin indices are implicit and we have introduced the
noninteracting partition function $Z_0$.
Explicit expressions for 
the Green's function in coordinate and momentum
space can be found in Ref.~\cite{Negele:1988vy}.
The linked-cluster theorem \cite{Negele:1988vy} shows that the difference of
the interacting and noninteracting thermodynamic
potentials $\Omega$ and $\Omega_0$ is given by 
the sum of connected diagrams from the expansion of $Z$, with
the external sources $\eta^\dagger$ and $\eta$ set to zero at the end:
\beq
  \Omega(\mu,T,V) - \Omega_0(\mu,T,V) =  
     \frac{1}{\beta}(W[0,0;\mu] - W_0[0,0;\mu]) \ .
     \label{eq:Omega}
\eeq

In evaluating the Feynman diagrams for $W[J]$ and new diagrams
for $\Gamma[\rho]$ order by order in the expansion
(e.g., EFT power counting),
the source $J_0({\bf x})$ is now the background field.
This means that
propagators (lines) in the background field $J_0({\bf x})$ are
 \beq
   G^0_{\rm KS}({\bf x},{\bf x'}; \omega)
   = \sum_i \phi_i({\bf x})\phi^\ast_i({\bf x'})
   \left[
    \frac{\theta(\varepsilon_i - \varepsilon_{\rm F})}
         {\omega - \varepsilon_i + i\eta}
       +
    \frac{\theta(\varepsilon_{\rm F} - \varepsilon_i)}
         {\omega - \varepsilon_i - i\eta}
   \right]
   \; ,
 \eeq
where $\phi_i(\bfx)$ satisfies:
  \beq
   \bigl[ -\frac{{\nabla}^2}{2M}  +  \Vext({\bf x}) - J_0({\bf x})
   \bigr]\, \phi_i(\bfx) = \varepsilon_i \phi_i(\bfx)
   \; .
  \eeq
For example, if we apply this prescription
to the short-range LO contribution (i.e., Hartree-Fock),
we obtain
    \bea
      W_1[J_0] &=& \frac{1}{2} \nu(\nu-1) C_0
        \int\! d^3{\bf x}\, 
	\int_{-\infty}^{\infty}\!\frac{d\omega}{2\pi}
	\int_{-\infty}^{\infty}\!\frac{d\omega'}{2\pi}\
	  G^0_{\rm KS} ({\bf x},{\bf x}; \omega)
	  G^0_{\rm KS} ({\bf x},{\bf x}; \omega')
	\nonumber \\
	&=&
	-\frac{1}{2}\frac{(\nu-1)}{\nu} C_0
	\int\! d^3{\bf x}\, [\rho_{J_0}({\bf x})]^2
	\; ,
    \eea
where $\nu$ is the spin-isospin degeneracy and
\beq
  \rho_{J_0}({\bf x}) \equiv \nu\sum_i^{\varepsilon_{\rm F}}
	  |\phi_i({\bf x})|^2 \; .
\eeq
Expressions for the other $W_i$'s proceed directly from conventional Feynman
rules using the new propagator.	  

Given $W[J]$ as an EFT expansion, we perform the Legendre
transformation, 
 \beqn
   \Gamma[\rho] = W[J] - \int\! J\rho \; ,
 \eeqn
by using the EFT power counting with the inversion method as described
above,
which gives us the means to invert to find
$J[\rho]$ and to make
an order-by-order inversion from
$W[J]$ to $\Gamma[\rho]$~\cite{Fukuda:1995im,Valiev:1997bb,Valiev:1997aa}.  
It proceeds by
decomposing $J(\bfx) = J_0(\bfx) + J_{\rm LO}(\bfx) + J_{\rm NLO}(\bfx) + \ldots $
as described earlier and fixing $J_0$ using
  \beqn
    \rho(\bfx) = \frac{\delta W_0[J_0]}{\delta J_0(\bfx)}
    \quad\mbox{and}\quad
    \left.J_0(\bfx)\right|_{\rho = \rho_{\rm gs}} = 
      \left.\frac{\delta\Gamma_{\rm interacting}[\rho]}
        {\delta \rho(\bfx)}\right|_{\rho = \rho_{\rm gs}}
	\; .
  \label{eq:J0d}
  \eeqn 
Evaluating the functional derivatives is immediate
if $\Gamma$ is approximated so that the dependence on the density
is explicit, as with the LDA or DME (see below).
Otherwise we apply the OEP chain rule.

\begin{figure}[t]
\centering
      \includegraphics*[width=3.2in]{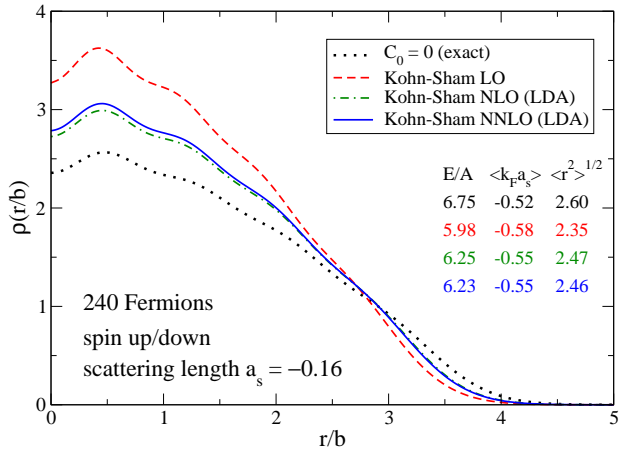}
 \caption{Density profile of a dilute system of fermions in a 
  trap~\cite{Puglia:2002vk}.}
 \label{fig:13}       
\end{figure}

Consider the $T=0$ local density approximation (LDA)
for a dilute fermion gas with natural short-ranged interactions
(meaning the scattering length $a_0$ is comparable in
magnitude to the effective range $r_0$ and the $p$-wave
scattering length $a_p$~\cite{Hammer:2000xg}).
The energy density in the uniform system evaluates to:
 \bea
  &&  \frac{E}{V}   =  { \rho} \frac{{ \kf^2}}{2M}
   \biggl[ \frac{3}{5} 
   {\null + (\nu-1)\frac{2}{3\pi}{ (\kf a_0)}}
   {\null +(\nu-1)
     \frac{4}{35\pi^2}(11-2\ln 2){ (\kf a_0)^2}}
     \\[0pt] & &  \quad \null
     {\null + (\nu-1)\bigl(0.076 
     + 0.057(\nu-3)\bigr){ (\kf a_0)^3}} 
     {\null 
     + (\nu-1)\frac{1}{10\pi}{ (\kf r_0)(\kf a_0)^2}} 
     {\null 
     + (\nu+1)\frac{1}{5\pi}{ (\kf a_p)^3} + \cdots}
     \nonumber
   \biggr]
   \; .  
 \eea
where $\kf a_0$, $\kf r_0$, and $\kf a_p$ are expansion parameters
with $\kf = (6\pi^2\rho/\nu)^{1/3}$.  Using this relation to
replace $\kf$ everywhere by $\rho(\xvec)$,
we directly obtain the LDA expression for $\Gamma[\rho]$,
 \bea
   \Gamma[\rho]  &=&  \int\!d^3 x\,\biggl[ T_{\rm KS}({\bfx})
   {+\frac12\frac{(\nu-1)}{\nu}
      \frac{4\pi a_0}{M} [\rho(\bfx)]^2} 
   {\null + d_1
         \frac{a_0^2}{2M}[\rho(\bfx)]^{7/3}}
    \nonumber \\[0pt] & &   \qquad\qquad\null
     {\null 
     + d_2\,  a_0^3 [\rho(\bfx)]^{8/3}}
     {\null 
     + d_3\, a_0^2\, r_0[\rho(\bfx)]^{8/3}} 
     + d_4\, a_p^3[\rho(\bfx)]^{8/3} + \cdots \biggr]
     \; .
     \label{eq:Gammalda}
 \eea
The Kohn-Sham $J_0$ according to the EFT expansion       
follows immediately in the LDA from (\ref{eq:J0d}):
      \beqn
        J_0(\bfx)  =   \biggl[ 
        -\frac{(\nu-1)}{\nu}
           \frac{4\pi a_0}{M} \rho(\bfx) 
         - c_1 \frac{a_0^2}{2M}[\rho(\bfx)]^{4/3}
          - c_2\,  a_0^3 [\rho(\bfx)]^{5/3}
          - c_3\, a_0^2\, r_0[\rho(\bfx)]^{5/3} 
          - c_4\, a_p^3[\rho(\bfx)]^{5/3} + \cdots \biggr]
  \; .
    \label{eq:J0lda} 
      \eeqn
where the $\{d_i\}$'s and $\{c_i\}$'s are given in Ref.~\cite{Furnstahl:2007xm}.
Sample results at different EFT orders 
for a dilute Fermi gas in a harmonic oscillator
trap are shown in Fig.~\ref{fig:13}.
Note the systematic convergence with successive orders 
in the LDA average $\langle\kf a_s\rangle$ (and $a_s \equiv a_0$) for both
the energy and density.
 
\begin{figure}[t]
\centering
      \includegraphics*[width=2.3in]{error_plot2}
      \qquad
      \includegraphics*[width=2.8in]{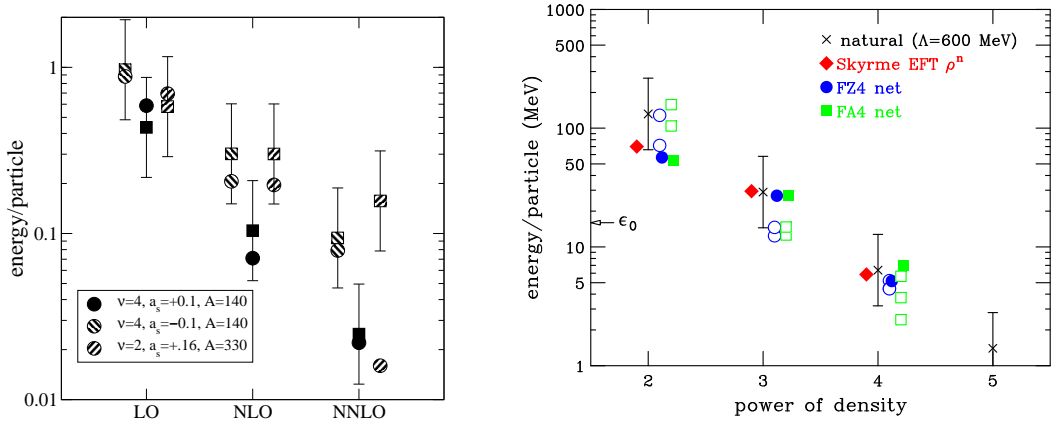}
 \caption{Estimates for energy functionals for a dilute
 fermions in a harmonic trap (left) and for phenomenological
 energy functionals for nuclei (right).}
 \label{fig:14}       
\end{figure}

An important consequence of the systematic EFT approach
is that we can also estimate individual terms in energy functionals.
If we scale contributions to the energy per particle
according to the average density or $\langle\kf\rangle$,
we can make such estimates \cite{Puglia:2002vk,Bhattacharyya:2004qm}.
This is shown in Fig.~\ref{fig:14} for both the dilute trapped
fermions, which is under complete control, and for phenomenological
energy functionals for nuclei, to which a postulated QCD power counting is
applied~\cite{Furnstahl:1997hq,Furnstahl:2003cd}.
In both cases the estimates%
\footnote{The symbols with error bars are natural estimates with a factor
of two spread in the order unity coefficient.}
agree well with the actual numbers
(sometimes overestimating the contribution because of 
accidental cancellations), which means that
truncation errors are understood.
Understanding how such power counting can emerge from
the MBPT-based DFT for nuclei is an important problem for
future study.

\subsection{Additional comments}

The perspective of Legendre 
transforms clarifies many DFT issues~\cite{Kutzelnigg:2006aa}.  
For example: 
\bi
 
 \I The Legendre transform
formulation adds mathematical rigor to DFT claims.  For example,
one finds that  a bijective mapping between $v$ and $\rho$
 does not exist in general.  However, 
 it is not necessary for there to be such a mapping to have a Legendre
 transform;  only the uniqueness of $\rho$ given $v$ is needed.

 \I For
 non-degenerate states, it is true that the density is the functional
 derivative of the energy with respect to 
 the external potential.  But this is not
 actually needed and so is not a problem
 for degenerate states; only the concavity of $E[v]$ is necessary.
 
 \I 
DFT is only variational with respect to $\rho$ if we have 
 the exact functional (i.e., never).  Once there
  are approximations, it is no longer variational.  This
  is not a cause for concern, as other successful
  many-body approaches such as coupled cluster are also not variational.

\ei
Additional useful comments on DFT and Legrendre transforms can be found in
Ref.~\cite{Kutzelnigg:2006aa}.


\section{Topics for nuclear DFT}  \label{sec:nuclear}

In this section, we consider various topics specifically
relevant to implementations of DFT for nuclei.


\subsection{Pairing}
  \label{subsec:pairing}

Pairing is an important feature of finite nuclei that is absent
from isolated atoms and molecules.
Its inclusion in nuclear DFT based on microscopic interactions
is a topic of much current activity.  
While there is some work on
superconductivity in a Coulomb DFT framework,
it is based on using
non-local source terms to avoid divergences.
The use of local densities for pairing is generally preferred
for finite nuclei from a computational point of view.
However, we note that recent results from hybrid EDF calculations
using low-momentum potentials at lowest order, which 
suggest that an accurate \abinitio\ DFT treatment of pairing 
is feasible~\cite{Hebeler:2009dy}, use non-local pairing fields. 
To use local pairing densities, we need 
a consistent generalization of
the Skyrme-Hartree-Fock-Bogoliubov approach \cite{Stoitsov:2003pd}.
In fact, one can use
functionals with local pairing fields as 
in phenomenological Skyrme functionals (that are based
on zero-range effective interactions) 
by properly renormalizing~\cite{Bulgac:2001ei,Borycki:2006it}.
   
Pairing is an example of spontaneous symmetry breaking, which is
naturally accommodated in an effective action 
framework~\cite{Peskin:1995ev,Weinberg:1996II}.   
For example, consider testing for zero-field magnetization $M$ in a spin
system by 
introducing an external field $H$ to break the rotational symmetry.
Legendre transform the Helmholtz free energy $F(H)$:
     \beqn
      \mbox{\normalsize invert\ \ } M = -\partial F(H)/\partial H \quad
        \Longrightarrow \quad  \Gamma[M] = F[H(M)] + M H(M)
	\; .
     \eeqn
Because $H = \partial \Gamma/\partial M \longrightarrow 0$, we look
for the stationary points of 
$\Gamma$ to identify possible ground states, including whether
the symmetry broken state is lowest.
For pairing, the broken symmetry is a $U(1)$ [phase] symmetry
for fermion number.
We apply an external source that breaks the
number symmetry, forcing $\psi\psi$ to have an expectation
value.  Then we turn that source off and see whether the 
expectation value (condensate) persists.
(Note: we will only have actual spontaneous symmetry breaking in
an infinite system.)

The textbook effective action treatment of pairing 
in condensed matter 
is to introduce a contact interaction~\cite{Nagaosa:1999,Stone:2000}:
$g\,\psi^\dagger\psi^\dagger\psi\psi$,
and perform a  Hubbard-Stratonovich transformation with an auxiliary
pairing field $\hat\Delta(x)$ 
coupled to $\psi^\dagger\psi^\dagger$,
which eliminates the contact interaction.
Then one constructs the one-particle-irreducible 
effective action $\Gamma[\Delta]$ with
$\Delta = \langle \hat\Delta \rangle$,
and looks for values for which 
$\delta\Gamma/\delta\Delta = 0$.
To leading order in the loop expansion (which is a mean field
approximation), this yields the 
BCS weak-coupling gap equation with gap $\Delta$.
              
The natural alternative here
is to use the 
inversion method for effective actions again~\cite{Fukuda:1995im,Valiev:1997bb,Valiev:1997aa}.
Thus we introduce another external current $j(\bfx)$, which is coupled
to the fermion pair density $\psi\psi$ 
in order to explicitly break the 
phase symmetry.
This is a natural generalization of normal-state Kohn-Sham
DFT~\cite{Bulgac:2001ai,Bulgac:2001ei,Yu:2002kc}.
The generating functional has sources $J,\pairj$
coupled to the corresponding densities~\cite{Furnstahl:2006pa}:
 \beqn
     Z[J,\/\pairj] = e^{-W[J,\/\pairj]}
       = \int\! D(\psi^\dag\psi) 
       \, e^{-\!\int\! d^4x\ [{\cal L} +   J(x)
     \,\psi_\alpha^\dag \psi_\alpha
     + {\pairj(x)(\psidagup\psidagdown + \psidown\psiup)}]}
     \; .
 \eeqn
Densities are found by functional derivatives with respect to $J$
and $\pairj$:
 \beqn
   \rho(x)\equiv \langle \psi^\dag(x)\psi (x)\rangle_{J,\pairj}
   = \left.\frac{\delta W[J,\pairj]}{\delta J(x)}\right|_{\pairj} 
   \; ,
 \eeqn
and (note that $\kappa$ is called $\phi$ in Ref.~\cite{Furnstahl:2006pa})
 \beqn
  {\kappa(x) \equiv \langle 
     \psidagup(x)\psidagdown(x)+ \psidown(x)\psiup(x)
   \rangle_{J,\pairj}
   = \left.\frac{\delta W[J,\pairj]}{\delta \pairj(x)}\right|_{J} }
   \; .
 \eeqn
(The source $j$ would in general be complex, but it is sufficient for
our purposes to take it to be real.)
The effective action $\Gamma[\rho,\kappa]$
follows as in Section~\ref{sec:lt} by functional Legendre transformation: 
    \beqn
    \Gamma[\rho,\kappa]  = 
      W[J,\pairj] - \int\! d^4x\, J(x)\rho(x)
      - \int\! d^4x\, \pairj(x)\kappa(x)
      \; ,
    \eeqn
and is proportional to the (free) energy functional $E[\rho,\kappa]$; 
at finite temperature, the proportionality constant is $\beta$.
The sources are given by functional derivatives
wrt $\rho$ and $\kappa$:
     \beqn
         \frac{\delta E[\rho,\kappa]}{\delta\rho({\bf x})}
       = J({\bf x})
       \qquad\mbox{and}\qquad
         \frac{\delta E[\rho,\kappa]}{\delta\kappa({\bf x})}
       = \pairj({\bf x})
       \; .
     \eeqn
But the sources are zero in the ground state,
so we determine the ground-state $\rho({\bf x})$
     and $\kappa({\bf x})$ by stationarity:
      \beqn
         \left.
           \frac{\delta E[\rho,\kappa]}{\delta \rho({\bf x})}
          \right|_{{\rho=\rho_\grounds,\kappa=\kappa_\grounds}}
          =
         \left.
           \frac{\delta E[\rho,\kappa]}{\delta \kappa({\bf x})}
          \right|_{{\rho=\rho_\grounds,\kappa=\kappa_\grounds}}
          = 0
	  \; .
      \eeqn
This is Hohenberg-Kohn DFT extended to pairing!

We need a method to carry out the Legendre transforms
to get Kohn-Sham DFT; an obvious choice is to apply 
the Kohn-Sham inversion method again,
with order-by-order matching in the counting parameter $\lambda$. 
Once again,
  \bea
    \mbox{ diagrams}&\Longrightarrow&
    {W}[J,\pairj,\lambda] = 
    {{W}_0[J,\pairj]} + \lambda{W}_1[J,\pairj] +
    \lambda^2{W}_2[J,\pairj] 
    + \cdots 
    \\
   \mbox{ assume}&\Longrightarrow&
    J[\rho,\kappa,\lambda] = 
    {J_0[\rho,\kappa]} + \lambda J_1[\rho,\kappa] +
    \lambda^2J_2[\rho,\kappa] + \cdots 
     \\
   \mbox{ assume}&\Longrightarrow&
    \pairj[\rho,\kappa,\lambda] = {\pairj_0[\rho,\kappa]} + 
    \lambda\pairj_1[\rho,\kappa] +
    \lambda^2\pairj_2[\rho,\kappa] + \cdots 
     \\
   \mbox{ derive}&\Longrightarrow&
    {\Gamma}[\rho,\kappa,\lambda] = 
    {{\Gamma}_0[\rho,\kappa]} 
    + \lambda{\Gamma}_1[\rho,\kappa] + \lambda^2{\Gamma}_2[\rho,\kappa] +
    \cdots 
  \eea
 Start with the exact expressions for $\Gamma$ and $\rho$ 
 \beqn
   \Gamma[\rho,\kappa] = W[J,j] - \int\! J\,\rho 
      - \int\! j\,\kappa \;,
 \eeqn
 and
 \beqn     
    \rho(x)
      = \frac{\delta W[J,j]}{\delta J(x)}\,, \quad
      \kappa(x) = \frac{\delta W[J,j]}{\delta j(x)} \; ,   
 \eeqn     
and plug in the expansions, with $\rho,\kappa$ treated as order unity.
Zeroth order is the Kohn-Sham system with potentials 
$J_0({\bf x})$ and $\pairj_0({\bf x})$,
 \beqn
   \Gamma_0[\rho,\kappa] = W_0[J_0,j_0] - \int\! J_0\,\rho 
      - \int\! j_0\,\kappa \; ,
 \eeqn     
so the \emph{exact} densities $\rho({\bf x})$ and $\kappa({\bf x})$
are   \emph{by construction}
 \beqn
   \rho(x)
      = \frac{\delta W_0[J_0,j_0]}{\delta J_0(x)}\,, \quad
      \kappa(x) = \frac{\delta W_0[J_0,j_0]}{\delta j_0(x)} \; .   
 \eeqn     
Now introduce single-particle orbitals to diagonalize $\Gamma_0$,
which means solving
 \beqn
  \left(
    \begin{array}{cc}
     h_0(\xvec) - \mu_0 & \pairj_0(\xvec) \\
     \pairj_0(\xvec)         & -h_0(\xvec) + \mu_0 
    \end{array}
  \right)
  \left(
    \begin{array}{c}
    u_i(\xvec) \\ v_i(\xvec)
    \end{array}
  \right)
  = E_i
  \left(
    \begin{array}{c}
    u_i(\xvec) \\ v_i(\xvec)
    \end{array}
  \right)
 \eeqn
where
  \beqn
   h_0 (\xvec) \equiv -\frac{\bm{\nabla}^2}{2M} 
      + \Vext(\xvec)
      -J_0(\xvec) \; .
  \eeqn
As expected,
this is just like the Skyrme Hartree-Fock Bogoliubov (HFB)
approach~\cite{Ring:2005}.  
For \abinitio\ DFT based on wave-function methods, Bogoliubov transformations
lead to the same generalized Kohn-Sham equations.

The diagrammatic expansion of the $W_i$'s is the same as without pairing, 
except now
lines in diagrams are KS Nambu-Gor'kov Green's 
functions~\cite{Abrikosov:1963}, 
%
 \beqn
   \renewcommand{\arraycolsep}{2pt}
   \hspace*{-.1in}
   {\bf G} = 
     \left(
     \begin{array}{cc}
     \langle T\psiup(x)\psidagup(x') \rangle_0 &
     \langle T\psiup(x)\psidown(x') \rangle_0 \\
     \langle T\psidagdown(x)\psidagup(x') \rangle_0 &
     \langle T\psidagdown(x)\psidown(x') \rangle_0 
     \end{array}
     \right)
    \equiv
    \left(
     \begin{array}{cc}
     \GKS^0  &  \FKS^0 \\
     {\FKS^0}^\dagger & -\wt G_{\ks}^0
     \end{array}
    \right)
    \; , 
 \eeqn
where the time-ordering is with respect to Euclidean time.
In frequency space, the Kohn-Sham Green's functions are
  \beqn
     \GKS^0 (\xvec, \xvec'; \omega) = \sum_j\, \left[
      \frac{u_j (\xvec)\, u_j^* (\xvec')}{i\omega - E_j} 
      + \frac{v_j (\xvec')\, v_j^* (\xvec)}{i\omega + E_j} 
      \right] \; ,
  \eeqn
  \beqn
     \FKS^0 (\xvec, \xvec'; \omega) = -\sum_j\, \left[
      \frac{u_j (\xvec)\, v_j^* (\xvec')}{i\omega - E_j} 
      - \frac{u_j (\xvec')\, v_j^* (\xvec)}{i\omega + E_j} 
      \right] \; .
  \eeqn
The Kohn-Sham self-consistency procedure
involves the same iterations 
as in phenomenological Skyrme HF (or relativistic mean-field) 
when pairing is included.
In terms of the orbitals, the fermion density is
 \beqn
   \rho(\xvec) =
   2\sum_i\, |v_i(\xvec)|^2 \; ,
 \eeqn
 and the pair density is (warning: this is unrenormalized!)
 \beqn
   \kappa(\xvec) =
   \sum_i\, [ u_i^*(\xvec) v_i(\xvec) + u_i(\xvec) v_i^*(\xvec) ]
   \; .
 \eeqn
The chemical potential $\mu_0$
is fixed by $\int\!\rho(\xvec) = A$.
Diagrams for
$\Gamma[\rho,\kappa] \propto E_0[\rho,\kappa] + E_{\rm int}[\rho,\kappa]$ 
yield the Kohn-sham potentials
  \beqn
    J_0(\xvec)\Bigr|_{\rho=\rho_{\rm gs}} = 
      \left.\frac{\delta E_{\rm int}[\rho,\kappa]}{\delta \rho(\xvec)}
    \right|_{\rho=\rho_{\rm gs}}
  \  \mbox{and} \quad
    \pairj_0(\xvec)\Bigr|_{\kappa=\kappa_{\rm gs}} = 
      \left.\frac{\delta E_{\rm int}[\rho,\kappa]}{\delta \kappa(\xvec)}
    \right|_{\kappa=\kappa_{\rm gs}} \; .
  \eeqn
So it appears that everything goes through smoothly as in the
\abinitio\ DFT without pairing.

Potential problems arise, however, because
the use of  local composite
operators leads to new ultraviolet
divergences even at the
mean-field (Hartree-Fock) level when pairing is included.
Divergences at this order are not encountered when coupling an external
source to the fermion density $\psi^\dagger\psi$, but appear now because   
the composite operators
$\psi\psi$ and $\psi^\dagger\psi^\dagger$  need additional 
renormalization~\cite{COLLINS86}.
The divergences at leading order
are symptomatic of generic 
problems identified long ago by 
Banks and Raby \cite{BANKS76} that arise with effective 
potentials of local composite operators.
(It was to avoid these divergences in DFT
that non-local sources were used 
in Refs.~\cite{Oliveira:1988,Kurth:1999}.)
These problems inhibited for many years
the use of effective actions of composite
operators but more
recently Verschelde et al.\ \cite{VERSCHELDE95,VERSCHELDE97,KNECT01}
and Miransky et al.\ \cite{Miransky:1992bj,Miransky:1996pd,Gusynin:1998kz} have revived
their use for relativistic field theories.
In Ref.~\cite{Furnstahl:2006pa}, it was shown how to identify
and renormalize up to NLO the EFT from Section~\ref{subsec:eft}
when pairing was included.   

\begin{figure}[t]
 \begin{center}
  \raisebox{.8in}{\includegraphics*[width=3.0in]{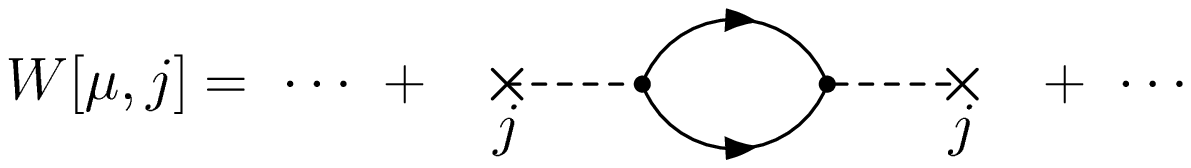}}
  \hspace*{.5in}
  \includegraphics*[width=2.8in]{convergence_plot_bw}
 \end{center}
 \caption{(a)~Feynman diagram at second order in a perturbative
 expansion of $W[\mu,j]$ in $j$.
 (b)~Converence of the integral for the pair density in the
 uniform system for two subtractions as a function of an
 energy cutoff $E_c$ in the integral.}
 \label{fig:pairing}
\end{figure}

The source of the divergences is found immediately
when we try to carry out the DFT pairing
calculation even for a \emph{uniform} dilute Fermi system.
The generating functional with constant sources ${\mu}$ 
and ${\pairj}$ for the EFT from Section~\ref{subsec:eft} is:
 \bea
    e^{-W[\mu,j]}
      &=& \int\! D(\psi^\dag\psi) 
      \exp\Bigl\{ -\!\int\! d^4x\  \bigl[\psi_\alpha^\dagger
        (\frac{\partial}{\partial\tau}  
      - \frac{\bm{\nabla}^{\,2}}{2M} - {\mu} )
             \psi_\alpha\, \nonumber
    \\
    & & \hspace*{0.2in}
  \null  + \frac{C_0}{2}\psidagup\psidagdown\psidown\psiup
    + \, {\pairj}(\psiup\psidown + \psidagdown\psidagup )
     {\bigr]\Bigr\}}
	         {+ \frac12\zeta\, j^2}\,\bigr]\Bigr\}
\eea
(cf.\ adding an integration over an auxiliary field 
$\int\! D(\Delta^\ast,\Delta)\ e^{-\frac{1}{|C_0|} \int\!
|\Delta|^2}$,
then shifting variables to eliminate
$\psidagup\psidagdown\psidown\psiup$
for $\Delta^\ast \psiup\psidown$).     
There are new divergences because of $\pairj$,
e.g., expand $W$ to ${\cal O}(\pairj^2)$ [see
Fig.~\ref{fig:pairing}(a)],
which has the same linear divergence found in 2-to-2 scattering.
(Equivalently, in coordinate space there is a $1/|{\bf r}_1-{\bf r}_2|$
divergence in $\sum_i v^\ast_i({\bf r}_1) u_i({\bf r}_2)$ as
it becomes local with $|{\bf r}_1-{\bf r}_2| \rightarrow 0$.)
To renormalize, we add the counterterm 
$\frac12\zeta |\pairj|^2$ to ${\cal L}$
(see~\cite{Zinnjustin:2002}), which is
additive to $W$ (cf. $|\Delta|^2$), so there is no effect on scattering.

We can follow a less formal and more numerically suitable
procedure to renormalize the pair (anomalous) density,
 \beq
   \kappa(\xvec) =
   \sum_i\, [ u_i^*(\xvec) v_i(\xvec) + u_i(\xvec) v_i^*(\xvec) ]
   \longrightarrow \infty \;,
 \eeq
which diverges for contact interactions in a finite system. 
This is to use the renormalized expression for $\kappa$ in the uniform system, 
 \beq
   \kappa = \int^{k_c}\! \frac{d^3k}{(2\pi)^3} \, j_0 
      \left(
        \frac{1}{\sqrt{(\epsilon_k^0-\mu_0)^2 + j_0^2}}
        - \frac{1}{\epsilon_k^0}
      \right)
  \stackrel{k_c\rightarrow \infty}{\longrightarrow} \mbox{finite,}
  \label{eq:phi1}
 \eeq
which is cut off at momentum $k_c$, 
and apply this in a local density approximation (i.e., Thomas-Fermi):
     \beq
       \kappa(\xvec) = 2\sum_i^{E_c} u_i(\xvec)v_i(\xvec)
          - j_0(\xvec) \frac{M\, {k_c(\xvec)}}{2\pi^2}
       \quad \mbox{with} \quad
      {E_c = \frac{k_c^2(\xvec)}{2M} + J(\xvec) - \mu_0 }
      \; .
     \eeq
This procedure was worked out by Bulgac and
collaborators in Refs.~\cite{Bulgac:2001ei,Bulgac:2001ai,Yu:2002kc,Yu:2003bp}.
Convergence is very slow as the energy cutoff $E_c$ is increased, so
Bulgac and Yu devised a different subtraction,
 \beq
   \kappa = \int^{k_c}\! \frac{d^3k}{(2\pi)^3} \, j_0 
      \left(
        \frac{1}{\sqrt{(\epsilon_k^0-\mu_0)^2 + j_0^2}}
      { - \frac{{\cal P}}{\epsilon_k^0-\mu_0} }
      \right)
  \stackrel{k_c\rightarrow \infty}{\longrightarrow} \mbox{finite}
  \; .
  \label{eq:phi2}
 \eeq
A comparison of convergence in the uniform system for the two
subtraction schemes (\ref{eq:phi1}) and (\ref{eq:phi2})
[see Fig.~\ref{fig:pairing}]  
shows dramatic improvement for the Bulgac/Yu subtraction.
Bulgac et al.\ have demonstrated that this works in 
finite systems~\cite{Yu:2002kc} and 
it has been adopted for some Skyrme HFB implementations.

\subsection{Broken symmetries}
  \label{sec:symmetry}

Ordinary nuclei are self-bound, which presents conceptual issues
about whether Kohn-Sham DFT is well defined and practical problems
on how to deal with the consequences of symmetry breaking by the
KS potentials, which will not have 
all of the symmetries of the Hamiltonian~\cite{Blaizot:1985}.%
\footnote{A state is one of broken
symmetry if it does not have the quantum numbers of the
eigenstates of the Hamiltonian (parity, particle number,
angular momentum, linear momentum, isospin, and so on).
Note that this does not mean it needs to be \emph{invariant}, just that
it can be labeled by definite quantum numbers~\cite{Blaizot:1985}.}
These broken symmetries include the $U(1)$ phase symmetry for fermion
number and translational and rotational invariance.
While restoring broken symmetries is a topic well-explored for mean-field
approximations~\cite{Blaizot:1985,Ring:2005}, it has only recently
been considered in a context relevant to \abinitio\ DFT.
Because no proven practical approaches are yet available, we simply
point out the issues and current references.

The textbook discussions of how to restore mean-field broken
symmetries tend to follow one of these two related lines of discussion:
\bi
 \item
States related by a unitary transformation $U(\alpha)$
corresponding to a broken symmetry are degenerate:
\beqn
   | \phi\,\alpha \ra = U(\alpha) | \phi \ra
\eeqn
with $|\phi\ra$ a  ``deformed'' state, implies
\beqn
  \la \phi\,\alpha | H_N | \phi\,\alpha\ra = \la \phi | H_N | \phi \ra 
  \;.
\eeqn
The degeneracy can be removed by diagonalizing in the subspace
spanned by the degenerate states.
The group parameter
$\alpha$ for continuous groups can be considered a
\emph{collective coordinate}, which specifies the
orientation in gauge space of the deformed state $|\phi \ra$.
A general strategy is to transform from $3A$ particle coordinates
into collective and internal coordinates~\cite{Ring:2005}. 
 
 \item
In finite systems, broken symmetries arise only as a result of 
approximations.  This usually happens with variational
calculations over trial wavefunctions that are too restricted;
a mean-field approximation is an example.
The symmetry can be restored by using a linear
superposition of degenerate states:
\beqn
  | \psi \ra = \int\! d\alpha\, f(\alpha) | \phi\,\alpha\ra
  \;,
\eeqn
which when minimized with respect to the $f(\alpha)$'s projects
states of good symmetry~\cite{Blaizot:1985}.  
(Because minimizing with respect to 
$|\phi\ra$ and with respect to $f(\alpha)$ do not commute, there
are two types of projection.  It is most accurate to project
first and then find the best deformed state corresponding
to a given quantum number.)
For example, particle number projection for EDF's
is described in Refs.~\cite{Stoitsov:2006tv,Dobaczewski:2007ch}.
\ei
When implemented, these approaches  are sometimes considered to
be beyond EDF, where there
are only densities and not a wavefunction.
From a different perspective, the restoration of broken symmetries
of GCM-type configuration mixings should be considered as
a ``multi-reference'' extension of the usual ``single-reference''
EDF implementation 
(see Refs.~\cite{Lacroix:2008rj,Bender:2008rn,Duguet:2008rr}).
What about \abinitio\ DFT as we have considered it?

For nuclear DFT, the conceptual question was highlighted 
by J.\ Engel~\cite{Engel:2006qu}, who pointed out that the
ground state of a self-bound system, with a plane wave describing
the center of mass, has a density distributed uniformly over
space.  Clearly this is not the density one wants to find from DFT, 
so there is a question of how to proceed.
There are two separate considerations:  i) Does Kohn-Sham
DFT exist in a useful form for self-bound systems? 
ii) If so, how does one formulate and implement it?   
Engel and other authors have addressed this 
issue 
recently~\cite{Engel:2006qu,Giraud:2007pe,Barnea:2007jx,Giraud:2008zz,Giraud:2008yw,Messud:2009jh},
with a consensus that HK existence proofs for DFT are still well
founded, but for \emph{internal} densities (meaning independent of the
center-of-mass motion when considering broken translational symmetry).%
\footnote{In other contexts, such densities are called ``intrinsic'',
but this has a different meaning in the context of symmetry breaking,
so ``internal'' is typically used instead.}
These authors also propose various schemes to carry out 
Kohn-Sham DFT that should be testable in the near future.

We have considered \abinitio\ DFT from two viewpoints:  MBPT using
wave-function methods and effective actions with path integrals.
How are these symmetry issues dealt with in these approaches?
Wave function methods have
several related strategies for dealing with
the ``center of mass'' (COM) problem:
  \begin{enumerate}
    \I Isolate the ``internal'' degrees of freedom, typically by
    introducing Jacobi coordinates.  Then the observables are by
    construction independent of the COM.
    This gets increasingly
    cumbersome with greater numbers of particles.
    \I Work in a harmonic-oscillator Slater determinant basis,
    for which the COM decouples,
    and introduce a potential for the COM that allows its contribution
    to be subtracted.
    \I Work with the internal Hamiltonian (i.e.,
    subtract the COM kinetic energy $T_{\rm CM}$)
    so that the COM part factors and does not contribute to
    observables to good approximation (see
    in particular Ref.~\cite{Hagen:2009pq} for coupled
    cluster calculations). 
  \end{enumerate}
Versions of the first two possibilities are in fact
among the ideas considered for DFT in
Refs.~\cite{Engel:2006qu,Giraud:2007pe,Barnea:2007jx,Giraud:2008zz,Giraud:2008yw,Messud:2009jh}.

For the effective action approach, 
the issue of broken symmetry is familiar from
the study of solitons~\cite{Rajaraman:1982,Negele:1988vy}, where it
arises as the problem of dealing with zero modes when calculating quantum
fluctuations.  The methods found in the literature are similar
to the textbook approaches cited above.
One compelling approach for \abinitio\ DFT
in the effective action formalism
uses the Fadeev-Popov construction
(or BRST invariance~\cite{Bes:1990}) to
introduce collective coordinates through ghost degrees of freedom.
This is worked out to some degree by
Calzetta and Hu for broken particle number symmetry 
in Ref.~\cite{Calzetta:2005}, but a concrete implementation
appropriate for nuclei remains to be formulated.

Thus, while dealing with broken symmetries in the DFT of self-bound nuclei
is a topic of active investigation, the best way forward
is not clear.

\subsection{Single-particle energies}
\label{subsec:gf}

The only quantities obtained from Kohn-Sham DFT for 
ground states that are guaranteed to be the same as those obtained from
many-body wave function calculations (at an equivalent level of approximation)
are the total energies and the densities associated
with the functionals.%
\footnote{Note that these densities
are not necessarily observables; e.g., the nuclear proton density
is related to the measured charge density by a model dependent
prescription, although this model dependence is generally
considered to be negligible.} 
Of course, measurable quantities that can be expressed as the 
differences of ground state energies are also reliable.    
For nuclear applications, one would like to establish 
a robust connection between KS eigenvalues and nuclear single-particle
energies, but
it is often observed that Kohn-Sham eigenvalues,
except at the Fermi surface, are not physical.  
On the other hand, Bartlett et al.\ claim that with a sufficiently
rich Coulomb
DFT functional, the KS orbital eigenvalues can be good approximations
to removal (ionization) energies~\cite{Bartlett:2005aa,Bartlett:2006aa}. 
  
\begin{figure}[th]
  \begin{center} 
   \includegraphics*[width=2.7in]{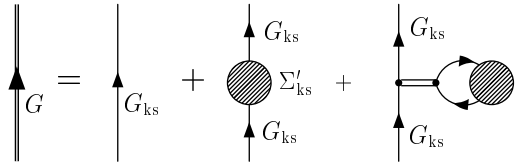}
 \end{center}
 \vspace*{-.1in}
 \caption{Full Green's function $G$ in terms of the
  Kohn-Sham Green's function $G_{\rm ks}$.}
  \label{fig:20a}
\end{figure}

We can understand how improved approximations 
can happen by considering how the full
self-consistent one-particle Green's function $G$, whose spectral
density \emph{is} physical, can be expressed in terms of the
KS Green's function $G_{\rm KS}$.   
We saw this earlier in the form of the Sham-Schl\"uter equation,
Eq.~\eqref{eq:shamshuluter}; here we consider an alternative
functional integral derivation and a diagrammatic representation
to make some additional points.
We add a non-local source $\xi(x',x)$ coupled to
$\psi(x)\psi^\dag(x')$ [we're in Minkowski space now with $x = (\bfx,t)$ so that
we get real-time Green's functions]:
 \beqn
    Z[J,\xi] = e^{iW[J,\xi]}
      = \int\! D\psi D\psi^\dag \ e^{i\int\! d^4x\ [{\cal L}\,+\,   J(x)
    \psi^\dag(x) \psi(x)\, { +\,
    \int\! d^4x'\, \psi(x)\xi(x,x')\psi^\dag(x')}]}
    \; .
\eeqn
Writing $\Gamma[\rho,\xi] = \Gamma_0[\rho,\xi] 
     + \Gamma_{\rm int}[\rho,\xi]$ and using functional properties
of Legendre transforms,
 \beqn
   G(x,x') = \left. \frac{\delta W}{\delta \xi}\right|_J
     = \left. \frac{\delta \Gamma}{\delta \xi}\right|_\rho
     = G_{\rm ks}(x,x') + G_{\rm ks}\Bigl[ 
     \frac{1}{i}\frac{\delta\Gamma_{\rm int}}{\delta G_{\rm ks}}
     + \frac{\delta\Gamma_{\rm int}}{\delta \rho}
     \Bigr] G_{\rm ks} \; ,
 \eeqn
which is represented diagrammatically in 
Fig.~\ref{fig:20a}~\cite{Bhattacharyya:2004aw}.
(Note that these are the reducible self-energies here; so this
is actually a rearranged Dyson-like equation that is equivalent
to Eq.~\eqref{eq:shamshuluter}.)   

The Green's functions $G$ and $G_{\rm ks}$ yield the same density 
\emph{by construction}; that is,
the Kohn-Sham density $\rho_{\rm ks}(\bfx) = -i\nu G_{\rm KS}^0(x,x^+)$
    equals the full density $\rho(\bfx) = -i\nu G(x,x^+)$.
Here is a simple diagrammatic demonstration
(the double line is minus the inverse of a single particle-hole ring):
\begin{center}
   \includegraphics*[width=4.0in]{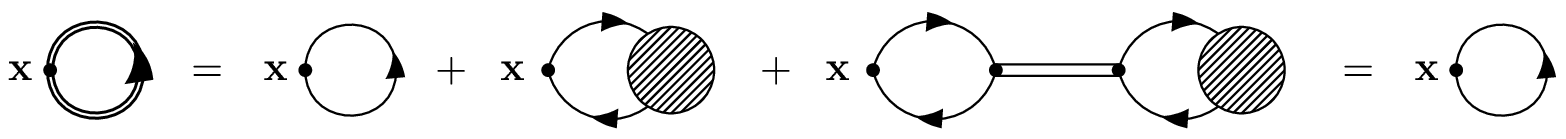}
\end{center}
But other single-particle properties (e.g., the spectrum)
are generally different,
because the last two diagrams in Fig.~\ref{fig:20a} will not cancel
exactly.
However, we can make them cancel more closely by redefining
the Kohn-Sham system as described in Section~\ref{subsec:improvedpt}.
In the effective action approach, we find that adding sources 
does exactly this.
For example, in the dilute Fermi gas EFT from
Section~\ref{subsec:eft}, the single-particle spectrum from a
functional with $\rho$ and $\tau$ densities was shown to be
closer to the exact spectrum (e.g., at the HF level) than
a functional based on $\rho$ alone 
(see Ref.~\cite{Bhattacharyya:2004qm} for details).
This is consistent with the Bartlett et al.\ claim.
More generally, we can use the Kohn-Sham basis in constructing
the full $G$.


\begin{figure}[tb] 
  \begin{center}
    \includegraphics*[width=2.5in]{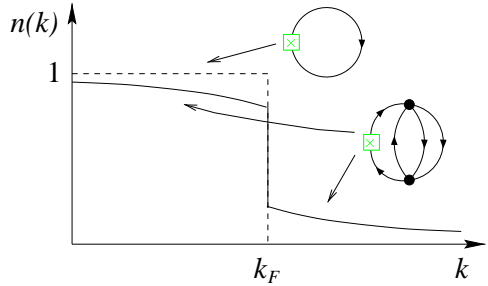}
  \end{center}
  \vspace*{-.1in}
  \caption{Schematic momentum occupation number $n(k)$
  for mean-field (Hartree-Fock) and with correlations.}
  \label{fig:occupf2}
\end{figure}
   
One final note regarding Green's functions.
The comparison of   
Kohn-Sham DFT and ``mean-field'' models often leads
to misunderstandings, as when considering ``occupation numbers'',
because of a confusion between $G$ and $G_{\rm KS}$.
Figure~\ref{fig:occupf2} suggests that occupation numbers are
equal to 0 or 1 if and only if correlations are \emph{not} included.  
The Kohn-Sham propagator \emph{always} has a ``mean-field'' structure,
which means that (in the absence of pairing) the Kohn-Sham occupation
numbers in the normal state are always 0 or 1.
But correlations are certainly included
in $\Gamma[\rho]$!
(In principle, all correlations can be included; in practice, certain
types like long-range particle-hole correlations may be largely omitted
because of limitations of the functional.)
Further, 
$n({\bf k}) = \langle a^\dagger_{\bf k} a^{\phantom{\dagger}}_{\bf k}
            \rangle$ is resolution dependent (not an observable!);
the operator related to experiment is more complicated.
Additional discussion on these issues can be found in
\cite{Furnstahl:2001xq}.

\subsection{Improving empirical EDF's}
\label{subsec:empirical}

\begin{figure}[tbh]
\centering
 \includegraphics*[width=6.in,angle=0.]{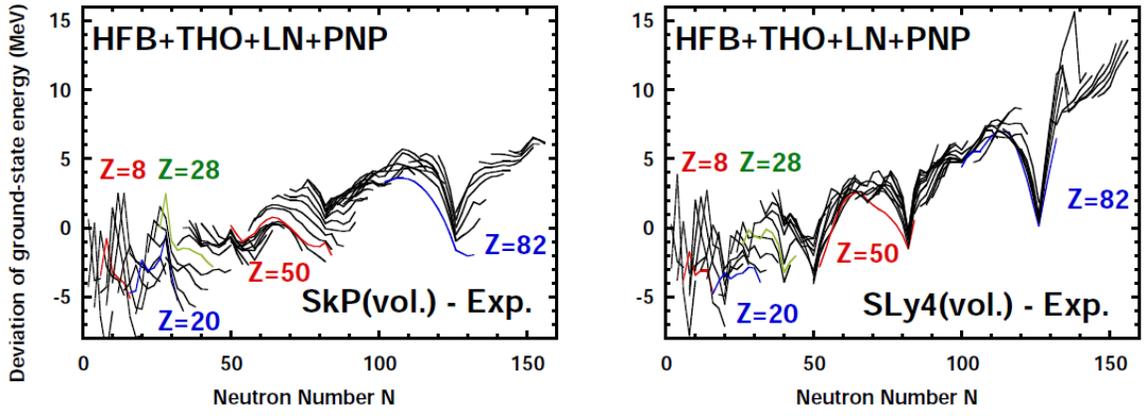}
 \caption{Deviations from experiment of 
 Skyrme-Hartree-Fock EDF predictions for ground-state
 energies by two functionals~\cite{Witekprivate}.}
 \label{fig:skyrmedeviate}       
\end{figure}

\begin{figure}[tbh]
\centering
 \includegraphics*[width=4.in,angle=0.]{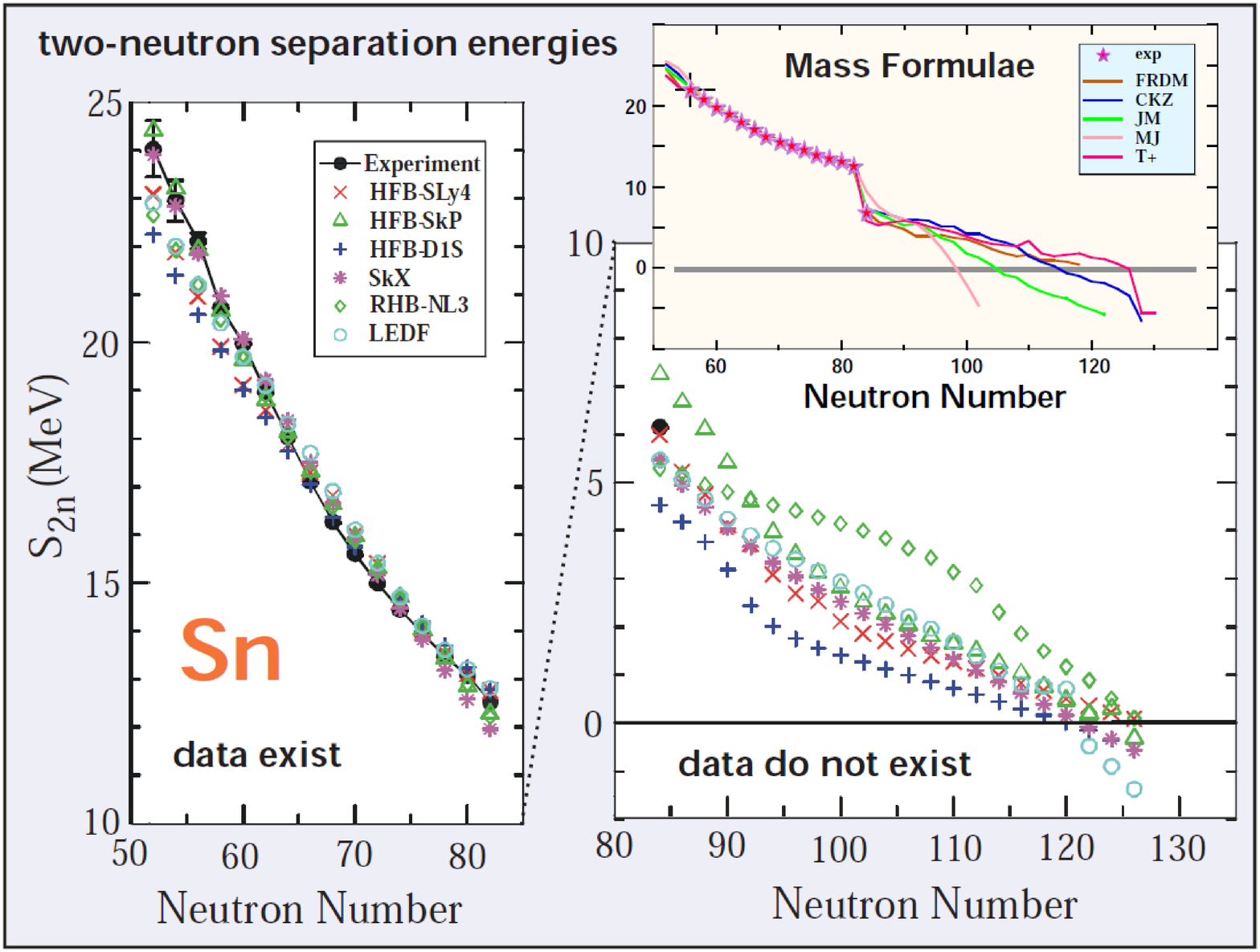}
 \caption{Extrapolation of predictions from
 Skyrme-Hartree-Fock functionals and mass formulas
 for two-neutron separation energies in tin isotopes~\cite{Witekprivate}.}
 \label{fig:tin}       
\end{figure}

The technology for calculating with phenomenological energy density 
functionals such as those
of the Skyrme form is already well developed and still improving.
For example, machinery exists to calculate the entire mass table using
a Skyrme form of the energy functional in a single day
(examples of current capabilities are given in 
Refs.~\cite{Stoitsov:2003pd,unedf:2007,Bertsch:2008yc}).
These calculations
are quite successful, achieving root-mean-squared errors 
better than 2~MeV for the measured nuclides~\cite{Bender:2003jk}.
(See Ref.~\cite{Goriely:2009zz} for a state-of-the-art Gogny
EDF.)
Figures~\ref{fig:skyrmedeviate} and \ref{fig:tin}
show examples of both the successes and limitations of
current Skyrme functionals.
In particular, the trend is toward good reproduction of experimental
data where it exists, but extrapolations toward the
driplines where there is no data become uncertain and sensitive to poorly
constrained parts of the functional.
Improving the reliability of extrapolation is, of course,
a prime motivation for \abinitio\ DFT.

There are substantial and diverse 
efforts in progress to improve the functionals
in use.  Most of these, even if not directly motivated by ties to
microscopic \abinitio\ input, will make such connections more likely.
In particular these efforts include:
\bi
  \I Generalizing the Skyrme functional.  By making the functional a more
  complete representation of the possible physics, one gets closer to 
  the systematics and model independence characteristic of effective
  field theory.  Work in this direction includes
  adding tensor 
  interactions~\cite{Brown:2006cc,Lesinski:2007zz,Zalewski:2008is},
  time-odd components~\cite{Dobaczewski:1995zz,Carlsson:2008gm},
  and gradient and density corrections~\cite{Carlsson:2008gm}.

  \I Correlation studies.
  By studying systematically the errors from empirical functionals,
  clues for the
  important microscopic physics are unearthed.
  One example is the study of odd-even mass 
  differences~\cite{Bertsch:2008yc,Friedman:2008fi}
  and other examples are described in 
  Refs.~\cite{Bertsch:2004us,Kortelainen:2008rp,Toivanen:2008im}.

  \I Non-empirical approaches to pairing.  At present these are
  hybrid calculations using a Skyrme EDF for the particle-hole
  part but a separable low-momentum potential at leading order to describe
  pairing~\cite{Duguet:2007be}.  The predictions of pairing gaps 
  are remarkably
  consistent with those extracted from experiments (from three-point
  formulas using energies of adjacent nuclei as
  input)~\cite{Duguet:2007be,Hebeler:2009dy,Duguet:2009gc}.  A systematic
  investigation of theoretical corrections is underway.

  \I New constraints from \abinitio\ theory.  As noted earlier, 
  microscopic many-body calculations of symmetric nuclear matter are
  less valuable than constraints from stable nuclei.  However, the
  isovector parts of the empirical functionals are much less
  constrained by data and this is where the microscopic calculations
  should be most reliable.   For example, Monte Carlo calculations
  of energies and densities for neutrons in traps (generally
  with AFMC).  The trap
 serves as a variable $\Vext(\rvec)$, so these calculations become benchmarks
 for testing or fitting empirical energy functionals.
 Note that 
 constraints can be significantly more than realized by
 comparison to uniform neutron matter.  Such a
 program is being carried out with
 GFMC/AFMC and NCSM calculations as part of the UNEDF 
 project~\cite{unedfweb}.

 \I Development of EDF without
 explicit underlying Hamiltonians.  The strategy and distinction
 of EDF from DFT are described in Ref.~\cite{Lacroix:2008rj}, as well
 as challenges~\cite{Duguet:2006jg}.
 The insights
 gained and techniques 
 being developed~\cite{Lacroix:2008rj,Bender:2008rn,Duguet:2008rr} 
 will carry over to \abinitio\ DFT.
 
 \I Long-range correlations.  
 With current EDF's, the incorporation of long-range
 correlation effects is handled separately from the functional
 calculations~\cite{Bender:2003jk,Bender:2005vy}.
 The outlook for directly including this physics
 in an \abinitio\ approach with OEP is unclear, but
 there are precedents in Coulomb OEP (e.g., the
 Coulomb ring sum describing the
 high-density expansion of the electron gas can be included).

\I
Merging a variant of the density matrix expansion (DME) and empirical functionals.
The comparison of the DME-I and DME-II curves in Fig.~\ref{fig:ABC_coeffs} 
gives us an estimate of
the truncation error in the expansion applied to the NNN
terms because these prescriptions
differ in the contributions of higher-order terms in the expansion.
Indeed, it has been verified that suppressing these terms by hand brings
the predictions for the $A$ and $B$ coefficients into 
agreement~\cite{Bogner:2008kj}.
The qualitative difference for the NNN-only contribution to $B$ is
large, but the actual coefficient itself is small, so this should
not be alarming.  However, because the combination of $A$ and $B$ and 
the kinetic energy to obtain the nuclear matter energy per particle
involves strong cancellations, the spread in Fig.~\ref{fig:ABC_coeffs}
is large on the scale of nuclear binding energies.

These differences motivate a generalization of the Negele-Vautherin
DME following the discussion in Ref.~\cite{Dobaczewski:2003cy}.
In this approach, the expansion of the scalar density matrix
takes the factorized form
\beqn
  \rho(\Rvec + \frac{\svec}{2}, \Rvec - \frac{\svec}{2})
    = \sum_n \Pi_n(\kf s) \langle \mathcal{O}_n(\Rvec) \rangle
    \;,
\eeqn
where 
\beqn
   \langle \mathcal{O}_n(\Rvec) \rangle = 
   \{
    \rho(\Rvec), \tau(\Rvec), \nabla^2\rho(\Rvec), \cdots
   \}
   \;,
\eeqn
and $\kf$ is a momentum scale typically taken to be
$\kf(\Rvec)$ as in Eq.~\eqref{eq:standardLDA}.
Similar expansions are made for the other components of the density
matrix.
Input from finite nuclei can be used to determine the $\Pi_n$ functions,
which can be viewed as general resummations of the DME expansion. 
A program to merge these developments with empirical Skyrme functionals
is underway~\cite{Biruk}.

\ei
Almost all of the major developments are only recently begun,
so many new results can be expected in the coming years.


\section{Summary and outlook} \label{sec:summary}

The quest for an understanding of \emph{all} nuclei based
on microscopic few-body interactions among nucleons is being attacked
from multiple directions.  
In this review, we have considered \abinitio\ density functional theory
for nuclei, defined as DFT 
based on a realistic nuclear Hamiltonian
(meaning one that accurately reproduces few-body data)
and orbitals that satisfy equations with local Kohn-Sham potentials.
This is an ambitious undertaking and one which will not be
realized without significant further developments in nuclear
structure calculations. 
Nevertheless,
the prospects are good for making microscopic connections to
empirical functionals in the short term and building
on them steadily.
 
We base this optimism in part on the successful advances
in \abinitio\ DFT for Coulomb-based systems by
quantum chemists and condensed matter physicists.  We
believe nuclear physicists will profit from greater attention paid 
to Coulomb DFT developments.
At the same time, we recognize that these advances have taken
many years to realize and may not be readily transferred to
the nuclear domain.
Indeed, we have argued that strong analogies between Coulomb
and nuclear systems are really only apparent
when one considers low-momentum interactions, so that the
correlations induced by more conventional (local or almost local)
interactions are tamed from the outset.
As one strives for greater accuracy, the differences between
atoms or molecules and nuclei are likely to become more significant.

Perdew and collaborators describe for Coulomb DFT a ``Jacob's Ladder''
of increasingly sophisticated ``nonempirical''
functionals stretching toward the heaven of chemical accuracy.
In Ref.~\cite{Perdew:2003}, five rungs of the ladder are 
identified (see also \cite{Perdew:2005aa}):
\begin{enumerate}
  \item The local density approximation (LDA) with both spin
  densities $\rho_\uparrow(\rvec)$ and $\rho_\downarrow(\rvec)$
  as ingredients (usually called LSDA).
  \item The generalized gradient approximation (GGA), which 
  adds dependence on
  $\nabla\rho_\uparrow(\rvec)$ and $\nabla\rho_\downarrow(\rvec)$
  to the LSDA functional.
  \item The meta-GGA adds dependence on (some subset of) 
  $\nabla^2\rho_\uparrow(\rvec)$, $\nabla^2\rho_\downarrow(\rvec)$,
  $\tau_\uparrow(\rvec)$, and $\tau_\downarrow(\rvec)$.
  Note that the kinetic energy densities are actually non-local
  functionals of the ordinary density, although only semi-local
  functionals of the oocupied KS orbitals. 
  \item Hyper-GGA, which includes the exact exchange energy density
  calculated with the (occupied) orbitals.
  \item Full orbital-based DFT, which in addition to exact
  exchange uses unoccupied orbitals.
  For example, this might include the random phase approximation (RPA)
  with Kohn-Sham orbitals to address long-range correlations.
\end{enumerate} 
An important aspect of the overall approach is ``constraint
satisfaction'' as opposed to data fitting, which is why the
functionals are called nonempirical~\cite{Perdew:2005aa}.
Climbing this ladder has been a decades-long effort by quantum chemists
with much development in progress on the last rung.

 We can imagine a corresponding ladder for nuclear
DFT climbing toward a universal energy density functional
tied to microscopic nuclear Hamiltonians (and ultimately to
QCD) that predicts known nuclear properties more accurately than
currently possible and robustly predicts unknown properties
with plausible theoretical error estimates. 
The rungs might look like this:
\begin{enumerate}
 \item Present-day Skyrme EDF's, which are mostly empirical
 (i.e., fit to properties of selected medium-to-heavy nuclei),
   such as described in Ref.~\cite{Bender:2003jk}.
 \item Generalized Skyrme interactions,
     with additional gradient and density dependences,
     with new constraints from microscopic theory (e.g., neutron drops).
 \item Functionals that merge the long-range parts of the microscopic
 NN and NNN interactions
 prescribed by chiral EFT, converted to semi-local form with
 a variant of the density matrix expansion (DME),
 with a Skyrme functional.  The short-range parameters
 should be refit to nuclear properties and theoretical constraints.
 \item A complete functional based on a variant of DME applied 
 to a low-momentum
 potential that is evolved from chiral EFT NN and NNN interactions.
 \item Full orbital-based DFT based on low-momentum interactions.
\end{enumerate}
This ladder is tied to our restriction to local Kohn-Sham potentials
and so naturally builds on Skyrme EDF's.  An alternative ladder
could build on non-local potentials (e.g., on Gogny-type EDF's)
that arise from derivatives with respect to density
matrices rather than densities;
see Refs.~\cite{Duguet:2006jg,Duguet:2009gc} for discussions
along these lines. 
Just as in the Coulomb ladder, one hopes for monotonic improvements
at each rung, but this may not always be the case. 
(Of course, rungs will shift or more will be added as progress
is made.)  

Rather than climb slowly over the course of decades, nuclear physicists
are trying to catch up rapidly by attacking all of the higher
rungs in parallel~\cite{unedf:2007,unedfweb,fidiproweb}.
As described in Section~\ref{subsec:empirical},
there are extensive ongoing efforts 
to explore the second rung, including
more general density dependence and higher-order
gradients~\cite{unedfweb,fidiproweb},
while projects on both rungs three and four have recently 
been started as part of the UNEDF project~\cite{unedf:2007}.
The last rung is in the exploratory stage with tests started
in one dimension with nuclear-like interactions for both
self-bound and trapped systems to explore the
accuracy of the KLI (and related) approximations compared to a
full OEP treatment of such systems
and to establish the limitations of semi-local expansions such as the DME. 

While the last two rungs can be purely predictive if
only few-body data is used to determine the input Hamiltonian, it
is likely that fine-tuning to heavier nuclei
will be needed to reach the accuracy goals.
This is because
even the best \abinitio\ nuclear structure calculations
at present only achieve about 1\% accuracy in ground-state energies
for a small fraction of the table of nuclides, while
nuclear EDF's such as Skyrme functionals easily surpass this
over most of the table.  
Indeed, the comparatively high accuracy of
nuclear EDF may imply that a different organization of the problem
(e.g., based on a finite-density effective field theory of nuclei)
may be needed.


There are many important open questions to be addressed in
the course of these projects (see also Ref.~\cite{Duguet:2006jg}).
Among them are those highlighted in Section~\ref{sec:nuclear} on symmetry
breaking and restoration, single-particle levels, and pairing; these
will be relevant for all rungs of the ladder.
At the top of the ladder we can ask:
Which of the problems encountered in Coulomb DFT that motivate
orbital-based functionals have analogs in nuclear DFT?  E.g., 
what is the impact of self-interaction and missing
derivative discontinuities?
Another class of open questions concerns the input Hamiltonian:
For low-momentum potentials, how accurate will many-body perturbation
theory be?  What
  corrections/summations will
  be needed to reach the desired accuracy for nuclear DFT?
Are four-body forces necessary?
The list of problems yet to solve might be intimidating, especially
given the many decades already spent attacking nuclear structure, 
but the coherence of the current effort and the
dramatic advances of just the last few years
give hope that they can be overcome.

We have taken a broad view at what \abinitio\ density functional
theory could be but
our discussion has been far from an exhaustive treatment.
By focusing our discussion on orbital-based DFT for nonrelativistic
Hamiltonians we have excluded various alternative paths
to \abinitio\ DFT.
Here are some of the other approaches 
that are relevant to the wider nuclear DFT effort:
\bi
  \I The perturbative in-medium results from low-momentum potentials suggest
  that an alternative EFT power counting may be 
  appropriate at nuclear matter densities.
Kaiser and collaborators
have proposed a perturbative chiral EFT approach to nuclear matter
and then to finite nuclei through an energy
functional~\cite{Kaiser:2001jx,Kaiser:2002jz,Fritsch:2004nx,Finelli:2005ni,Finelli:2006xh}
(see also \cite{Lutz:1999vc}).
They consider Lagrangians both for nucleons and pions and
for nucleons, pions, and $\Delta$'s, and fit parameters to nuclear
saturation properties.  
They construct a loop expansion for the nuclear matter energy
per particle, which leads to an energy expansion of the form
\beqn 
  E(\kf) = \sum_{n=2}^\infty \kf^n \, f_n(\kf/m_\pi,\Delta/m_\pi) \ ,
  \qquad [\Delta = M_\Delta - M_N \approx 300\,\mbox{MeV}]
\eeqn
where each $f_n$ is determined from a finite number of
in-medium Feynman diagrams, which incorporate the long-distance
physics.
All powers of $\kf/m_\pi$ and $\Delta/m_\pi$ are kept in the $f_n$'s
because these ratios are not small quantities \cite{Kaiser:2006tu}.
A semi-quantitative description of nuclear matter is
found even with just the lowest two terms without 
$\Delta$'s and adding $\Delta$'s brings uniform improvement
(e.g., in the neutron matter equation of state).
There are open questions about power counting and convergence,
but many promising avenues to pursue.
By applying the DME in momentum
space to this expansion they 
derive a Skyrme-like energy functional
for nuclei, which has also been merged
with a treatment of strong scalar and vector
mean fields.~\cite{Kaiser:2002jz,Kaiser:2003uh,Fritsch:2004nx,Finelli:2005ni,Finelli:2006xh}.

\I
The Superfluid Local Density Approximation (SLDA) and extensions developed
by Bulgac and collaborators~\cite{Yu:2002kc,Bulgac:2003ti,Bulgac:2007wm,Bulgac:2008qx}
have been applied with great success to systems of trapped cold
fermionic atoms with large scattering lengths.  
The SLDA is an extension of DFT to superfluid
system that uses a semi-local energy density with
parameters determined by matching to numerically accurate Monte Carlo
simulations of uniform systems.
Extensive further development of the SLDA for nuclear systems is in progress.
  
  \I 
We have noted that the Kohn-Sham DFT emphasis on locality
in coordinate space is not so natural in many-body formulations.
In particular, the nuclear shell model is naturally considered
in the space of orbitals.
Papenbrock has applied the Legendre transform
formulation of DFT  to energy functionals based on
shell-model occupation numbers.  That is, a functional of the
density now becomes a function of the occupation numbers in the model
space.  He and his collaborators have shown the usefulness of this in solvable
models (the pairing Hamiltonian~\cite{Papenbrock:2006hg}
and the three-level Lipkin model~\cite{Bertolli:2008uq}).
The functional is orbital-based and thus non-local in density.
For applications to nuclei, the idea is to generalize the Duflo-Zuker
mass formula~\cite{Duflo:1994tr,Duflo:1995ep} to a functional of occupation numbers, with values
determined from the minimization of the functional. 
With 10 parameters, an RMS value of deviation for masses of
about 2000 nuclei is just over 1\,MeV.

  \I 
Schwenk and Polonyi have proposed
an alternative approach to \abinitio\ but orbit-free DFT 
(i.e., not Kohn-Sham) using a clever
renormalization group (RG) evolution~\cite{Schwenk:2004hm}.  The idea is to
introduce an effective action for a nucleus with a low-momentum
interaction included with a multiplicative factor $\lambda$ 
and a confining background potential (e.g., a harmonic oscillator
trap) with a factor $(1-\lambda)$.  As $\lambda$ flows from 0 to 1,
the background potential is turned off and the interactions, with
associated many-body correlations, are turned on.  This evolution
is dictated by an RG equation in $\lambda$. 
Work is in progress to implement this approach for
a realistic nucleus~\cite{Braun:2009aa}. 

\I There is a large body of work on covariant approaches to nuclear
DFT, which 
cannot be adequately summarized here.  Fortunately, there
are many reviews that highlight DFT connections, 
including
Refs.~\cite{Ring:1996qi,Serot:1997xg,Furnstahl:2000in,Furnstahl:2003cd,%
Vretenar:2003bt,Vretenar:2008uq}, to supplement our brief
discussion. 
Covariant DFT invokes a different organization of the nuclear
many-body problem that is particularly compelling because of the
coupling of spin-orbit and central components dictated by 
relativity.
In particular, the characteristic feature of relativistic approaches
to nuclei is large isoscalar Lorentz scalar and vector mean fields, 
several hundred MeV in magnitude at nuclear densities, which
cancel to provide nuclear binding but add to produce spin-orbit
splittings.

Relativistic mean-field models were originally motivated as the
Hartree approximation to a more complete theory, 
but a more recent
interpretation of the largely empirical functionals
is as relativistic versions of Kohn-Sham DFT~\cite{Serot:1997xg}.
The mean fields serve as local Kohn-Sham potentials in Dirac
equations for the orbitals.
This picture is consistent with a microscopic derivation within
an effective action formulation as in Section~\ref{sec:lt}
with sources coupled to each of the relativistic
densities or currents or to corresponding meson-like auxiliary
fields~\cite{Furnstahl:2003cd}
(this is in contrast to implementations of
relativistic electronic DFT for heavy atoms, which include only
the vector four-current~\cite{Fiolhais:2003}). 
Application of a loop expansion with proper renormalization
gives the possibility of a systematic microscopic 
expansion~\cite{McIntire:2007ud,Hu:2007na}.
Other ongoing efforts to provide more \abinitio\ aspects
including a connection to free-space NN scattering derive
relativistic EDF with density dependent
meson-baryon vertex functionals, derived from the
Dirac-Brueckner
theory (see Refs.~\cite{Lenske:2006zz,Lenske:2007zz,Hirose:2007ar}
and references cited therein).

These more microscopic approaches do not at present have the close connection
to few-body data and to \abinitio\ structure calculations of light
nuclei that is possible with the non-relativistic approaches
we have considered.  
However, the wide phenomenological successes and complementary
nature of covariant nuclear
DFT motivate further work toward incorporating additional 
microscopic constraints.
In the process, most of the challenges confronting nonrelativistic
DFT discussed above are being attacked in parallel for covariant DFT.
This includes the issues of realistic pairing
interactions~\cite{Serra:2002pv,Tian:2009zzc,Tian:2009zw}, 
coupling to particle-hole vibrations~\cite{Litvinova:2007gg},
and symmetry breaking/restoration~\cite{Sheikh:2002zz,Niksic:2006kv,Yao:2009ry}.

\ei
Each of these avenues has the potential to make a significant
impact on nuclear DFT.

\section*{Acknowledgments}
We thank S. Bogner, T.~Duguet, and A. Schwenk for useful discussions.
This work was supported in part by the National Science 
Foundation under Grant No.~PHY--0653312 and  
the UNEDF SciDAC Collaboration under DOE Grant 
DE-FC02-07ER41457.

\addcontentsline{toc}{section}{References}

\bibliographystyle{h-elsevier_new} 

\bibliography{dft_refs} 

\end{document}